\documentclass  [twocolumn, amsmath,amssymb,aps,showpacs, superscriptaddress,longbibliography]{revtex4-1}

\usepackage{graphicx}
\usepackage{dcolumn}
\usepackage{bm}
\usepackage{braket,amsmath,amssymb,bm}
\usepackage{color}
\usepackage{siunitx}
\usepackage{float}
\usepackage{hyperref}
\usepackage{lineno}
\usepackage{multirow}
\usepackage{tabularx}
\usepackage{enumitem}
\usepackage{etoolbox}
\usepackage[version=4]{mhchem}

\newcommand{\Rev}[1]{\textcolor{black}{#1}}

\begin{document}

\title{Glass-patternable notch-shaped microwave architecture for on-chip spin detection in biological samples}

\author{Keisuke Oshimi}
 \affiliation{Department of Chemistry, Graduate School of Natural Science and Technology, Okayama University, Okayama 700-8530, Japan}
 \affiliation{Department of Chemistry, Graduate School of Science, Osaka City University, Osaka 558-8585, Japan}
 
\author{Yushi Nishimura}
 \affiliation{Department of Chemistry, Graduate School of Science, Osaka City University, Osaka 558-8585, Japan}
 \affiliation{Institute for Quantum Life Science, Quantum Life and Medical Science Directorate, National Institutes for Quantum Science and Technology, Anagawa, Inage-ku, Chiba, 263-8555, Japan.}

\author{Tsutomu Matsubara}
 \affiliation{Department of Anatomy and Regenerative Biology, Graduate School of Medicine, Osaka City University, 1-4-3 Asahi-machi, Abeno, Osaka, 545-8585, Japan}

\author{Masuaki Tanaka}
 \affiliation{Department of Electrical and Information Engineering, Graduate School of Engineering, Osaka City University, Osaka 558-8585, Japan}

\author{Eiji Shikoh}
 \affiliation{Department of Electrical and Information Engineering, Graduate School of Engineering, Osaka City University, Osaka 558-8585, Japan}

\author{Li Zhao}
 \affiliation{State Key Laboratory of Radiation Medicine and Protection, School for Radiological and Interdisciplinary Sciences (RAD-X) and Collaborative Innovation Center of Radiation Medicine of Jiangsu Higher Education Institutions, Soochow University, Suzhou 215123, P. R. China.}

\author{Yajuan Zou}
 \affiliation{Department of Chemistry, Graduate School of Natural Science and Technology, Okayama University, Okayama 700-8530, Japan}
 \affiliation{Graduate School of Human and Environmental Studies, Kyoto University, Kyoto 606-8501, Japan}

\author{Naoki Komatsu}
 \affiliation{Graduate School of Human and Environmental Studies, Kyoto University, Kyoto 606-8501, Japan}

\author{Yuta Ikado}
 \affiliation{Department of Chemistry, Graduate School of Natural Science and Technology, Okayama University, Okayama 700-8530, Japan}
 
\author{Yuka Takezawa}
 \affiliation{Department of Human Life Science, Graduate School of Food and Human Life Science, Osaka City University, Osaka 558-8585, Japan}

\author{Eriko Kage-Nakadai}
 \affiliation{Department of Human Life Science, Graduate School of Food and Human Life Science, Osaka City University, Osaka 558-8585, Japan}

\author{Yumi Izutsu}
 \affiliation{Department of Biology, Faculty of Science, Niigata University, Niigata 950-2181, Japan}

\author{Katsutoshi Yoshizato}
 \affiliation{Synthetic biology laboratory, Graduate school of medicine, Osaka City University, Osaka, 545-8585, Japan}

\author{Saho Morita}
 \affiliation{Department of Biomolecular Engineering, Graduate School of Engineering, Nagoya University, Nagoya 464-8603, Japan}

\author{Masato Tokunaga}
 \affiliation{Department of Biomolecular Engineering, Graduate School of Engineering, Nagoya University, Nagoya 464-8603, Japan}

\author{Hiroshi Yukawa}
 \affiliation{Department of Biomolecular Engineering, Graduate School of Engineering, Nagoya University, Nagoya 464-8603, Japan}
 \affiliation{Institute of Nano-Life-Systems, Institutes of Innovation for Future Society, Nagoya University, Nagoya 464-8603, Japan}
 \affiliation{Institute for Quantum Life Science, Quantum Life and Medical Science Directorate, National Institutes for Quantum Science and Technology, Anagawa, Inage-ku, Chiba, 263-8555, Japan.}

\author{Yoshinobu Baba}
 \affiliation{Department of Biomolecular Engineering, Graduate School of Engineering, Nagoya University, Nagoya 464-8603, Japan}
 \affiliation{Institute of Nano-Life-Systems, Institutes of Innovation for Future Society, Nagoya University, Nagoya 464-8603, Japan}
 \affiliation{Institute for Quantum Life Science, Quantum Life and Medical Science Directorate, National Institutes for Quantum Science and Technology, Anagawa, Inage-ku, Chiba, 263-8555, Japan.}

\author{Yoshio Teki}
 \affiliation{Department of Chemistry, Graduate School of Science, Osaka City University, Osaka 558-8585, Japan}

\author{Masazumi Fujiwara}
 \affiliation{Department of Chemistry, Graduate School of Natural Science and Technology, Okayama University, Okayama 700-8530, Japan}
 \affiliation{Department of Chemistry, Graduate School of Science, Osaka City University, Osaka 558-8585, Japan}

\begin{abstract}
We report a notch-shaped coplanar microwave waveguide antenna on a glass plate designed for on-chip detection of optically detected magnetic resonance (ODMR) of fluorescent nanodiamonds (NDs). 
A lithographically patterned thin wire at the center of the notch area in the coplanar waveguide realizes a millimeter-scale ODMR detection area ($1.5\times2.0 \ \si{\mm}^2$) and gigahertz-broadband characteristics with low reflection ($\sim 8$\%).
The ODMR signal intensity in the detection area is quantitatively predictable by numerical simulation. 
Using this chip device, we demonstrate a uniform ODMR signal intensity over the detection area for cells, tissue, and worms. 
The present demonstration of a chip-based microwave architecture will enable scalable chip integration of ODMR-based quantum sensing technology into various bioassay platforms.
\end{abstract}
\maketitle

\twocolumngrid

\section{Introduction}

The miniaturization and compactification of analytical systems into chip devices are crucial for achieving efficient high-throughput and highly sensitive bioassays with significantly reduced sample numbers and volumes~\cite{ngo2014dna, tani2004chip, rothbauer2018recent}. 
Implementing small, sensitive, and multimodal sensors into a detection area is key for realizing miniaturized assay devices.
Ultrasensitive multimodal nanometer-sized quantum sensors based on fluorescent nanodiamonds (NDs) have been applied to various biological systems, including biomolecules~\cite{ferrier2009microwave, ziem2013highly, wackerlig2016applications, rendler2017optical, miller2020spin, haziza2017fluorescent}, cells~\cite{kucsko2013nanometre, simpson2017non, toraille2018optical, claveau2018fluorescent, nie2021quantum}, and small organisms~\cite{davis2018mapping, van2020evaluation, fujiwara2020real}. 
Nanodiamonds have a low cytotoxicity~\cite{zhu2012biocompatibility, krueger2008new, mohan2010vivo}, and their surfaces can be functionalized for biological targeting and labeling~\cite{hemelaar2017generally, sotoma2018highly, reina2019chemical}. 
In most workflows, they first bind to target molecules or cells to produce labeled samples that are subsequently introduced into the detection area for multimodal quantum sensing.
The sensing multimodality of NDs results from the dependence of electron-spin-resonance frequencies of nitrogen vacancy (NV) centers on the magnetic field~\cite{rondin2014magnetometry, maclaurin2013nanoscale, horowitz2012electron}, electric field~\cite{dolde2011electric, iwasaki2017direct, bian2021nanoscale}, and temperature~\cite{neumann2013high, fujiwara2021diamond, wang2018magnetic}. 
It is detected through the modulation of fluorescence intensity by microwave excitation and referred to as optically detected magnetic resonance (ODMR).
As ODMR uses fluorescence detection and microwave excitation, scaling down an ODMR-based bioassay into a chip device relies on the miniaturization of both these technologies. 
By exploiting the suitability of optical methods at the submillimeter scale, several fluorescence-detection techniques and measurement systems have been used in chip devices~\cite{mazurczyk2006novel, dochow2013raman, qi2018rotational}, which are readily utilized for ODMR. 
For microwave manipulation, several architectures have been reported for on-chip thermal control, analyte sensing, and spin detection~\cite{marchiarullo2013low, yesiloz2015label,wong2016microwave, chen2020universal}.
However, the miniaturization of microwave excitation circuits into ODMR chip devices has been challenging, owing to difficulties in 
\Rev{providing broad-band and large-area microwave excitation in a chip-integrable configuration, while having} 
high-efficiency fluorescence photon collection in aqueous environments.

Several types of microwave circuitry have been used as emission antennas for ODMR measurements, including thin wires~\cite{rendler2017optical, fukushige2020identification}, coplanar waveguides~\cite{sadzak2018coupling, jia2018ultra}, coils~\cite{igarashi2012real, ajoy2020room}, and omega-shaped patterns~\cite{choi2020probing, miller2020spin, bolshedvorskii2017single}.
Thin wires or coplanar waveguides produce a strong excitation intensity around the transmission wires or  gaps; hence, the ODMR detection area is limited to $\sim 100 \ \si{\um}$.
Coils can provide a more spatially uniform excitation area, but their excitation intensity is limited.
Further, the observed ODMR signal intensity substantially depends on the coupling losses and sample-antenna distances that vary experiment by experiment, hindering the scalable engineering of ODMR chip devices.
Omega-shaped patterns have recently been reported as an efficient and designable platform for microwave delivery, enabling paper-based lateral-flow assays with portable microwave devices~\cite{miller2020spin}.
However, their frequency bandwidth is approximately 70 MHz, and the 5 cm antenna size cannot be further scaled down owing to the cavity resonance effect, which limits ODMR applications that require a broad bandwidth of up to 400 MHz, such as vector magnetometry~\cite{schloss2018simultaneous, clevenson2018robust, igarashi2020threeD, tsukamoto2021vector} and nanoscale NMR spectroscopy~\cite{devience2015nanoscale, smits2019two, holzgrafe2020nanoscale}.
None of these circuity patterns fully satisfy the requirements of a millimeter-scale detection area, gigahertz bandwidth, and scalability for multiple detection areas within a small volume, all of which are necessary for the chip integration of diamond ODMR technology.

This study proposes a notch-shaped coplanar antenna on a glass plate that achieves a millimeter-scale ODMR detection area and broadband low-reflection microwave characteristics. 
A lithographically patterned thin wire at the center of the notch area in the coplanar waveguide generates spatially uniform and broadband excitation pattern of microwave magnetic field.
We demonstrate ODMR measurements over a $1.5\times2.0 \ \si{\mm}^2$ chip area, and even quantitatively predicting the ODMR signal intensity in the antenna detection area.
By integrating this antenna within multiple bioassay platforms (including dishes and plates), uniform ODMR detection is observed for NDs labeled in cultured cells, tissue, and nematode worms.
The scalability of the present microwave architecture facilitates the integration of ODMR detection areas within large-scale multi-well plates. 
Thus, we expect the present concept of a notch-shaped coplanar antenna to greatly extend the applicability of diamond quantum sensing to chip-based bioassays. 

\section{Methods}

\subsection{Numerical modeling of chip devices}
The chip device was numerically designed using the COMSOL finite element method (FEM) software package with an RF module. 
It consists of an antenna-patterned coverslip with plastic supports. 
Numerical simulations were performed for the coverslip, neglecting the effect of the plastic supports on the overall microwave characteristics (as confirmed in Fig. S1). 
We assumed a coverslip thickness of 0.17 mm and an infinitely thin gold-patterned layer on one side. 
The borosilicate glass coverslip was simulated with a relative permittivity $\epsilon = 4.6$, relative permeability $\mu = 1.0$, and electrical conductivity $\sigma = 0.0~\si{S/m}$.
The thin gold layer was considered as a perfect electric conductor (PEC).
A 70 mm radius sphere, filled with air ($\epsilon = 1.0, \mu = 1.0, \sigma = 0.0~\si{S/m}$), surrounded the coverslip, and a perfectly matched layer with an absorption constant of $10^{-6}$ was set at the sphere boundary. 
The mesh was created using the software's physics-controlled mode. 
The mesh size was chosen to ensure the convergence of the simulation results (Fig. S2).

\subsection{Device fabrication}
Borosilicate glass coverslips (Matsunami, with dimensions of $22 \times 22~\si{mm^2}$ for Part No. C022221 and $30 \times 40~\si{mm^2}$ sizes for Part No. C030401, and a thickness of 0.13--0.17 mm) were cleaned with an alkali detergent.
Photomasks were fabricated and used for transferring the patterns to coverslips using standard photolithography, as follows.
Gold was deposited on coverslips with a 100 nm thickness (chromium buffer layer of $\sim$ 3 nm).
A photoresist (MicroChemicals, AZ 1500) was spin-coated at 4000--4500 rpm and baked at 95$\si{\degreeCelsius}$ for 90 s.
The gold-coated coverslips were then exposed to a mercury lamp for 18 s, and the resist was removed. 
Gold and chromium were removed using appropriate etchants (AURUM-302 for gold and Cr-201 for chromium, both from Kanto Chemical Co.). 
The coverslips were washed with acetone to remove residual photoresist and bonded with dishes or multiwell plates using polydimethylsiloxane (75~$\si{\degreeCelsius}$, 1 h).
Most of the experiments described below used 35 mm plastic dishes with an identical central hole diameter of 14 mm for ease of handling. 

\subsection{Scattering parameter characterization}
We characterized the transmission and reflection of the antenna in the microwave frequency range by measuring the scattering parameters (S-parameters). 
Of the four two-port S-parameters, $S_{11}$ and $S_{21}$ were determined in both the simulation and the experiments.
In this study, we expressed the reflection and transmission properties of the devices in dB units as: 
\begin{equation}
    \begin{aligned}
    S_{11} &= 10\log_{10}\frac{P_{\rm reflected}}{P_{\rm incident}} \\ 
    S_{21} &= 10\log_{10}\frac{P_{\rm transmission}}{P_{\rm incident}}
    \end{aligned}
    \label{eq:s-para}
\end{equation}
\noindent
where $P_{\rm incident}$, $P_{\rm reflected}$, and $P_{\rm transmission}$ denote the incident, reflected, and transmitted powers, respectively~\cite{pozar2005}.
These parameters were determined using COMSOL in the simulation and experimentally using a network analyzer (MS46122B, Anritsu).
    
\begin{figure*}[th!]
 \centering
 \includegraphics[width=15.8cm]{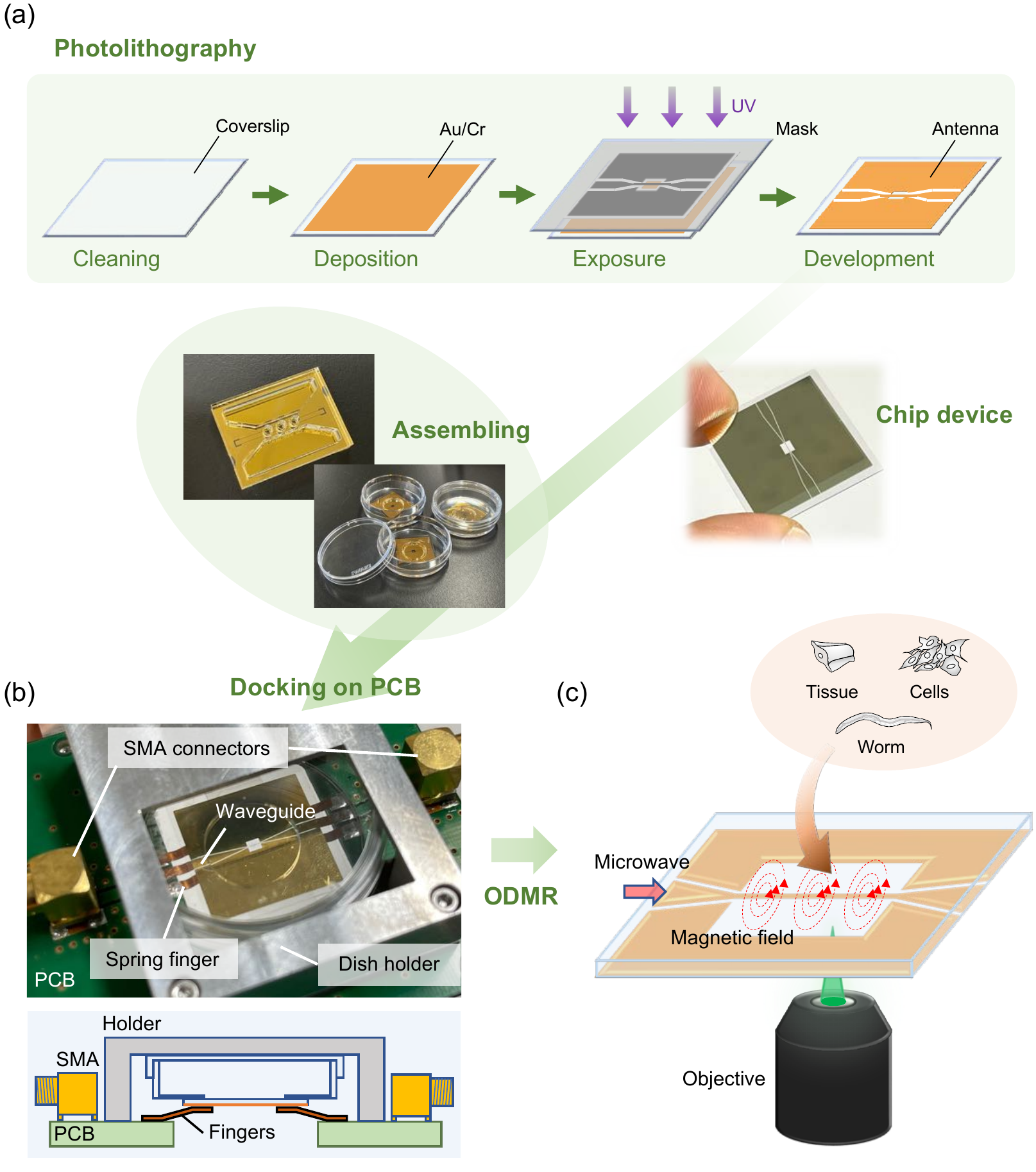}
 \caption{(a) Chip fabrication scheme, including photolithography and bonding to bioassay platforms (35 mm dishes and three-well plates). 
 (b) Photograph of a dish-style chip device docked to a PCB circuit board \Rev{(top), and a schematic side-view diagram of the board and copper spring fingers (bottom).}
 (c) Schematic spatial arrangement of biological samples relative to the waveguide pattern and microscope objective.
}
 \label{figure 1}
\end{figure*}

\subsection{ND spin-coating on the coverslips}
To perform ODMR measurements on NDs directly deposited onto the coverslips, we spin-coated 20 $\si{\uL}$ of a commercially available ND suspension (Adámas Nanotechnologies, ND-NV-100 nm-Hi) onto the antenna-patterned coverslips at 2000 rpm for approximately 30 s.
The NDs were spin-coated on either the Au-Cr layer side or the reverse glass side to determine the microwave magnetic field magnitude $|\boldsymbol{B}|$.
\subsection{ND labeling of biological samples}
For the cell experiments, HeLa cells and adipose tissue-derived stem cells (ASCs) were cultivated in dish-style chip devices. 
HeLa cells and ASCs were labeled with NDs, as previously described ~\cite{nishimura2021wide,yukawa2020quantum}.
The NDs with concentrations of 10 $\si{\ug/\mL}$ and 20 $\si{\ug/\mL}$ (Adámas Nanotechnologies, ND-NV-100 nm-Hi) were added to the culture media of HeLa cells and ASCs, respectively. 
The cells were incubated at 37 $\si{\degreeCelsius}$, 5\% $\mathrm{CO_2}$ for 24 h, washed three times with phosphate-buffered saline, and immersed in the respective culture media. 

Tadpoles were anesthetized using MS222~\cite{izutsu2019skin, izutsu2005analyses}, and polyglycerol-grafted NDs (PG-NDs)~\cite{reina2019chemical,zhao2011chromatographic} were introduced into the tail by intramuscular injection using a syringe. 
The tail tips were isolated and placed in the measurement area of the chip device.

For the worm experiments, young adult \textit{Caenorhabditis elegans} (\textit{C. elegans}) were used. 
The wild-type \textit{C. elegans} strain Bristol N2 was obtained from the Caenorhabditis Genetics Center (Minneapolis, MN, USA) and maintained following the standard protocol~\cite{brenner1974genetics}. 
Next, PG-NDs were microinjected into the gonads as previously described ~\cite{fujiwara2020real}. 
The ND-labeled \textit{C. elegans} were placed on the measurement area of the chip devices with agar pads.

\subsection{ODMR experiments}
The ODMR was measured using our home-built confocal fluorescence microscope equipped with a microwave excitation system~\cite{yukawa2020quantum, tsukahara2019removing, fujiwara2020real, fujiwara2020real-1}.
Microwaves were generated by a signal generator (Rohde \& Schwarz, SMB100A) and sent to a radiofrequency switch (Mini-circuit, ZYSWA-2-50DRS and General Microwave, F9160) triggered by a bit-pattern generator (SpinCore, PBESR-PRO-300). 
The signal was then amplified using a 45 dB amplifier (Mini-circuit, ZHL-16W-43+).
\Rev{The fluorescence signal was detected by a single photon counting module (Excelitas, SPCM-AQRH-14) using conventional confocal microscopy~\cite{fujiwara2019monitoring,nishimura2021wide}.}
The ODMR signal was measured in both the continuous wave (CW) and pulsed modes. 
In the CW mode, the microwave excitation was gated ON and OFF to suppress noise. 
The CW ODMR measurements were performed on the NDs placed directly on coverslips and for the NDs in biological samples, as described above. 
In the pulsed mode, Rabi measurements were performed to determine the duration of the $\pi$-pulse for NV electron spins.
The pulsed ODMR measurements were performed only on the NDs on the coverslips.
In addition, in the pulsed measurements, external magnetic fields were applied using neodymium magnets to lift the degeneracy of the magnetic sublevels to allow clear Rabi-nutation detection.

\begin{figure*}[th!]
 \centering
 \includegraphics[width=15.8cm]{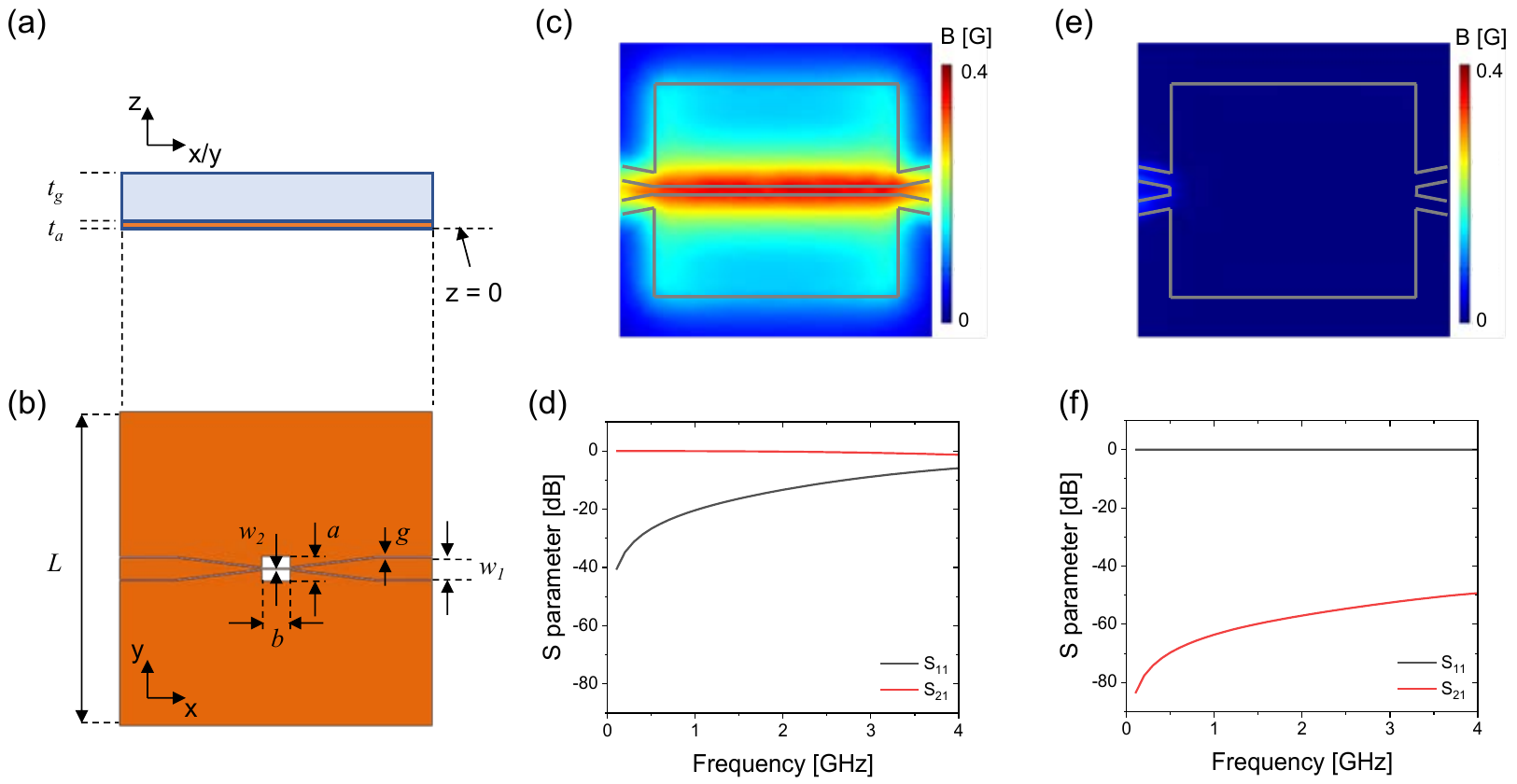}
 \caption{Geometrical structure of the antenna coverslip in the (a) $xz$ and (b) $xy$ planes.  The coverslip thickness is denoted as $t_g$, and $t_a$ is that of the Au-Cr layer. $L$ is the side length, and $w_1$ and $w_2$ are the widths of the transmission line at the coverslip edges and at the central thin wire, respectively.
 The notch height and width are $a$ and $b$, respectively.  
 In this particular case, $L = 22 \ \si{\mm}$, $w_1 = 1.5 \ \si{\mm}$, $w_2 = 50 \ \si{\um}$, $a = 1.75 \ \si{\mm}$, $b = 2.0 \ \si{\mm}$, $t_g = 0.17 \ \si{\mm}$, and $t_a = 100 \ \si{\nm}$. The Au-Cr layer is approximated as an infinitely thin PEC in the simulation (i.e., $t_a \sim 0 \ \si{\nm}$).
 (c) The computed spatial distribution of $\left| B \right|$ at 2.87 GHz on the Au-Cr layer ($z = 0$ plane) and (d)  corresponding simulated $S_{\rm 11}$ (black) and $S_{\rm 21}$ (red) spectra in the present notch-shaped coplanar waveguide structure.
 The $\left| B \right|$ distribution and $S$-parameter spectra with the central thin-wire transmission line removed are shown in (e) and (f). 
 }
 \label{figure 2}
\end{figure*}

\begin{figure}[th!]
 \centering
 \includegraphics[width=8.4 cm]{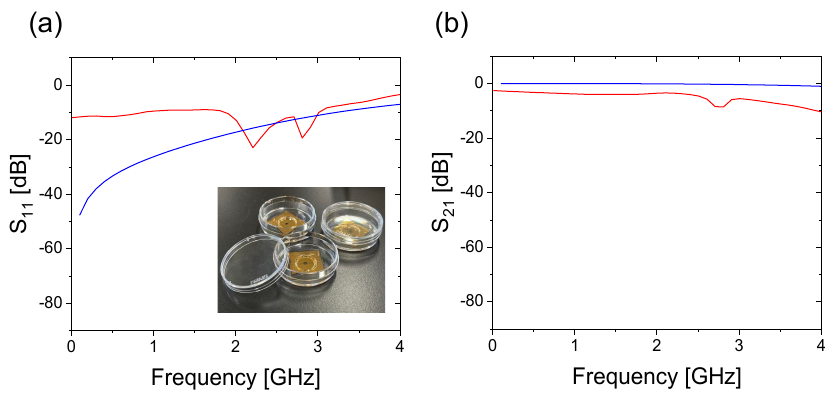}
 \caption{S-parameter spectra of the fabricated devices for (a) $S_{\rm 11}$ and (b) $S_{\rm 21}$  (blue: simulation; red: experiment).
 }
 \label{figure 3}
\end{figure}

\section{Results}
\label{sec3}
\subsection{Numerical modeling of the chip devices}
\label{sec3.1}

Figure~\ref{figure 1}(a) shows the fabrication workflow for our chip devices, from photolithography to coverslip bonding to various bioassay platforms (35 mm dishes and three-well plates in this figure).
We first designed a microwave antenna pattern on a coverslip by numerical simulation and transferred the designed pattern onto coverslips using standard photolithography. 
The fabricated antenna-patterned coverslips were then bonded with conventional bioassay platforms such as dishes, multi-well plates, and microscope slides designed for efficient handling.
Finally, these chip devices were docked~\cite{fujiwara2020real} on our custom printed circuit board (PCB) comprising a coplanar waveguide and connectors (see Figs.~\ref{figure 1}(b) and (c)).
We fabricated the chip devices in a dish, multi-well plate, and microscope slide. We primarily used the dish-style devices in the following experiments because of their ease of handling.

We simulated microwave antenna characteristics using FEM to optimize the structures for microwave irradiation.
The basic design of the proposed antenna is illustrated in Figs.~\ref{figure 2}(a) and (b). 
The device is a notch-shaped coplanar waveguide with a tapered transmission line.
\Rev{The peripheral gold layer coating serves to adequately ground the device to the PCB.} 
The multiple geometry parameters were determined for efficient input and large-area microwave irradiation at 2.87 GHz. 
The borosilicate glass coverslips had a typical thickness of $t_g = 0.17$ mm, and a thin gold layer was deposited with a thickness of $t_a \sim 100$ nm (with a chromium buffer layer of a few nanometers).
The simulation assumed a coverslip relative permittivity $\epsilon = 4.6$ and an infinitesimally thin PEC layer for the thin gold film (see Methods for more information on the material parameters).
With these experimental constraints, we determined the width of the transmission wire at the coverslip edges to be $w_1 = 1.5$ mm and the gap width between the transmission wire and lateral ground region as $g = 0.1$ mm, giving a 50 \si{\ohm} impedance to match that of the coaxial cables.
The notch shape is characterized by its width $a$ and height $b$ and by the width of the central thin transmission wire $w_2$. 
For the purpose of ODMR detection in biological samples, we find that $a = 1.5$ mm, $b = 2.0$ mm, and $w_2 = 50~\si{\um}$ is a good trade-off between the area size and magnetic field intensity, as described below.
The tapered structures connecting the coplanar waveguides and the notch area act as an impedance transformer that raises the 50 \si{\ohm} impedance to the higher impedance of the central thin transmission wire (approximately 300 \si{\ohm})~\cite{sefa2011analysis,perez2018simplified}.

With this geometry, we calculated the spatial distribution of the $\left| B \right|$ and $S$-parameter spectra of the coverslip antenna, as shown in Figs.~\ref{figure 2}(c) and (d).
In the spatial pattern of $\left| B \right|$ at 2.87~GHz, microwaves emitted from the transmission wire reach the ground plane of the notch area and a uniform intensity $\left| B \right|$ = 0.3 G is produced over 1.5 mm along the $x$ axis (see values in Fig.~\ref{fig4}(e)). This enables the observation of ODMR over the entire area, as described in Sec.~\ref{sec3.2}.
A strong magnetic field of up to $\left| B \right|$ = 6~G was also formed around the thin wire.
While maintaining a large area of microwave excitation, the antenna conserves the broadband characteristics typical of coplanar waveguides, as shown in the $S_{11}$ and $S_{21}$ spectra (Fig.~\ref{figure 2} (d)). 

We found that this spatially uniform and broadband excitation pattern originated from the thin wire that remained at the center of the notch area.
Figures~\ref{figure 2}(e) and (f) show, respectively, the simulated spatial pattern of $|\boldsymbol{B}|$ at 2.87 GHz and the $S$-parameter spectra when the thin wire is removed.
In the absence of the thin wire, $\left| B \right|$ nearly vanishes in the notch area, and $S_{11}$ shows near-perfect reflection ($S_{11} = 0$), indicating that no microwaves enter the notch area.
These results clearly demonstrate the effectiveness of the thin wire structure in the notch area for simultaneously realizing a strong magnetic field intensity, broadband frequency width, and large-area excitation.
For more details on the notch-shaped antenna, we present parameter sweeps of the notch structures, that is, $a$, $b$, and $w_2$. 
As the notch size decreases (in terms of either $a$ or $b$), the microwave transmission increases (a smaller $S_{11}$ and larger $S_{21}$), as shown in Figs. S3 (a)--(f). 
The thin-wire width $w_{2}$ also affects the S-parameters: as $w_2$ increases, $S_{11}$ ($S_{21}$) decreases (increases), as shown in Figs. S3 (g)--(i). 
We also analyzed the effect of the taper structures.
As shown in Figs. S3 (j)--(l), the taper structure acts as an impedance transformer and improves $S_{11}$ by $\sim 5$ dB and $S_{21}$ by $\sim 0.3$ dB at 2.9 GHz.

\subsection{Experimental characterization of the $S$ parameters}
\label{sec3.2}

We next experimentally characterized the simulation results by measuring the $S$ parameters.
Figures~\ref{figure 3}(a) and (b) compare the simulated and experimental results measured by the vector network analyzer for $S_{\rm 11}$ and $S_{\rm 21}$, respectively. 
For $S_{\rm 11}$, the reflection at 2.87 GHz is $S_{\rm 11} = - 8.13$ dB ($R = 15.4$ \%) in the simulation and $S_{\rm 11} = - 11.0$ dB ($R = 7.94$ \%) in the experiments.
For $S_{\rm 21}$, the transmission is $S_{\rm 21} = - 0.726$ dB ($T = 84.6$ \%) in the simulation and $S_{\rm 21} = - 6.32$ dB ($T = 23.4$ \%) in the experiments.
Here, $R$ (the reflection microwave power) and $T$ (the reflection transmission power) are converted from $S_{\rm 11}$ and $S_{\rm 21}$, respectively, using Eq.~\ref{eq:s-para}.
In both spectra, the experimental results show one or two dips, as also observed in experiments on other antenna structures, including a simple coplanar antenna and omega-shaped thin-wire antennas (see Fig. S4).
At the junctions, we used 8-mm-long copper spring fingers (thickness 0.3 mm) to make electrical connections, as shown in Fig.~\ref{figure 1} (b). 
We associate these dips with a cavity resonance produced by impedance-mismatched reflections at the junctions between the coverslip antenna and the PCB.
Indeed, these dips were found to disappear when an anisotropic conductive rubber connector was used instead of the spring fingers (see Fig. S5). 
It is therefore likely that these dips arise from impedance mismatching at the junctions between the coverslip antenna and the PCB, and not from imperfections in the device fabrication.

\subsection{Experimental characterization of the magnetic field by ODMR measurements}
\label{sec3.3}

\begin{figure*}[th!]
    \centering
    \includegraphics[width=15.8cm]{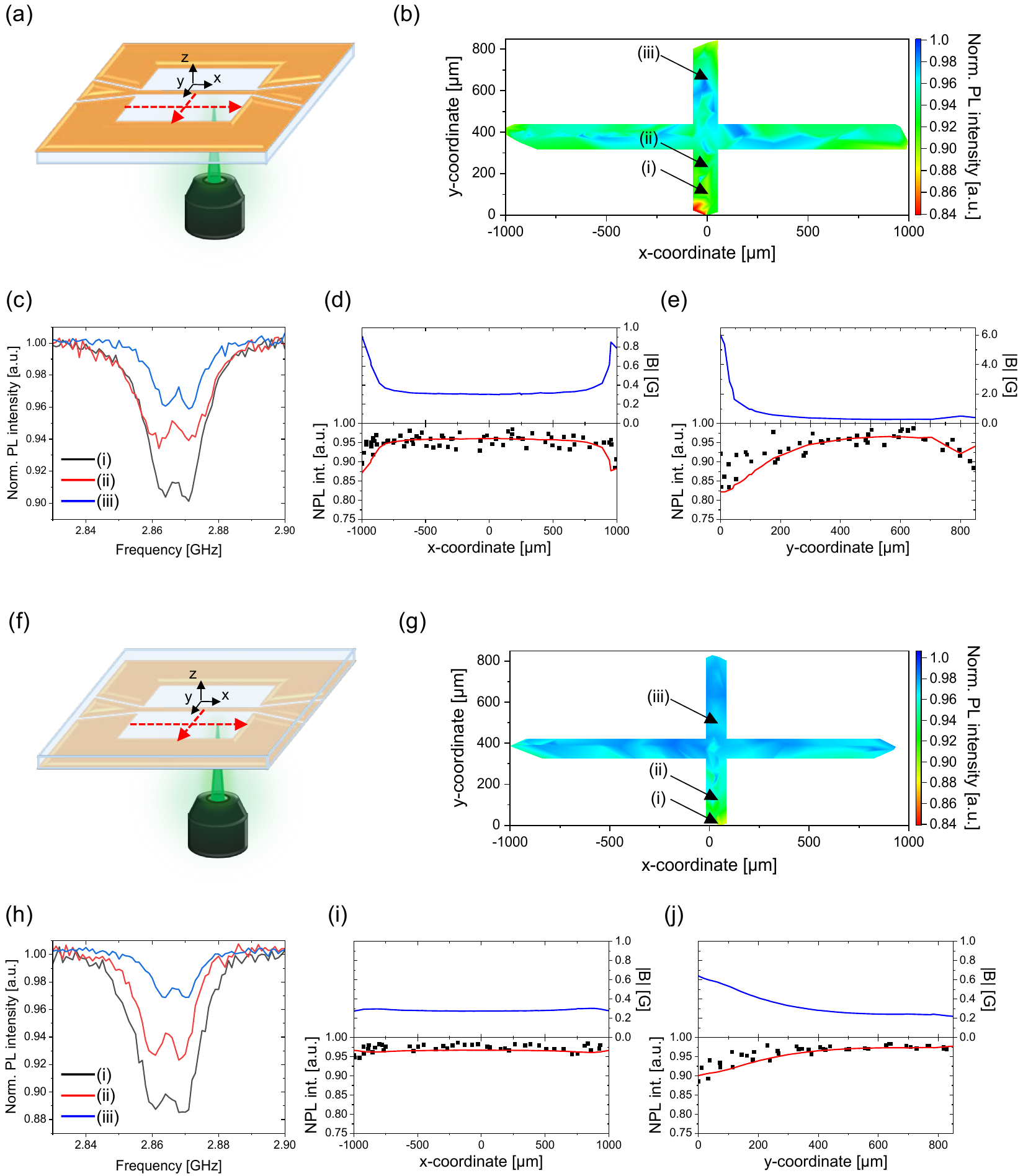}
    \caption{(a) Schematic diagram for Case 1, where NDs are deposited onto the Au-Cr layer side. \Rev{(b) Heat map of experimental ODMR depth for Case 1, and (c) the ODMR signals at positions (i)--(iii) in Fig.~\ref{fig4}(b). (d) Dependence of the simulated $\left| B \right|$ (top; blue line) and measured ODMR depth (bottom; black squares) with the simulated line calculated from the simulated $\left| B \right|$ (bottom; red line) on the $x$- and (e) $y$-coordinates ("NPL int." stands for "Normalized PL intensity").} (f) Schematic diagram for Case 2, where the antenna layer was located beneath the coverslip, while the NDs were spin-coated onto it. 
    \Rev{(g) Heat map of experimental ODMR depth for Case 2, and (h) the ODMR signals at positions (i)--(iii) in Fig.~\ref{fig4}(g)}
    (i) Dependence of $\left| B \right|$ and ODMR depth on the $x$- and (j) $y$-coordinates.
    \Rev{The normalized PL intensity ($\bar{I}$) is defined as $\bar{I} = I_{\rm ON} / I_{\rm OFF}$, where $I_{\rm ON}$ and $I_{\rm OFF}$ are the PL intensities when the microwave is on and off, respectively.}
}
    \label{fig4}
\end{figure*}

Having designed the microwave circuit structure, we measured the ODMR for NDs spin-coated onto the antenna coverslips to experimentally confirm the simulations of the spatial distribution of $|\boldsymbol{B}|$. 
We measured the ODMR depth for two cases. 
In Case 1, the NDs and antenna were situated on top of the coverslip  (Fig.~\ref{fig4}(a)), \Rev{and we measured the ODMR along the line $x = -8.94~\si{\um}$ and $y = 384~\si{\um}$.
Figure~\ref{fig4}(b) shows the heat map of the experimental ODMR depth, and the positional dependence of its depth at positions (i)--(iii) is shown in Fig.~\ref{fig4}(c).}
Figures~\ref{fig4}(d) and (e) show graphs of $\left| B \right|$ calculated by the numerical simulation (top panels) and the corresponding plots of the ODMR depth as functions of $x$ and $y$ (bottom panels).
In the bottom panels, the black squares represent the experimental data and the red lines indicate the theoretical ODMR depth calculated from the simulated $|\boldsymbol{B}|$~\cite{dreau2011avoiding} (see Supporting Information for the details of the calculations).
In Case 2, the antenna layer was located beneath the coverslip, while the NDs were spin-coated onto it, that is, separated from the antenna by the coverslip thickness (Fig.~\ref{fig4}(f)).
\Rev{We measured the ODMR along $x = 31.3~\si{\um}$ and $y = 389~\si{\um}$ like the measurements in Case 1.
Figures~\ref{fig4} (g) and (h) show the heat map of the experimental ODMR depth, and the positional dependence of its depth at positions (i)--(iii). Figs.~\ref{fig4} (i) and (j) show the graphs of the simulated and experimental results, respectively.}
The experimental results of the ODMR depth closely matched the simulated ODMR depth in most regions. 


\begin{figure}[th!]
    \centering
    \includegraphics[width=8.4cm]{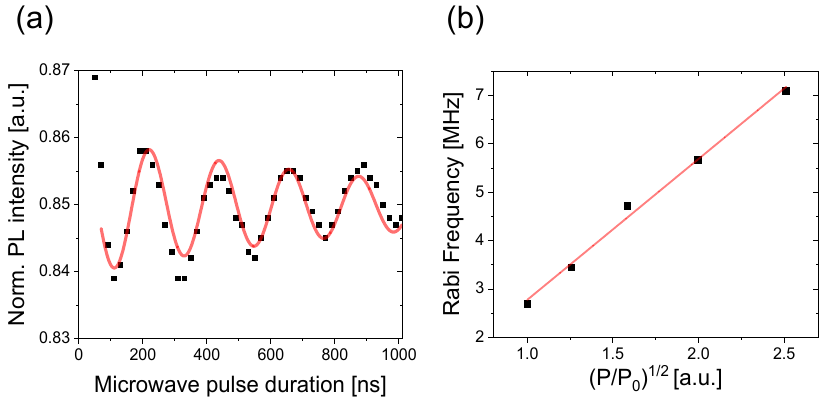}
    \caption{(a) Representative Rabi nutation profile of NDs deposited close to the central thin wire.
    (black dots: experiment; red line: fitting based on the theoretical sine-damping equation).
    (b) Rabi nutation frequency as a function of the normalized microwave power ($(P/P_0)^{1/2}$), where $P$ and $P_0$ are the input microwave power and the minimum ODMR observable input microwave power, respectively 
    (black dots: experiment; red line: linear fit).
    }
    \label{fig5}
\end{figure}

To further confirm this comparison between the simulations and experiments in  $|\boldsymbol{B}|$~\cite{bayat2014efficient}, we measured Rabi nutations for NDs located close to the central thin-wire transmission line. 
Figures~\ref{fig5}(a) and (b) show a representative Rabi nutation and its square-root microwave power dependence for an ND located at ($x$, $y$). 
We obtained the Rabi nutation frequencies ($f_R$) by fitting the data to the theoretical equation~\cite{camparo1988observation, bertaina2014rabi}, giving $f_R = 4.5$ MHz for this particular ND with 63 mW microwave power (see Supporting Information for the detailed estimation of microwave input power) and obtained a mean of $f_R = 4.4 \pm 0.8$ MHz for four NDs measured near the transmission wire (they were located 25--57~\si{\um} from the transmission line).
This Rabi nutation frequency corresponds to $|\boldsymbol{B}|= 2.2 \pm 0.41$ G based on the following equation~\cite{hanson2008coherent, sasaki2016broadband, vallabhapurapu2021fast}: 
\begin{equation}
    f_R = \frac{\gamma |\boldsymbol{B}|}{\sqrt{2}}, 
    \label{eq:rabi}
\end{equation}
where $\gamma$ = 2.8 MHz/G is the NV gyromagnetic ratio.
This value is consistent with the numerical simulation in the corresponding region ($|\boldsymbol{B}| \approx 2.5$ G).
Note that Eq.~\ref{eq:rabi} assumes that $B$ is perpendicular to the NV axes, which may cause an uncertainty in estimating $|\boldsymbol{B}|$ (see Supporting Information).

\subsection{Biological application: cultured cells}
We used the notch-shaped coplanar waveguide chip devices to perform ODMR measurements of NDs in cultured cells.
We glued the glass face of the devices (not the Au-Cr face) onto 35 mm plastic dishes and seeded the cells in the dishes.
\Rev{Figure~\ref{fig6}(a) shows the flow diagram of ODMR measurement preparation using cells.}
The glass face of the chip device was coated with collagen to improve the cell adhesion before seeding.
We then introduced carboxylated NDs into HeLa cells by endocytosis, as previously described ~\cite{nishimura2021wide, slegerova2014nanodiamonds}. 
Figure~\ref{fig6}(b) shows a low-magnification bright-field image of HeLa cells in the chip devices.
Figure~\ref{fig6}(c) merges a bright-field image with red fluorescence acquired in our home-built microscope.
We measured the ODMR of an isolated ND, indicated by a yellow arrow, and successfully obtained the ODMR spectrum, as shown in Fig.~\ref{fig6}(d).
\Rev{Comparable ODMR depths can be observed for NDs in other cells distributed in the center of the notch area (see Fig.~S7(c)).
For a mean depth of 0.951, we estimated $|\boldsymbol{B}| = 0.35 \pm 0.01$ G using Eq.~S3 and its error propagation,}
which reflects the uniformity of $|\boldsymbol{B}|$ in the notch area.
\Rev{The estimation uncertainty of 0.01 G was comparable to that obtained for NDs on coverslip in the central notch region (from $x = -600~\si{\um}$ to $+ 600~\si{\um}$ in Fig.~\ref{fig4}(i)).}

\subsection{Biological application: tadpole tissue}
Our devices provide an observation area ($1.75 \times 2.0 \rm~mm^2$ square) that is sufficiently large for various tissue experiments, beyond the reach of previous waveguide-based antennas. 
As a demonstration, we used tail tissue from \textit{Xenopus laevis} tadpoles. 
\textit{Xenopus} is a common animal model used in developmental biology~\cite{sater2017using, straka2012xenopus, nakai2017mechanisms} for studying regeneration and inflammation~\cite{slack2008xenopus, phipps2020model, king2012developing, davaapil2017conserved}.
For example, microscopy studies have observed the dynamics of regeneration, degeneration, and metamorphosis in amputated tail tissue~\cite{weber1969isolated, niki1982epidermal, izutsu1996adult, nakajima2019unique, tsujioka2017interleukin}.
Therefore, we amputated tail fragments from stage-54 \textit{Xenopus} tadpoles whose tails were labeled with PG-NDs by intramuscular injection with syringes before the final stage of tail degeneration (stage 66)~\cite{mukaigasa2009keratin}.
The fragments were placed on dish-type chip devices and subjected to ODMR detection, as shown in Fig.~\ref{fig6}(e).
Figure~\ref{fig6}(f) shows a merged bright-field and red-fluorescence image of the tissue fragments.
The low background red-wavelength fluorescence enabled the observation of multiple ND fluorescence spots in the tail tissue.
In contrast to the cell experiments described above, the NDs in the tissue underwent noticeable Brownian motion, which disturbed the stable ODMR measurements.
Therefore, we reduced the ODMR measurement time by decreasing the sampling frequency and successfully acquired the ODMR spectra of NDs in the tissue, as shown in Fig.~\ref{fig6}(g).
\Rev{The ODMR depth was 0.936 and the NDs in the other tissue (Fig.~S7(f)) also showed ODMR depths within the variation comparable to the coverslip case (Fig.~\ref{fig4}).}

\subsection{Biological application: nematode worms in vivo}
\begin{figure*}[th!]
    \centering
    \includegraphics[width=14.8cm]{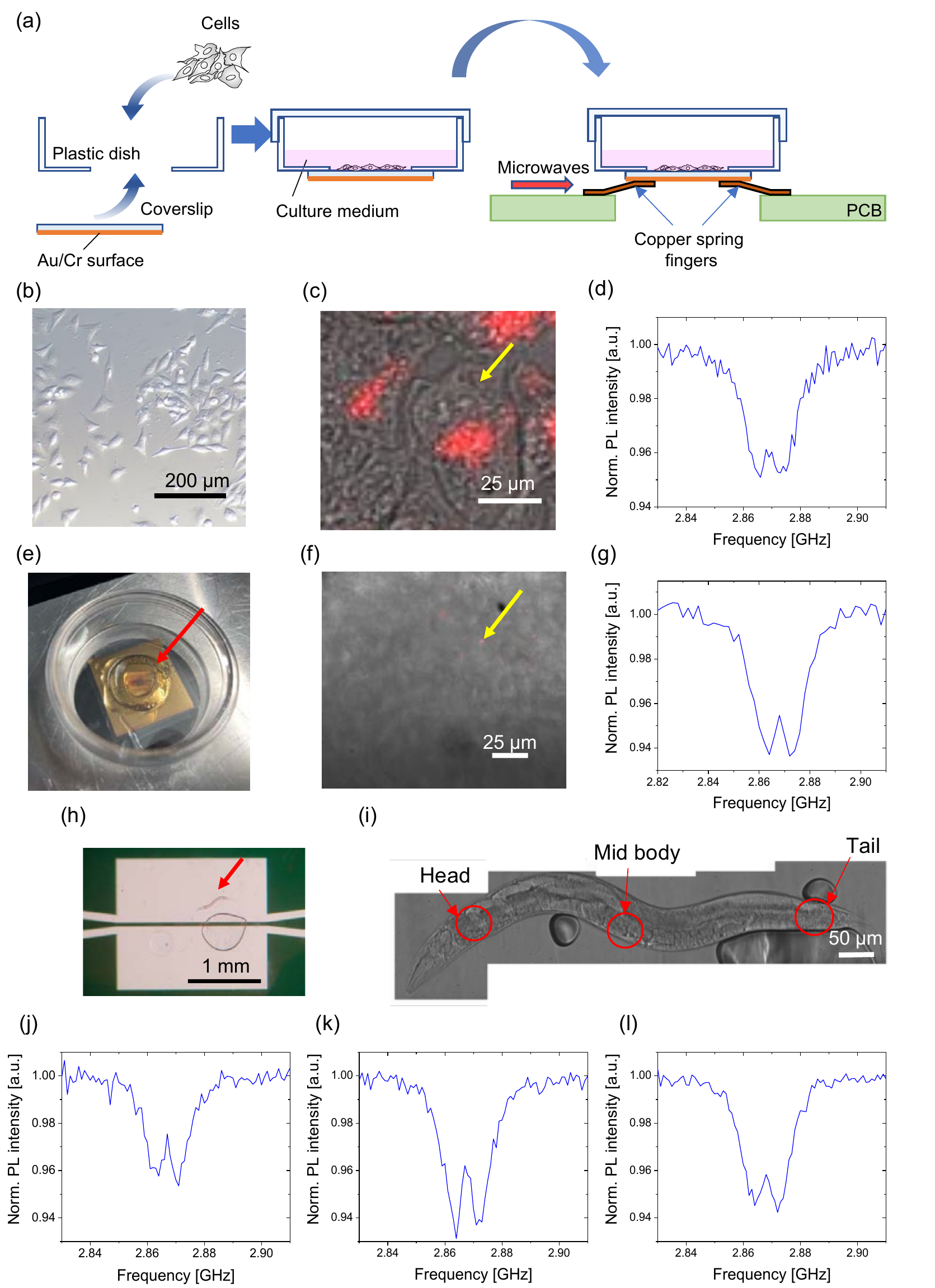}
    \caption{\Rev{(a) Flow diagram of the device assembly and ODMR measurements using cells.}
    (b) Bright-field image of living HeLa cells with a low magnification objective. 
    (c) Merged bright-field image with red fluorescence of ND-labeled HeLa cells acquired using the ODMR microscope. 
    (d) ODMR spectrum of the ND, indicated by the yellow arrow in Fig.~\ref{fig6} (c).
    (e) Photograph of a piece of tissue fragment of a stage-54 \textit{Xenopus} tadpole tail in the dish-type chip devices.
    (f) Merged bright-field image with red fluorescence of the ND-injected tissue acquired using the ODMR microscope and (g) ODMR spectrum of the ND indicated by the yellow arrow in Fig.~\ref{fig6} (f).
    (h) Microscopic photo of \textit{C. elegans} in the notch area of the chip devices. An air bubble was included below the worm. 
    (i) Microscopic photomontage of another worm in the notch area, indicating the body parts measured for ODMR.
    (j), (k), (l) Representative ODMR spectra of NDs found in three body parts: head, middle body, and tail, respectively.
}
    \label{fig6}
\end{figure*}

Exploiting the large observation area provided by our devices, we performed ODMR measurements over the entire body of \textit{C. elegans} in vivo.
We microinjected PG-NDs into the gonads of \textit{C. elegans}, as previously reported ~\cite{zhao2011chromatographic, mohan2010vivo}. 
After a 24 h recovery, we placed a worm on the agar pads for anesthetization and transferred the entire pad to the notch area of the device.
Figure~\ref{fig6}(h) shows a microscope photograph of the \textit{C. elegans} immobilized in the notch area.
The worm was oriented parallel to the transmission line as much as possible to ensure uniform excitation of the magnetic field along the worm axis.
We then measured the ODMR spectra of representative NDs located in the following three body parts (neck, midbody, and tail), as shown in Fig.~\ref{fig6}(i).
The observed ODMR depths have variations of 0.963, 0.942, and 0.953 for Figs.~\ref{fig6}(j)--(l). 
This variation is consistent with the intrinsic variation observed for NDs directly deposited on the glass surface of the devices (see Fig.~\ref{figure 3}), which confirms the uniformity of the magnetic field intensity over the entire worm body.
\Rev{The uniform ODMR depths in the central notch region were confirmed in different worms, as illustrated in Fig.~S7(i).}

\section{Discussion}
This study presents a glass-patternable notch-shaped coplanar waveguide microwave antenna that provides a millimeter-scale observation area and a gigahertz frequency bandwidth for performing ODMR measurements on diamond NV centers in biological samples.
Conventional antennas used in biological ODMR experiments involving NDs include thin wires, coils, and omega-shaped antennas ~\cite{yukawa2020quantum, miller2020spin, fukushige2020identification, igarashi2012real, choi2020probing, bolshedvorskii2017single}.
Thin-wire antennas have been widely used with samples such as cells and worms because of their simple fabrication and wide frequency bandwidth. 
However, they display strong microwave reflection up to $S_{11} \sim -3$ dB and have a limited irradiation area at a distance of $\sim$ $100 ~\si{\um}$ from the wire~\cite{yukawa2020quantum}. 
Coils have been frequently used in biological ODMR experiments to provide microwave radiation to nearby samples.
Figures S8 (a)--(c) show the microwave irradiation characteristics of conventional coils.
Coil miniaturization is a significant technical challenge. 
Because the ODMR of NV centers is usually performed in the S-band (2--4 GHz), impedance matching to 50 $\si{\ohm}$ coaxial cables requires approximately 1 mm in diameter, resulting in low $|B|$ intensity(see Fig. S8(b)).  
This difficulty of balancing the trade-off between the impedance matching and high $|B|$ intensity may hinder compatibility with the measurement systems. 
Accurate prediction of the ODMR signal intensity is also challenging because the distance between the NDs and coils cannot be defined with 10 \si{\um} resolution.
Omega-shaped antennas have become an appealing alternative for biological ODMR applications. 
These antennas are of two types: impedance-matched fully resonant omega shapes ~\cite{miller2020spin} and non-resonant omega shapes ~\cite{choi2020probing}. 
Impedance-matched omega-shapes have been recently employed for portable ODMR detection systems because of their high input efficiency and millimeter-scale spatial uniformity of $|\boldsymbol{B}|$~\cite{miller2020spin}.
Conversely, their frequency response shows resonance, which is undesirable for broad-bandwidth applications such as vector magnetometry \Rev{that reads multiple ODMR dips from four NV axes spanning over  200-MHz width~\cite{le2013optical, schloss2018simultaneous, clevenson2018robust, igarashi2020threeD, tsukamoto2021vector}}. 
We compare the microwave characteristics of a representative non-resonant omega shape and the present notch-shaped coplanar antenna in Figs. S8 (d)--(f). 
The non-resonant omega shape provides a uniform $\left| B \right|$ over a length scale on the order of 10 $\si{\um}$ by perturbing the steep $|\boldsymbol{B}|$ distribution around the thin wire line.
It also affords a broad bandwidth that is comparable to the coplanar waveguides.
Alternatively, the effect of the omega shape appears within 100 $\si{\um}$, and becomes insignificant in the opposite-plane configuration (Case 2 in Fig.~\ref{fig4}), where the NDs are separated from the Au-Cr layer by 0.17 mm across the coverslip, approaching the same microwave characteristics as the present notch-shaped coplanar structure.
In practice, the size of the omega shape has an upper limit of $500 ~\si{\um}$.

In addition to the large detection area and broad bandwidth, the notch-shaped coplanar structure is biocompatible and scalable.
For biocompatibility, by bringing the Au-Cr layer to the bottom side of the coverslip, the device can be used in exactly the same way as a conventional glass-bottom dish. 
The plain glass surface inside the dish-style device allows various surface coatings for culturing cells and tissue.
Indeed, we confirmed that adipose-tissue-derived stem cells~\cite{yukawa2020quantum}, which are more delicate than HeLa cells, can be successfully cultured on our collagen-coated devices (see Fig. S9).
The concept of docking the devices onto a PCB for making circuit connections also helps to conserve the sample quality by shortening the handling time.
For the scalability, the flat S-parameter spectral profiles enable the integration of multiple observation areas (triple wells in this example) within 15 mm on a single chip without incurring a significant deterioration of the microwave characteristics, as shown in Fig. S5. 
Such scalability is not easily attainable with S-band microwave circuits because of the size limitation originating from the centimeter-long wavelength in this band (3 GHz corresponds to 10 cm wavelength and its quarter is 2.5 cm).  
It is anticipated that such triple-well chip devices can be readily extended to large-scale multi-well plates for plate reading in ODMR-based assays.

\section{Conclusion}
Quantum nano-sensors involving fluorescent NDs are a promising new technology for multimodal bioassays that are particularly well suited to on-chip device miniaturization.
However, the miniaturization of microwave excitation circuits used with optical detection is challenging because they operate at centimeter wavelengths.
To address this issue, we developed a method to quantitatively predict the ODMR signal intensity by employing the FEM-based numerical simulations and proposed a notch-shaped coplanar antenna on a glass coverslip that can be assembled into various bioassay platforms, including dishes, microscope slides, and multi-well plates. 
This notch-shaped structure provided a millimeter-scale ODMR detection area over $1.5\times2.0 \ \si{\mm}^2$ and broadband low-reflection characteristics of microwaves ($R \approx 8$\% over a few gigahertz).
This device achieved uniform ODMR detection within cells, tadpole tail tissue, and \textit{C. elegans}. 
The concept of this notch-shaped coplanar antenna and methodology for the quantitative modeling of ODMR characteristics will facilitate the integration of diamond-based quantum sensing technology into chip-based bioassays.

\section*{Author Contributions}
K.O. and M.F. designed the research.
K.O. performed the numerical simulation.
K.O, Y.N., M.Tnk., and E.S. performed the device fabrication. 
L.Z., Y.Z. and N.K. synthesized the PG-NDs.
K.O., Y.N., T.M., S.M., M.Tkng., H.Y., and Y.B. performed the cell experiments.
Y.T., Y.Z., Y.I. and E.K.-N prepared the worms and their ND labeling.
K.O., K.Y. and Y.I. performed the tadpole experiments.
K.O., Y.N., Y.I., Y.T., and M.F. performed the ODMR experiments and analyses.
All the authors participated in the discussion and writing of the paper.

\section*{Conflicts of interest}
There are no conflicts to declare.

\begin{acknowledgements}
We thank S. Sakakihara, Y. Umehara, and Y. Shikano for their support in the device fabrication, bio-sample preparation and discussion, respectively.
A part of this work was supported by the “Nanotechnology Platform Project (Nanotechnology Open Facilities in Osaka University)” of MEXT (JPMXP09F21OS0055).
We acknowledge funding from JSPS-KAKENHI (19K21935, 20H00335, 20KK031703), AMED (JP21zf0127004), JST (JPMJMI21G1), the MEXT-LEADER program, the Mazda Foundation, and Osaka City University Strategic Research Grant 2017--2020, MEXT Quantum Leap Flagship Program (MEXT Q-LEAP, JPMXS0120330644).
\end{acknowledgements}

\clearpage
\newpage
\bibliography{achemso-demo}

\begin{thebibliography}{87}%
\makeatletter
\providecommand \@ifxundefined [1]{%
 \@ifx{#1\undefined}
}%
\providecommand \@ifnum [1]{%
 \ifnum #1\expandafter \@firstoftwo
 \else \expandafter \@secondoftwo
 \fi
}%
\providecommand \@ifx [1]{%
 \ifx #1\expandafter \@firstoftwo
 \else \expandafter \@secondoftwo
 \fi
}%
\providecommand \natexlab [1]{#1}%
\providecommand \enquote  [1]{``#1''}%
\providecommand \bibnamefont  [1]{#1}%
\providecommand \bibfnamefont [1]{#1}%
\providecommand \citenamefont [1]{#1}%
\providecommand \href@noop [0]{\@secondoftwo}%
\providecommand \href [0]{\begingroup \@sanitize@url \@href}%
\providecommand \@href[1]{\@@startlink{#1}\@@href}%
\providecommand \@@href[1]{\endgroup#1\@@endlink}%
\providecommand \@sanitize@url [0]{\catcode `\\12\catcode `\$12\catcode
  `\&12\catcode `\#12\catcode `\^12\catcode `\_12\catcode `\%12\relax}%
\providecommand \@@startlink[1]{}%
\providecommand \@@endlink[0]{}%
\providecommand \url  [0]{\begingroup\@sanitize@url \@url }%
\providecommand \@url [1]{\endgroup\@href {#1}{\urlprefix }}%
\providecommand \urlprefix  [0]{URL }%
\providecommand \Eprint [0]{\href }%
\providecommand \doibase [0]{http://dx.doi.org/}%
\providecommand \selectlanguage [0]{\@gobble}%
\providecommand \bibinfo  [0]{\@secondoftwo}%
\providecommand \bibfield  [0]{\@secondoftwo}%
\providecommand \translation [1]{[#1]}%
\providecommand \BibitemOpen [0]{}%
\providecommand \bibitemStop [0]{}%
\providecommand \bibitemNoStop [0]{.\EOS\space}%
\providecommand \EOS [0]{\spacefactor3000\relax}%
\providecommand \BibitemShut  [1]{\csname bibitem#1\endcsname}%
\let\auto@bib@innerbib\@empty
\bibitem [{\citenamefont {Ngo}\ \emph {et~al.}(2014)\citenamefont {Ngo},
  \citenamefont {Wang}, \citenamefont {Fales}, \citenamefont {Nicholson},
  \citenamefont {Woods},\ and\ \citenamefont {Vo-Dinh}}]{ngo2014dna}%
  \BibitemOpen
  \bibfield  {author} {\bibinfo {author} {\bibfnamefont {Hoan~T}\ \bibnamefont
  {Ngo}}, \bibinfo {author} {\bibfnamefont {Hsin-Neng}\ \bibnamefont {Wang}},
  \bibinfo {author} {\bibfnamefont {Andrew~M}\ \bibnamefont {Fales}}, \bibinfo
  {author} {\bibfnamefont {Bradly~P}\ \bibnamefont {Nicholson}}, \bibinfo
  {author} {\bibfnamefont {Christopher~W}\ \bibnamefont {Woods}}, \ and\
  \bibinfo {author} {\bibfnamefont {Tuan}\ \bibnamefont {Vo-Dinh}},\ }\bibfield
   {title} {\enquote {\bibinfo {title} {Dna bioassay-on-chip using sers
  detection for dengue diagnosis},}\ }\href@noop {} {\bibfield  {journal}
  {\bibinfo  {journal} {Analyst}\ }\textbf {\bibinfo {volume} {139}},\ \bibinfo
  {pages} {5655--5659} (\bibinfo {year} {2014})}\BibitemShut {NoStop}%
\bibitem [{\citenamefont {Tani}\ \emph {et~al.}(2004)\citenamefont {Tani},
  \citenamefont {Maehana},\ and\ \citenamefont {Kamidate}}]{tani2004chip}%
  \BibitemOpen
  \bibfield  {author} {\bibinfo {author} {\bibfnamefont {Hirofumi}\
  \bibnamefont {Tani}}, \bibinfo {author} {\bibfnamefont {Koji}\ \bibnamefont
  {Maehana}}, \ and\ \bibinfo {author} {\bibfnamefont {Tamio}\ \bibnamefont
  {Kamidate}},\ }\bibfield  {title} {\enquote {\bibinfo {title} {Chip-based
  bioassay using bacterial sensor strains immobilized in three-dimensional
  microfluidic network},}\ }\href {\doibase 10.1021/ac049401d} {\bibfield
  {journal} {\bibinfo  {journal} {Anal. Chem.}\ }\textbf {\bibinfo {volume}
  {76}},\ \bibinfo {pages} {6693--6697} (\bibinfo {year} {2004})}\BibitemShut
  {NoStop}%
\bibitem [{\citenamefont {Rothbauer}\ \emph {et~al.}(2018)\citenamefont
  {Rothbauer}, \citenamefont {Zirath},\ and\ \citenamefont
  {Ertl}}]{rothbauer2018recent}%
  \BibitemOpen
  \bibfield  {author} {\bibinfo {author} {\bibfnamefont {Mario}\ \bibnamefont
  {Rothbauer}}, \bibinfo {author} {\bibfnamefont {Helene}\ \bibnamefont
  {Zirath}}, \ and\ \bibinfo {author} {\bibfnamefont {Peter}\ \bibnamefont
  {Ertl}},\ }\bibfield  {title} {\enquote {\bibinfo {title} {Recent advances in
  microfluidic technologies for cell-to-cell interaction studies},}\
  }\href@noop {} {\bibfield  {journal} {\bibinfo  {journal} {Lab Chip}\
  }\textbf {\bibinfo {volume} {18}},\ \bibinfo {pages} {249--270} (\bibinfo
  {year} {2018})}\BibitemShut {NoStop}%
\bibitem [{\citenamefont {Ferrier}\ \emph {et~al.}(2009)\citenamefont
  {Ferrier}, \citenamefont {Romanuik}, \citenamefont {Thomson}, \citenamefont
  {Bridges},\ and\ \citenamefont {Freeman}}]{ferrier2009microwave}%
  \BibitemOpen
  \bibfield  {author} {\bibinfo {author} {\bibfnamefont {Graham~A}\
  \bibnamefont {Ferrier}}, \bibinfo {author} {\bibfnamefont {Sean~F}\
  \bibnamefont {Romanuik}}, \bibinfo {author} {\bibfnamefont {Douglas~J}\
  \bibnamefont {Thomson}}, \bibinfo {author} {\bibfnamefont {Greg~E}\
  \bibnamefont {Bridges}}, \ and\ \bibinfo {author} {\bibfnamefont {Mark~R}\
  \bibnamefont {Freeman}},\ }\bibfield  {title} {\enquote {\bibinfo {title} {A
  microwave interferometric system for simultaneous actuation and detection of
  single biological cells},}\ }\href@noop {} {\bibfield  {journal} {\bibinfo
  {journal} {Lab Chip}\ }\textbf {\bibinfo {volume} {9}},\ \bibinfo {pages}
  {3406--3412} (\bibinfo {year} {2009})}\BibitemShut {NoStop}%
\bibitem [{\citenamefont {Ziem}\ \emph {et~al.}(2013)\citenamefont {Ziem},
  \citenamefont {Götz}, \citenamefont {Zappe}, \citenamefont {Steinert},\ and\
  \citenamefont {Wrachtrup}}]{ziem2013highly}%
  \BibitemOpen
  \bibfield  {author} {\bibinfo {author} {\bibfnamefont {Florestan~C}\
  \bibnamefont {Ziem}}, \bibinfo {author} {\bibfnamefont {Nicolas~S}\
  \bibnamefont {Götz}}, \bibinfo {author} {\bibfnamefont {Andrea}\
  \bibnamefont {Zappe}}, \bibinfo {author} {\bibfnamefont {Steffen}\
  \bibnamefont {Steinert}}, \ and\ \bibinfo {author} {\bibfnamefont {Jörg}\
  \bibnamefont {Wrachtrup}},\ }\bibfield  {title} {\enquote {\bibinfo {title}
  {Highly sensitive detection of physiological spins in a microfluidic
  device},}\ }\href@noop {} {\bibfield  {journal} {\bibinfo  {journal} {Nano
  Lett.}\ }\textbf {\bibinfo {volume} {13}},\ \bibinfo {pages} {4093--4098}
  (\bibinfo {year} {2013})}\BibitemShut {NoStop}%
\bibitem [{\citenamefont {Wackerlig}\ and\ \citenamefont
  {Schirhagl}(2016)}]{wackerlig2016applications}%
  \BibitemOpen
  \bibfield  {author} {\bibinfo {author} {\bibfnamefont {Judith}\ \bibnamefont
  {Wackerlig}}\ and\ \bibinfo {author} {\bibfnamefont {Romana}\ \bibnamefont
  {Schirhagl}},\ }\bibfield  {title} {\enquote {\bibinfo {title} {Applications
  of molecularly imprinted polymer nanoparticles and their advances toward
  industrial use: a review},}\ }\href@noop {} {\bibfield  {journal} {\bibinfo
  {journal} {Anal. Chem.}\ }\textbf {\bibinfo {volume} {88}},\ \bibinfo {pages}
  {250--261} (\bibinfo {year} {2016})}\BibitemShut {NoStop}%
\bibitem [{\citenamefont {Rendler}\ \emph {et~al.}(2017)\citenamefont
  {Rendler}, \citenamefont {Neburkova}, \citenamefont {Zemek}, \citenamefont
  {Kotek}, \citenamefont {Zappe}, \citenamefont {Chu}, \citenamefont {Cigler},\
  and\ \citenamefont {Wrachtrup}}]{rendler2017optical}%
  \BibitemOpen
  \bibfield  {author} {\bibinfo {author} {\bibfnamefont {Torsten}\ \bibnamefont
  {Rendler}}, \bibinfo {author} {\bibfnamefont {Jitka}\ \bibnamefont
  {Neburkova}}, \bibinfo {author} {\bibfnamefont {Ondrej}\ \bibnamefont
  {Zemek}}, \bibinfo {author} {\bibfnamefont {Jan}\ \bibnamefont {Kotek}},
  \bibinfo {author} {\bibfnamefont {Andrea}\ \bibnamefont {Zappe}}, \bibinfo
  {author} {\bibfnamefont {Zhiqin}\ \bibnamefont {Chu}}, \bibinfo {author}
  {\bibfnamefont {Petr}\ \bibnamefont {Cigler}}, \ and\ \bibinfo {author}
  {\bibfnamefont {J{\"o}rg}\ \bibnamefont {Wrachtrup}},\ }\bibfield  {title}
  {\enquote {\bibinfo {title} {Optical imaging of localized chemical events
  using programmable diamond quantum nanosensors},}\ }\href@noop {} {\bibfield
  {journal} {\bibinfo  {journal} {Nat. Commun.}\ }\textbf {\bibinfo {volume}
  {8}},\ \bibinfo {pages} {1--9} (\bibinfo {year} {2017})}\BibitemShut
  {NoStop}%
\bibitem [{\citenamefont {Miller}\ \emph {et~al.}(2020)\citenamefont {Miller},
  \citenamefont {Bezinge}, \citenamefont {Gliddon}, \citenamefont {Huang},
  \citenamefont {Dold}, \citenamefont {Gray}, \citenamefont {Heaney},
  \citenamefont {Dobson}, \citenamefont {Nastouli}, \citenamefont {Morton}
  \emph {et~al.}}]{miller2020spin}%
  \BibitemOpen
  \bibfield  {author} {\bibinfo {author} {\bibfnamefont {Benjamin~S}\
  \bibnamefont {Miller}}, \bibinfo {author} {\bibfnamefont {L{\'e}onard}\
  \bibnamefont {Bezinge}}, \bibinfo {author} {\bibfnamefont {Harriet~D}\
  \bibnamefont {Gliddon}}, \bibinfo {author} {\bibfnamefont {Da}~\bibnamefont
  {Huang}}, \bibinfo {author} {\bibfnamefont {Gavin}\ \bibnamefont {Dold}},
  \bibinfo {author} {\bibfnamefont {Eleanor~R}\ \bibnamefont {Gray}}, \bibinfo
  {author} {\bibfnamefont {Judith}\ \bibnamefont {Heaney}}, \bibinfo {author}
  {\bibfnamefont {Peter~J}\ \bibnamefont {Dobson}}, \bibinfo {author}
  {\bibfnamefont {Eleni}\ \bibnamefont {Nastouli}}, \bibinfo {author}
  {\bibfnamefont {John~JL}\ \bibnamefont {Morton}},  \emph {et~al.},\
  }\bibfield  {title} {\enquote {\bibinfo {title} {Spin-enhanced nanodiamond
  biosensing for ultrasensitive diagnostics},}\ }\href@noop {} {\bibfield
  {journal} {\bibinfo  {journal} {Nature}\ }\textbf {\bibinfo {volume} {587}},\
  \bibinfo {pages} {588--593} (\bibinfo {year} {2020})}\BibitemShut {NoStop}%
\bibitem [{\citenamefont {Haziza}\ \emph {et~al.}(2017)\citenamefont {Haziza},
  \citenamefont {Mohan}, \citenamefont {Loe-Mie}, \citenamefont
  {Lepagnol-Bestel}, \citenamefont {Massou}, \citenamefont {Adam},
  \citenamefont {Le}, \citenamefont {Viard}, \citenamefont {Plancon},
  \citenamefont {Daudin} \emph {et~al.}}]{haziza2017fluorescent}%
  \BibitemOpen
  \bibfield  {author} {\bibinfo {author} {\bibfnamefont {Simon}\ \bibnamefont
  {Haziza}}, \bibinfo {author} {\bibfnamefont {Nitin}\ \bibnamefont {Mohan}},
  \bibinfo {author} {\bibfnamefont {Yann}\ \bibnamefont {Loe-Mie}}, \bibinfo
  {author} {\bibfnamefont {Aude-Marie}\ \bibnamefont {Lepagnol-Bestel}},
  \bibinfo {author} {\bibfnamefont {Sophie}\ \bibnamefont {Massou}}, \bibinfo
  {author} {\bibfnamefont {Marie-Pierre}\ \bibnamefont {Adam}}, \bibinfo
  {author} {\bibfnamefont {Xuan~Loc}\ \bibnamefont {Le}}, \bibinfo {author}
  {\bibfnamefont {Julia}\ \bibnamefont {Viard}}, \bibinfo {author}
  {\bibfnamefont {Christine}\ \bibnamefont {Plancon}}, \bibinfo {author}
  {\bibfnamefont {Rachel}\ \bibnamefont {Daudin}},  \emph {et~al.},\ }\bibfield
   {title} {\enquote {\bibinfo {title} {Fluorescent nanodiamond tracking
  reveals intraneuronal transport abnormalities induced by
  brain-disease-related genetic risk factors},}\ }\href@noop {} {\bibfield
  {journal} {\bibinfo  {journal} {Nat. Nanotechnol.}\ }\textbf {\bibinfo
  {volume} {12}},\ \bibinfo {pages} {322--328} (\bibinfo {year}
  {2017})}\BibitemShut {NoStop}%
\bibitem [{\citenamefont {Kucsko}\ \emph {et~al.}(2013)\citenamefont {Kucsko},
  \citenamefont {Maurer}, \citenamefont {Yao}, \citenamefont {Kubo},
  \citenamefont {Noh}, \citenamefont {Lo}, \citenamefont {Park},\ and\
  \citenamefont {Lukin}}]{kucsko2013nanometre}%
  \BibitemOpen
  \bibfield  {author} {\bibinfo {author} {\bibfnamefont {Georg}\ \bibnamefont
  {Kucsko}}, \bibinfo {author} {\bibfnamefont {Peter~C}\ \bibnamefont
  {Maurer}}, \bibinfo {author} {\bibfnamefont {Norman~Ying}\ \bibnamefont
  {Yao}}, \bibinfo {author} {\bibfnamefont {MICHAEL}\ \bibnamefont {Kubo}},
  \bibinfo {author} {\bibfnamefont {Hyun~Jong}\ \bibnamefont {Noh}}, \bibinfo
  {author} {\bibfnamefont {Po~Kam}\ \bibnamefont {Lo}}, \bibinfo {author}
  {\bibfnamefont {Hongkun}\ \bibnamefont {Park}}, \ and\ \bibinfo {author}
  {\bibfnamefont {Mikhail~D}\ \bibnamefont {Lukin}},\ }\bibfield  {title}
  {\enquote {\bibinfo {title} {Nanometre-scale thermometry in a living cell},}\
  }\href@noop {} {\bibfield  {journal} {\bibinfo  {journal} {Nature}\ }\textbf
  {\bibinfo {volume} {500}},\ \bibinfo {pages} {54--58} (\bibinfo {year}
  {2013})}\BibitemShut {NoStop}%
\bibitem [{\citenamefont {Simpson}\ \emph {et~al.}(2017)\citenamefont
  {Simpson}, \citenamefont {Morrisroe}, \citenamefont {McCoey}, \citenamefont
  {Lombard}, \citenamefont {Mendis}, \citenamefont {Treussart}, \citenamefont
  {Hall}, \citenamefont {Petrou},\ and\ \citenamefont
  {Hollenberg}}]{simpson2017non}%
  \BibitemOpen
  \bibfield  {author} {\bibinfo {author} {\bibfnamefont {David~A}\ \bibnamefont
  {Simpson}}, \bibinfo {author} {\bibfnamefont {Emma}\ \bibnamefont
  {Morrisroe}}, \bibinfo {author} {\bibfnamefont {Julia~M}\ \bibnamefont
  {McCoey}}, \bibinfo {author} {\bibfnamefont {Alain~H}\ \bibnamefont
  {Lombard}}, \bibinfo {author} {\bibfnamefont {Dulini~C}\ \bibnamefont
  {Mendis}}, \bibinfo {author} {\bibfnamefont {Francois}\ \bibnamefont
  {Treussart}}, \bibinfo {author} {\bibfnamefont {Liam~T}\ \bibnamefont
  {Hall}}, \bibinfo {author} {\bibfnamefont {Steven}\ \bibnamefont {Petrou}}, \
  and\ \bibinfo {author} {\bibfnamefont {Lloyd~CL}\ \bibnamefont
  {Hollenberg}},\ }\bibfield  {title} {\enquote {\bibinfo {title}
  {Non-neurotoxic nanodiamond probes for intraneuronal temperature mapping},}\
  }\href@noop {} {\bibfield  {journal} {\bibinfo  {journal} {ACS Nano}\
  }\textbf {\bibinfo {volume} {11}},\ \bibinfo {pages} {12077--12086} (\bibinfo
  {year} {2017})}\BibitemShut {NoStop}%
\bibitem [{\citenamefont {Toraille}\ \emph {et~al.}(2018)\citenamefont
  {Toraille}, \citenamefont {A{\"\i}zel}, \citenamefont {Balloul},
  \citenamefont {Vicario}, \citenamefont {Monzel}, \citenamefont {Coppey},
  \citenamefont {Secret}, \citenamefont {Siaugue}, \citenamefont {Sampaio},
  \citenamefont {Rohart} \emph {et~al.}}]{toraille2018optical}%
  \BibitemOpen
  \bibfield  {author} {\bibinfo {author} {\bibfnamefont {Lo{\"\i}c}\
  \bibnamefont {Toraille}}, \bibinfo {author} {\bibfnamefont {Koceila}\
  \bibnamefont {A{\"\i}zel}}, \bibinfo {author} {\bibfnamefont {{\'E}lie}\
  \bibnamefont {Balloul}}, \bibinfo {author} {\bibfnamefont {Chiara}\
  \bibnamefont {Vicario}}, \bibinfo {author} {\bibfnamefont {Cornelia}\
  \bibnamefont {Monzel}}, \bibinfo {author} {\bibfnamefont {Mathieu}\
  \bibnamefont {Coppey}}, \bibinfo {author} {\bibfnamefont {Emilie}\
  \bibnamefont {Secret}}, \bibinfo {author} {\bibfnamefont {Jean-Michel}\
  \bibnamefont {Siaugue}}, \bibinfo {author} {\bibfnamefont {Joao}\
  \bibnamefont {Sampaio}}, \bibinfo {author} {\bibfnamefont {Stanislas}\
  \bibnamefont {Rohart}},  \emph {et~al.},\ }\bibfield  {title} {\enquote
  {\bibinfo {title} {Optical magnetometry of single biocompatible micromagnets
  for quantitative magnetogenetic and magnetomechanical assays},}\ }\href@noop
  {} {\bibfield  {journal} {\bibinfo  {journal} {Nano Lett.}\ }\textbf
  {\bibinfo {volume} {18}},\ \bibinfo {pages} {7635--7641} (\bibinfo {year}
  {2018})}\BibitemShut {NoStop}%
\bibitem [{\citenamefont {Claveau}\ \emph {et~al.}(2018)\citenamefont
  {Claveau}, \citenamefont {Bertrand},\ and\ \citenamefont
  {Treussart}}]{claveau2018fluorescent}%
  \BibitemOpen
  \bibfield  {author} {\bibinfo {author} {\bibfnamefont {Sandra}\ \bibnamefont
  {Claveau}}, \bibinfo {author} {\bibfnamefont {Jean-R{\'e}mi}\ \bibnamefont
  {Bertrand}}, \ and\ \bibinfo {author} {\bibfnamefont {Fran{\c{c}}ois}\
  \bibnamefont {Treussart}},\ }\bibfield  {title} {\enquote {\bibinfo {title}
  {Fluorescent nanodiamond applications for cellular process sensing and cell
  tracking},}\ }\href@noop {} {\bibfield  {journal} {\bibinfo  {journal}
  {Micromachines}\ }\textbf {\bibinfo {volume} {9}},\ \bibinfo {pages} {247}
  (\bibinfo {year} {2018})}\BibitemShut {NoStop}%
\bibitem [{\citenamefont {Nie}\ \emph {et~al.}(2021)\citenamefont {Nie},
  \citenamefont {Nusantara}, \citenamefont {Damle}, \citenamefont {Sharmin},
  \citenamefont {Evans}, \citenamefont {Hemelaar}, \citenamefont {van~der
  Laan}, \citenamefont {Li}, \citenamefont {Martinez}, \citenamefont {Vedelaar}
  \emph {et~al.}}]{nie2021quantum}%
  \BibitemOpen
  \bibfield  {author} {\bibinfo {author} {\bibfnamefont {L}~\bibnamefont
  {Nie}}, \bibinfo {author} {\bibfnamefont {AC}~\bibnamefont {Nusantara}},
  \bibinfo {author} {\bibfnamefont {VG}~\bibnamefont {Damle}}, \bibinfo
  {author} {\bibfnamefont {R}~\bibnamefont {Sharmin}}, \bibinfo {author}
  {\bibfnamefont {EPP}\ \bibnamefont {Evans}}, \bibinfo {author} {\bibfnamefont
  {SR}~\bibnamefont {Hemelaar}}, \bibinfo {author} {\bibfnamefont
  {KJ}~\bibnamefont {van~der Laan}}, \bibinfo {author} {\bibfnamefont
  {R}~\bibnamefont {Li}}, \bibinfo {author} {\bibfnamefont {FP~Perona}\
  \bibnamefont {Martinez}}, \bibinfo {author} {\bibfnamefont {T}~\bibnamefont
  {Vedelaar}},  \emph {et~al.},\ }\bibfield  {title} {\enquote {\bibinfo
  {title} {Quantum monitoring of cellular metabolic activities in single
  mitochondria},}\ }\href@noop {} {\bibfield  {journal} {\bibinfo  {journal}
  {Sci. Adv.}\ }\textbf {\bibinfo {volume} {7}},\ \bibinfo {pages} {eabf0573}
  (\bibinfo {year} {2021})}\BibitemShut {NoStop}%
\bibitem [{\citenamefont {Davis}\ \emph {et~al.}(2018)\citenamefont {Davis},
  \citenamefont {Ramesh}, \citenamefont {Bhatnagar}, \citenamefont
  {Lee-Gosselin}, \citenamefont {Barry}, \citenamefont {Glenn}, \citenamefont
  {Walsworth},\ and\ \citenamefont {Shapiro}}]{davis2018mapping}%
  \BibitemOpen
  \bibfield  {author} {\bibinfo {author} {\bibfnamefont {Hunter~C}\
  \bibnamefont {Davis}}, \bibinfo {author} {\bibfnamefont {Pradeep}\
  \bibnamefont {Ramesh}}, \bibinfo {author} {\bibfnamefont {Aadyot}\
  \bibnamefont {Bhatnagar}}, \bibinfo {author} {\bibfnamefont {Audrey}\
  \bibnamefont {Lee-Gosselin}}, \bibinfo {author} {\bibfnamefont {John~F}\
  \bibnamefont {Barry}}, \bibinfo {author} {\bibfnamefont {David~R}\
  \bibnamefont {Glenn}}, \bibinfo {author} {\bibfnamefont {Ronald~L}\
  \bibnamefont {Walsworth}}, \ and\ \bibinfo {author} {\bibfnamefont
  {Mikhail~G}\ \bibnamefont {Shapiro}},\ }\bibfield  {title} {\enquote
  {\bibinfo {title} {Mapping the microscale origins of magnetic resonance image
  contrast with subcellular diamond magnetometry},}\ }\href@noop {} {\bibfield
  {journal} {\bibinfo  {journal} {Nat. Commun.}\ }\textbf {\bibinfo {volume}
  {9}},\ \bibinfo {pages} {1--9} (\bibinfo {year} {2018})}\BibitemShut
  {NoStop}%
\bibitem [{\citenamefont {van~der Laan}\ \emph {et~al.}(2020)\citenamefont
  {van~der Laan}, \citenamefont {Morita}, \citenamefont {Perona-Martinez},\
  and\ \citenamefont {Schirhagl}}]{van2020evaluation}%
  \BibitemOpen
  \bibfield  {author} {\bibinfo {author} {\bibfnamefont {Kiran~J}\ \bibnamefont
  {van~der Laan}}, \bibinfo {author} {\bibfnamefont {Aryan}\ \bibnamefont
  {Morita}}, \bibinfo {author} {\bibfnamefont {Felipe~P}\ \bibnamefont
  {Perona-Martinez}}, \ and\ \bibinfo {author} {\bibfnamefont {Romana}\
  \bibnamefont {Schirhagl}},\ }\bibfield  {title} {\enquote {\bibinfo {title}
  {Evaluation of the oxidative stress response of aging yeast cells in response
  to internalization of fluorescent nanodiamond biosensors},}\ }\href@noop {}
  {\bibfield  {journal} {\bibinfo  {journal} {Nanomaterials}\ }\textbf
  {\bibinfo {volume} {10}},\ \bibinfo {pages} {372} (\bibinfo {year}
  {2020})}\BibitemShut {NoStop}%
\bibitem [{\citenamefont {Fujiwara}\ \emph
  {et~al.}(2020{\natexlab{a}})\citenamefont {Fujiwara}, \citenamefont {Sun},
  \citenamefont {Dohms}, \citenamefont {Nishimura}, \citenamefont {Suto},
  \citenamefont {Takezawa}, \citenamefont {Oshimi}, \citenamefont {Zhao},
  \citenamefont {Sadzak}, \citenamefont {Umehara} \emph
  {et~al.}}]{fujiwara2020real}%
  \BibitemOpen
  \bibfield  {author} {\bibinfo {author} {\bibfnamefont {Masazumi}\
  \bibnamefont {Fujiwara}}, \bibinfo {author} {\bibfnamefont {Simo}\
  \bibnamefont {Sun}}, \bibinfo {author} {\bibfnamefont {Alexander}\
  \bibnamefont {Dohms}}, \bibinfo {author} {\bibfnamefont {Yushi}\ \bibnamefont
  {Nishimura}}, \bibinfo {author} {\bibfnamefont {Ken}\ \bibnamefont {Suto}},
  \bibinfo {author} {\bibfnamefont {Yuka}\ \bibnamefont {Takezawa}}, \bibinfo
  {author} {\bibfnamefont {Keisuke}\ \bibnamefont {Oshimi}}, \bibinfo {author}
  {\bibfnamefont {Li}~\bibnamefont {Zhao}}, \bibinfo {author} {\bibfnamefont
  {Nikola}\ \bibnamefont {Sadzak}}, \bibinfo {author} {\bibfnamefont {Yumi}\
  \bibnamefont {Umehara}},  \emph {et~al.},\ }\bibfield  {title} {\enquote
  {\bibinfo {title} {Real-time nanodiamond thermometry probing in vivo
  thermogenic responses},}\ }\href@noop {} {\bibfield  {journal} {\bibinfo
  {journal} {Sci. Adv.}\ }\textbf {\bibinfo {volume} {6}},\ \bibinfo {pages}
  {eaba9636} (\bibinfo {year} {2020}{\natexlab{a}})}\BibitemShut {NoStop}%
\bibitem [{\citenamefont {Zhu}\ \emph {et~al.}(2012)\citenamefont {Zhu},
  \citenamefont {Li}, \citenamefont {Li}, \citenamefont {Zhang}, \citenamefont
  {Yang}, \citenamefont {Chen}, \citenamefont {Sun}, \citenamefont {Zhao},
  \citenamefont {Fan},\ and\ \citenamefont {Huang}}]{zhu2012biocompatibility}%
  \BibitemOpen
  \bibfield  {author} {\bibinfo {author} {\bibfnamefont {Ying}\ \bibnamefont
  {Zhu}}, \bibinfo {author} {\bibfnamefont {Jing}\ \bibnamefont {Li}}, \bibinfo
  {author} {\bibfnamefont {Wenxin}\ \bibnamefont {Li}}, \bibinfo {author}
  {\bibfnamefont {Yu}~\bibnamefont {Zhang}}, \bibinfo {author} {\bibfnamefont
  {Xiafeng}\ \bibnamefont {Yang}}, \bibinfo {author} {\bibfnamefont {Nan}\
  \bibnamefont {Chen}}, \bibinfo {author} {\bibfnamefont {Yanhong}\
  \bibnamefont {Sun}}, \bibinfo {author} {\bibfnamefont {Yun}\ \bibnamefont
  {Zhao}}, \bibinfo {author} {\bibfnamefont {Chunhai}\ \bibnamefont {Fan}}, \
  and\ \bibinfo {author} {\bibfnamefont {Qing}\ \bibnamefont {Huang}},\
  }\bibfield  {title} {\enquote {\bibinfo {title} {The biocompatibility of
  nanodiamonds and their application in drug delivery systems},}\ }\href@noop
  {} {\bibfield  {journal} {\bibinfo  {journal} {Theranostics}\ }\textbf
  {\bibinfo {volume} {2}},\ \bibinfo {pages} {302} (\bibinfo {year}
  {2012})}\BibitemShut {NoStop}%
\bibitem [{\citenamefont {Krueger}(2008)}]{krueger2008new}%
  \BibitemOpen
  \bibfield  {author} {\bibinfo {author} {\bibfnamefont {Anke}\ \bibnamefont
  {Krueger}},\ }\bibfield  {title} {\enquote {\bibinfo {title} {New carbon
  materials: biological applications of functionalized nanodiamond
  materials},}\ }\href@noop {} {\bibfield  {journal} {\bibinfo  {journal}
  {Chem. Eur. J.}\ }\textbf {\bibinfo {volume} {14}},\ \bibinfo {pages}
  {1382--1390} (\bibinfo {year} {2008})}\BibitemShut {NoStop}%
\bibitem [{\citenamefont {Mohan}\ \emph {et~al.}(2010)\citenamefont {Mohan},
  \citenamefont {Chen}, \citenamefont {Hsieh}, \citenamefont {Wu},\ and\
  \citenamefont {Chang}}]{mohan2010vivo}%
  \BibitemOpen
  \bibfield  {author} {\bibinfo {author} {\bibfnamefont {Nitin}\ \bibnamefont
  {Mohan}}, \bibinfo {author} {\bibfnamefont {Chao-Sheng}\ \bibnamefont
  {Chen}}, \bibinfo {author} {\bibfnamefont {Hsiao-Han}\ \bibnamefont {Hsieh}},
  \bibinfo {author} {\bibfnamefont {Yi-Chun}\ \bibnamefont {Wu}}, \ and\
  \bibinfo {author} {\bibfnamefont {Huan-Cheng}\ \bibnamefont {Chang}},\
  }\bibfield  {title} {\enquote {\bibinfo {title} {In vivo imaging and toxicity
  assessments of fluorescent nanodiamonds in caenorhabditis elegans},}\
  }\href@noop {} {\bibfield  {journal} {\bibinfo  {journal} {Nano Lett.}\
  }\textbf {\bibinfo {volume} {10}},\ \bibinfo {pages} {3692--3699} (\bibinfo
  {year} {2010})}\BibitemShut {NoStop}%
\bibitem [{\citenamefont {Hemelaar}\ \emph {et~al.}(2017)\citenamefont
  {Hemelaar}, \citenamefont {van~der Laan}, \citenamefont {Hinterding},
  \citenamefont {Koot}, \citenamefont {Ellermann}, \citenamefont
  {Perona-Martinez}, \citenamefont {Roig}, \citenamefont {Hommelet},
  \citenamefont {Novarina}, \citenamefont {Takahashi} \emph
  {et~al.}}]{hemelaar2017generally}%
  \BibitemOpen
  \bibfield  {author} {\bibinfo {author} {\bibfnamefont {Simon~R}\ \bibnamefont
  {Hemelaar}}, \bibinfo {author} {\bibfnamefont {Kiran~J}\ \bibnamefont
  {van~der Laan}}, \bibinfo {author} {\bibfnamefont {Sophie~R}\ \bibnamefont
  {Hinterding}}, \bibinfo {author} {\bibfnamefont {Manon~V}\ \bibnamefont
  {Koot}}, \bibinfo {author} {\bibfnamefont {Else}\ \bibnamefont {Ellermann}},
  \bibinfo {author} {\bibfnamefont {Felipe~P}\ \bibnamefont {Perona-Martinez}},
  \bibinfo {author} {\bibfnamefont {David}\ \bibnamefont {Roig}}, \bibinfo
  {author} {\bibfnamefont {Severin}\ \bibnamefont {Hommelet}}, \bibinfo
  {author} {\bibfnamefont {Daniele}\ \bibnamefont {Novarina}}, \bibinfo
  {author} {\bibfnamefont {Hiroki}\ \bibnamefont {Takahashi}},  \emph
  {et~al.},\ }\bibfield  {title} {\enquote {\bibinfo {title} {Generally
  applicable transformation protocols for fluorescent nanodiamond
  internalization into cells},}\ }\href@noop {} {\bibfield  {journal} {\bibinfo
   {journal} {Sci. Rep.}\ }\textbf {\bibinfo {volume} {7}},\ \bibinfo {pages}
  {1--7} (\bibinfo {year} {2017})}\BibitemShut {NoStop}%
\bibitem [{\citenamefont {Sotoma}\ \emph {et~al.}(2018)\citenamefont {Sotoma},
  \citenamefont {Hsieh}, \citenamefont {Chen}, \citenamefont {Tsai},\ and\
  \citenamefont {Chang}}]{sotoma2018highly}%
  \BibitemOpen
  \bibfield  {author} {\bibinfo {author} {\bibfnamefont {Shingo}\ \bibnamefont
  {Sotoma}}, \bibinfo {author} {\bibfnamefont {Feng-Jen}\ \bibnamefont
  {Hsieh}}, \bibinfo {author} {\bibfnamefont {Yen-Wei}\ \bibnamefont {Chen}},
  \bibinfo {author} {\bibfnamefont {Pei-Chang}\ \bibnamefont {Tsai}}, \ and\
  \bibinfo {author} {\bibfnamefont {Huan-Cheng}\ \bibnamefont {Chang}},\
  }\bibfield  {title} {\enquote {\bibinfo {title} {Highly stable
  lipid-encapsulation of fluorescent nanodiamonds for bioimaging
  applications},}\ }\href@noop {} {\bibfield  {journal} {\bibinfo  {journal}
  {Chem. Commun.}\ }\textbf {\bibinfo {volume} {54}},\ \bibinfo {pages}
  {1000--1003} (\bibinfo {year} {2018})}\BibitemShut {NoStop}%
\bibitem [{\citenamefont {Reina}\ \emph {et~al.}(2019)\citenamefont {Reina},
  \citenamefont {Zhao}, \citenamefont {Bianco},\ and\ \citenamefont
  {Komatsu}}]{reina2019chemical}%
  \BibitemOpen
  \bibfield  {author} {\bibinfo {author} {\bibfnamefont {Giacomo}\ \bibnamefont
  {Reina}}, \bibinfo {author} {\bibfnamefont {Li}~\bibnamefont {Zhao}},
  \bibinfo {author} {\bibfnamefont {Alberto}\ \bibnamefont {Bianco}}, \ and\
  \bibinfo {author} {\bibfnamefont {Naoki}\ \bibnamefont {Komatsu}},\
  }\bibfield  {title} {\enquote {\bibinfo {title} {Chemical functionalization
  of nanodiamonds: Opportunities and challenges ahead},}\ }\href@noop {}
  {\bibfield  {journal} {\bibinfo  {journal} {Angew. Chem. Int. Ed.}\ }\textbf
  {\bibinfo {volume} {58}},\ \bibinfo {pages} {17918--17929} (\bibinfo {year}
  {2019})}\BibitemShut {NoStop}%
\bibitem [{\citenamefont {Rondin}\ \emph {et~al.}(2014)\citenamefont {Rondin},
  \citenamefont {Tetienne}, \citenamefont {Hingant}, \citenamefont {Roch},
  \citenamefont {Maletinsky},\ and\ \citenamefont
  {Jacques}}]{rondin2014magnetometry}%
  \BibitemOpen
  \bibfield  {author} {\bibinfo {author} {\bibfnamefont {Lo{\"\i}c}\
  \bibnamefont {Rondin}}, \bibinfo {author} {\bibfnamefont {Jean-Philippe}\
  \bibnamefont {Tetienne}}, \bibinfo {author} {\bibfnamefont {Thomas}\
  \bibnamefont {Hingant}}, \bibinfo {author} {\bibfnamefont
  {Jean-Fran{\c{c}}ois}\ \bibnamefont {Roch}}, \bibinfo {author} {\bibfnamefont
  {Patrick}\ \bibnamefont {Maletinsky}}, \ and\ \bibinfo {author}
  {\bibfnamefont {Vincent}\ \bibnamefont {Jacques}},\ }\bibfield  {title}
  {\enquote {\bibinfo {title} {Magnetometry with nitrogen-vacancy defects in
  diamond},}\ }\href@noop {} {\bibfield  {journal} {\bibinfo  {journal} {Rep.
  Prog. Phys.}\ }\textbf {\bibinfo {volume} {77}},\ \bibinfo {pages} {056503}
  (\bibinfo {year} {2014})}\BibitemShut {NoStop}%
\bibitem [{\citenamefont {Maclaurin}\ \emph {et~al.}(2013)\citenamefont
  {Maclaurin}, \citenamefont {Hall}, \citenamefont {Martin},\ and\
  \citenamefont {Hollenberg}}]{maclaurin2013nanoscale}%
  \BibitemOpen
  \bibfield  {author} {\bibinfo {author} {\bibfnamefont {D}~\bibnamefont
  {Maclaurin}}, \bibinfo {author} {\bibfnamefont {LT}~\bibnamefont {Hall}},
  \bibinfo {author} {\bibfnamefont {AM}~\bibnamefont {Martin}}, \ and\ \bibinfo
  {author} {\bibfnamefont {LCL}\ \bibnamefont {Hollenberg}},\ }\bibfield
  {title} {\enquote {\bibinfo {title} {Nanoscale magnetometry through quantum
  control of nitrogen--vacancy centres in rotationally diffusing
  nanodiamonds},}\ }\href@noop {} {\bibfield  {journal} {\bibinfo  {journal}
  {New J. Phys.}\ }\textbf {\bibinfo {volume} {15}},\ \bibinfo {pages} {013041}
  (\bibinfo {year} {2013})}\BibitemShut {NoStop}%
\bibitem [{\citenamefont {Horowitz}\ \emph {et~al.}(2012)\citenamefont
  {Horowitz}, \citenamefont {Alem{\'a}n}, \citenamefont {Christle},
  \citenamefont {Cleland},\ and\ \citenamefont
  {Awschalom}}]{horowitz2012electron}%
  \BibitemOpen
  \bibfield  {author} {\bibinfo {author} {\bibfnamefont {Viva~R}\ \bibnamefont
  {Horowitz}}, \bibinfo {author} {\bibfnamefont {Benjam{\'\i}n~J}\ \bibnamefont
  {Alem{\'a}n}}, \bibinfo {author} {\bibfnamefont {David~J}\ \bibnamefont
  {Christle}}, \bibinfo {author} {\bibfnamefont {Andrew~N}\ \bibnamefont
  {Cleland}}, \ and\ \bibinfo {author} {\bibfnamefont {David~D}\ \bibnamefont
  {Awschalom}},\ }\bibfield  {title} {\enquote {\bibinfo {title} {Electron spin
  resonance of nitrogen-vacancy centers in optically trapped nanodiamonds},}\
  }\href@noop {} {\bibfield  {journal} {\bibinfo  {journal} {Proc. Natl. Acad.
  Sci. U.S.A.}\ }\textbf {\bibinfo {volume} {109}},\ \bibinfo {pages}
  {13493--13497} (\bibinfo {year} {2012})}\BibitemShut {NoStop}%
\bibitem [{\citenamefont {Dolde}\ \emph {et~al.}(2011)\citenamefont {Dolde},
  \citenamefont {Fedder}, \citenamefont {Doherty}, \citenamefont {N{\"o}bauer},
  \citenamefont {Rempp}, \citenamefont {Balasubramanian}, \citenamefont {Wolf},
  \citenamefont {Reinhard}, \citenamefont {Hollenberg}, \citenamefont {Jelezko}
  \emph {et~al.}}]{dolde2011electric}%
  \BibitemOpen
  \bibfield  {author} {\bibinfo {author} {\bibfnamefont {Florian}\ \bibnamefont
  {Dolde}}, \bibinfo {author} {\bibfnamefont {Helmut}\ \bibnamefont {Fedder}},
  \bibinfo {author} {\bibfnamefont {Marcus~W}\ \bibnamefont {Doherty}},
  \bibinfo {author} {\bibfnamefont {Tobias}\ \bibnamefont {N{\"o}bauer}},
  \bibinfo {author} {\bibfnamefont {Florian}\ \bibnamefont {Rempp}}, \bibinfo
  {author} {\bibfnamefont {Gopalakrishnan}\ \bibnamefont {Balasubramanian}},
  \bibinfo {author} {\bibfnamefont {Thomas}\ \bibnamefont {Wolf}}, \bibinfo
  {author} {\bibfnamefont {Friedemann}\ \bibnamefont {Reinhard}}, \bibinfo
  {author} {\bibfnamefont {Lloyd~CL}\ \bibnamefont {Hollenberg}}, \bibinfo
  {author} {\bibfnamefont {Fedor}\ \bibnamefont {Jelezko}},  \emph {et~al.},\
  }\bibfield  {title} {\enquote {\bibinfo {title} {Electric-field sensing using
  single diamond spins},}\ }\href@noop {} {\bibfield  {journal} {\bibinfo
  {journal} {Nat. Phys.}\ }\textbf {\bibinfo {volume} {7}},\ \bibinfo {pages}
  {459--463} (\bibinfo {year} {2011})}\BibitemShut {NoStop}%
\bibitem [{\citenamefont {Iwasaki}\ \emph {et~al.}(2017)\citenamefont
  {Iwasaki}, \citenamefont {Naruki}, \citenamefont {Tahara}, \citenamefont
  {Makino}, \citenamefont {Kato}, \citenamefont {Ogura}, \citenamefont
  {Takeuchi}, \citenamefont {Yamasaki},\ and\ \citenamefont
  {Hatano}}]{iwasaki2017direct}%
  \BibitemOpen
  \bibfield  {author} {\bibinfo {author} {\bibfnamefont {Takayuki}\
  \bibnamefont {Iwasaki}}, \bibinfo {author} {\bibfnamefont {Wataru}\
  \bibnamefont {Naruki}}, \bibinfo {author} {\bibfnamefont {Kosuke}\
  \bibnamefont {Tahara}}, \bibinfo {author} {\bibfnamefont {Toshiharu}\
  \bibnamefont {Makino}}, \bibinfo {author} {\bibfnamefont {Hiromitsu}\
  \bibnamefont {Kato}}, \bibinfo {author} {\bibfnamefont {Masahiko}\
  \bibnamefont {Ogura}}, \bibinfo {author} {\bibfnamefont {Daisuke}\
  \bibnamefont {Takeuchi}}, \bibinfo {author} {\bibfnamefont {Satoshi}\
  \bibnamefont {Yamasaki}}, \ and\ \bibinfo {author} {\bibfnamefont {Mutsuko}\
  \bibnamefont {Hatano}},\ }\bibfield  {title} {\enquote {\bibinfo {title}
  {Direct nanoscale sensing of the internal electric field in operating
  semiconductor devices using single electron spins},}\ }\href@noop {}
  {\bibfield  {journal} {\bibinfo  {journal} {ACS Nano}\ }\textbf {\bibinfo
  {volume} {11}},\ \bibinfo {pages} {1238--1245} (\bibinfo {year}
  {2017})}\BibitemShut {NoStop}%
\bibitem [{\citenamefont {Bian}\ \emph {et~al.}(2021)\citenamefont {Bian},
  \citenamefont {Zheng}, \citenamefont {Zeng}, \citenamefont {Chen},
  \citenamefont {St{\"o}hr}, \citenamefont {Denisenko}, \citenamefont {Yang},
  \citenamefont {Wrachtrup},\ and\ \citenamefont {Jiang}}]{bian2021nanoscale}%
  \BibitemOpen
  \bibfield  {author} {\bibinfo {author} {\bibfnamefont {Ke}~\bibnamefont
  {Bian}}, \bibinfo {author} {\bibfnamefont {Wentian}\ \bibnamefont {Zheng}},
  \bibinfo {author} {\bibfnamefont {Xianzhe}\ \bibnamefont {Zeng}}, \bibinfo
  {author} {\bibfnamefont {Xiakun}\ \bibnamefont {Chen}}, \bibinfo {author}
  {\bibfnamefont {Rainer}\ \bibnamefont {St{\"o}hr}}, \bibinfo {author}
  {\bibfnamefont {Andrej}\ \bibnamefont {Denisenko}}, \bibinfo {author}
  {\bibfnamefont {Sen}\ \bibnamefont {Yang}}, \bibinfo {author} {\bibfnamefont
  {J{\"o}rg}\ \bibnamefont {Wrachtrup}}, \ and\ \bibinfo {author}
  {\bibfnamefont {Ying}\ \bibnamefont {Jiang}},\ }\bibfield  {title} {\enquote
  {\bibinfo {title} {Nanoscale electric-field imaging based on a quantum sensor
  and its charge-state control under ambient condition},}\ }\href@noop {}
  {\bibfield  {journal} {\bibinfo  {journal} {Nat. Commun.}\ }\textbf {\bibinfo
  {volume} {12}},\ \bibinfo {pages} {2457} (\bibinfo {year}
  {2021})}\BibitemShut {NoStop}%
\bibitem [{\citenamefont {Neumann}\ \emph {et~al.}(2013)\citenamefont
  {Neumann}, \citenamefont {Jakobi}, \citenamefont {Dolde}, \citenamefont
  {Burk}, \citenamefont {Reuter}, \citenamefont {Waldherr}, \citenamefont
  {Honert}, \citenamefont {Wolf}, \citenamefont {Brunner}, \citenamefont {Shim}
  \emph {et~al.}}]{neumann2013high}%
  \BibitemOpen
  \bibfield  {author} {\bibinfo {author} {\bibfnamefont {Philipp}\ \bibnamefont
  {Neumann}}, \bibinfo {author} {\bibfnamefont {Ingmar}\ \bibnamefont
  {Jakobi}}, \bibinfo {author} {\bibfnamefont {Florian}\ \bibnamefont {Dolde}},
  \bibinfo {author} {\bibfnamefont {Christian}\ \bibnamefont {Burk}}, \bibinfo
  {author} {\bibfnamefont {Rolf}\ \bibnamefont {Reuter}}, \bibinfo {author}
  {\bibfnamefont {Gerald}\ \bibnamefont {Waldherr}}, \bibinfo {author}
  {\bibfnamefont {Jan}\ \bibnamefont {Honert}}, \bibinfo {author}
  {\bibfnamefont {Thomas}\ \bibnamefont {Wolf}}, \bibinfo {author}
  {\bibfnamefont {Andreas}\ \bibnamefont {Brunner}}, \bibinfo {author}
  {\bibfnamefont {Jeong~Hyun}\ \bibnamefont {Shim}},  \emph {et~al.},\
  }\bibfield  {title} {\enquote {\bibinfo {title} {High-precision nanoscale
  temperature sensing using single defects in diamond},}\ }\href@noop {}
  {\bibfield  {journal} {\bibinfo  {journal} {Nano Lett.}\ }\textbf {\bibinfo
  {volume} {13}},\ \bibinfo {pages} {2738--2742} (\bibinfo {year}
  {2013})}\BibitemShut {NoStop}%
\bibitem [{\citenamefont {Fujiwara}\ and\ \citenamefont
  {Shikano}(2021)}]{fujiwara2021diamond}%
  \BibitemOpen
  \bibfield  {author} {\bibinfo {author} {\bibfnamefont {Masazumi}\
  \bibnamefont {Fujiwara}}\ and\ \bibinfo {author} {\bibfnamefont {Yutaka}\
  \bibnamefont {Shikano}},\ }\bibfield  {title} {\enquote {\bibinfo {title}
  {Diamond quantum thermometry: From foundations to applications},}\
  }\href@noop {} {\bibfield  {journal} {\bibinfo  {journal} {Nanotechnology}\
  }\textbf {\bibinfo {volume} {32}},\ \bibinfo {pages} {482002} (\bibinfo
  {year} {2021})}\BibitemShut {NoStop}%
\bibitem [{\citenamefont {Wang}\ \emph {et~al.}(2018)\citenamefont {Wang},
  \citenamefont {Liu}, \citenamefont {Leong}, \citenamefont {Zeng},
  \citenamefont {Feng}, \citenamefont {Li}, \citenamefont {Dolde},
  \citenamefont {Fedder}, \citenamefont {Wrachtrup}, \citenamefont {Cui} \emph
  {et~al.}}]{wang2018magnetic}%
  \BibitemOpen
  \bibfield  {author} {\bibinfo {author} {\bibfnamefont {Ning}\ \bibnamefont
  {Wang}}, \bibinfo {author} {\bibfnamefont {Gang-Qin}\ \bibnamefont {Liu}},
  \bibinfo {author} {\bibfnamefont {Weng-Hang}\ \bibnamefont {Leong}}, \bibinfo
  {author} {\bibfnamefont {Hualing}\ \bibnamefont {Zeng}}, \bibinfo {author}
  {\bibfnamefont {Xi}~\bibnamefont {Feng}}, \bibinfo {author} {\bibfnamefont
  {Si-Hong}\ \bibnamefont {Li}}, \bibinfo {author} {\bibfnamefont {Florian}\
  \bibnamefont {Dolde}}, \bibinfo {author} {\bibfnamefont {Helmut}\
  \bibnamefont {Fedder}}, \bibinfo {author} {\bibfnamefont {J{\"o}rg}\
  \bibnamefont {Wrachtrup}}, \bibinfo {author} {\bibfnamefont {Xiao-Dong}\
  \bibnamefont {Cui}},  \emph {et~al.},\ }\bibfield  {title} {\enquote
  {\bibinfo {title} {Magnetic criticality enhanced hybrid nanodiamond
  thermometer under ambient conditions},}\ }\href@noop {} {\bibfield  {journal}
  {\bibinfo  {journal} {Phys. Rev. X.}\ }\textbf {\bibinfo {volume} {8}},\
  \bibinfo {pages} {011042} (\bibinfo {year} {2018})}\BibitemShut {NoStop}%
\bibitem [{\citenamefont {Mazurczyk}\ \emph {et~al.}(2006)\citenamefont
  {Mazurczyk}, \citenamefont {Vieillard}, \citenamefont {Bouchard},
  \citenamefont {Hannes},\ and\ \citenamefont {Krawczyk}}]{mazurczyk2006novel}%
  \BibitemOpen
  \bibfield  {author} {\bibinfo {author} {\bibfnamefont {Radoslaw}\
  \bibnamefont {Mazurczyk}}, \bibinfo {author} {\bibfnamefont {Julien}\
  \bibnamefont {Vieillard}}, \bibinfo {author} {\bibfnamefont {Aude}\
  \bibnamefont {Bouchard}}, \bibinfo {author} {\bibfnamefont {Benjamin}\
  \bibnamefont {Hannes}}, \ and\ \bibinfo {author} {\bibfnamefont {Stanislas}\
  \bibnamefont {Krawczyk}},\ }\bibfield  {title} {\enquote {\bibinfo {title} {A
  novel concept of the integrated fluorescence detection system and its
  application in a lab-on-a-chip microdevice},}\ }\href@noop {} {\bibfield
  {journal} {\bibinfo  {journal} {Sens. Actuators B Chem.}\ }\textbf {\bibinfo
  {volume} {118}},\ \bibinfo {pages} {11--19} (\bibinfo {year}
  {2006})}\BibitemShut {NoStop}%
\bibitem [{\citenamefont {Dochow}\ \emph {et~al.}(2013)\citenamefont {Dochow},
  \citenamefont {Becker}, \citenamefont {Spittel}, \citenamefont {Beleites},
  \citenamefont {Stanca}, \citenamefont {Latka}, \citenamefont {Schuster},
  \citenamefont {Kobelke}, \citenamefont {Unger}, \citenamefont {Henkel} \emph
  {et~al.}}]{dochow2013raman}%
  \BibitemOpen
  \bibfield  {author} {\bibinfo {author} {\bibfnamefont {Sebastian}\
  \bibnamefont {Dochow}}, \bibinfo {author} {\bibfnamefont {Martin}\
  \bibnamefont {Becker}}, \bibinfo {author} {\bibfnamefont {Ron}\ \bibnamefont
  {Spittel}}, \bibinfo {author} {\bibfnamefont {Claudia}\ \bibnamefont
  {Beleites}}, \bibinfo {author} {\bibfnamefont {Sarmiza}\ \bibnamefont
  {Stanca}}, \bibinfo {author} {\bibfnamefont {Ines}\ \bibnamefont {Latka}},
  \bibinfo {author} {\bibfnamefont {Kay}\ \bibnamefont {Schuster}}, \bibinfo
  {author} {\bibfnamefont {Jens}\ \bibnamefont {Kobelke}}, \bibinfo {author}
  {\bibfnamefont {Sonja}\ \bibnamefont {Unger}}, \bibinfo {author}
  {\bibfnamefont {Thomas}\ \bibnamefont {Henkel}},  \emph {et~al.},\ }\bibfield
   {title} {\enquote {\bibinfo {title} {Raman-on-chip device and detection
  fibres with fibre bragg grating for analysis of solutions and particles},}\
  }\href@noop {} {\bibfield  {journal} {\bibinfo  {journal} {Lab Chip}\
  }\textbf {\bibinfo {volume} {13}},\ \bibinfo {pages} {1109--1113} (\bibinfo
  {year} {2013})}\BibitemShut {NoStop}%
\bibitem [{\citenamefont {Qi}\ \emph {et~al.}(2018)\citenamefont {Qi},
  \citenamefont {Li}, \citenamefont {Wang}, \citenamefont {Fu}, \citenamefont
  {Luo},\ and\ \citenamefont {Chen}}]{qi2018rotational}%
  \BibitemOpen
  \bibfield  {author} {\bibinfo {author} {\bibfnamefont {Ji}~\bibnamefont
  {Qi}}, \bibinfo {author} {\bibfnamefont {Bowei}\ \bibnamefont {Li}}, \bibinfo
  {author} {\bibfnamefont {Xiaoyan}\ \bibnamefont {Wang}}, \bibinfo {author}
  {\bibfnamefont {Longwen}\ \bibnamefont {Fu}}, \bibinfo {author}
  {\bibfnamefont {Liqiang}\ \bibnamefont {Luo}}, \ and\ \bibinfo {author}
  {\bibfnamefont {Lingxin}\ \bibnamefont {Chen}},\ }\bibfield  {title}
  {\enquote {\bibinfo {title} {Rotational paper-based microfluidic-chip device
  for multiplexed and simultaneous fluorescence detection of phenolic
  pollutants based on a molecular-imprinting technique},}\ }\href@noop {}
  {\bibfield  {journal} {\bibinfo  {journal} {Anal. Chem.}\ }\textbf {\bibinfo
  {volume} {90}},\ \bibinfo {pages} {11827--11834} (\bibinfo {year}
  {2018})}\BibitemShut {NoStop}%
\bibitem [{\citenamefont {Marchiarullo}\ \emph {et~al.}(2013)\citenamefont
  {Marchiarullo}, \citenamefont {Sklavounos}, \citenamefont {Oh}, \citenamefont
  {Poe}, \citenamefont {Barker},\ and\ \citenamefont
  {Landers}}]{marchiarullo2013low}%
  \BibitemOpen
  \bibfield  {author} {\bibinfo {author} {\bibfnamefont {Daniel~J}\
  \bibnamefont {Marchiarullo}}, \bibinfo {author} {\bibfnamefont {Angelique~H}\
  \bibnamefont {Sklavounos}}, \bibinfo {author} {\bibfnamefont {Kyudam}\
  \bibnamefont {Oh}}, \bibinfo {author} {\bibfnamefont {Brian~L}\ \bibnamefont
  {Poe}}, \bibinfo {author} {\bibfnamefont {N~Scott}\ \bibnamefont {Barker}}, \
  and\ \bibinfo {author} {\bibfnamefont {James~P}\ \bibnamefont {Landers}},\
  }\bibfield  {title} {\enquote {\bibinfo {title} {Low-power microwave-mediated
  heating for microchip-based pcr},}\ }\href@noop {} {\bibfield  {journal}
  {\bibinfo  {journal} {Lab Chip}\ }\textbf {\bibinfo {volume} {13}},\ \bibinfo
  {pages} {3417--3425} (\bibinfo {year} {2013})}\BibitemShut {NoStop}%
\bibitem [{\citenamefont {Yesiloz}\ \emph {et~al.}(2015)\citenamefont
  {Yesiloz}, \citenamefont {Boybay},\ and\ \citenamefont
  {Ren}}]{yesiloz2015label}%
  \BibitemOpen
  \bibfield  {author} {\bibinfo {author} {\bibfnamefont {Gurkan}\ \bibnamefont
  {Yesiloz}}, \bibinfo {author} {\bibfnamefont {Muhammed~Said}\ \bibnamefont
  {Boybay}}, \ and\ \bibinfo {author} {\bibfnamefont {Carolyn~L}\ \bibnamefont
  {Ren}},\ }\bibfield  {title} {\enquote {\bibinfo {title} {Label-free
  high-throughput detection and content sensing of individual droplets in
  microfluidic systems},}\ }\href@noop {} {\bibfield  {journal} {\bibinfo
  {journal} {Lab Chip}\ }\textbf {\bibinfo {volume} {15}},\ \bibinfo {pages}
  {4008--4019} (\bibinfo {year} {2015})}\BibitemShut {NoStop}%
\bibitem [{\citenamefont {Wong}\ \emph {et~al.}(2016)\citenamefont {Wong},
  \citenamefont {Yesiloz}, \citenamefont {Boybay},\ and\ \citenamefont
  {Ren}}]{wong2016microwave}%
  \BibitemOpen
  \bibfield  {author} {\bibinfo {author} {\bibfnamefont {David}\ \bibnamefont
  {Wong}}, \bibinfo {author} {\bibfnamefont {Gurkan}\ \bibnamefont {Yesiloz}},
  \bibinfo {author} {\bibfnamefont {Muhammed~S}\ \bibnamefont {Boybay}}, \ and\
  \bibinfo {author} {\bibfnamefont {Carolyn~L}\ \bibnamefont {Ren}},\
  }\bibfield  {title} {\enquote {\bibinfo {title} {Microwave temperature
  measurement in microfluidic devices},}\ }\href@noop {} {\bibfield  {journal}
  {\bibinfo  {journal} {Lab Chip}\ }\textbf {\bibinfo {volume} {16}},\ \bibinfo
  {pages} {2192--2197} (\bibinfo {year} {2016})}\BibitemShut {NoStop}%
\bibitem [{\citenamefont {Chen}\ \emph {et~al.}(2020)\citenamefont {Chen},
  \citenamefont {Hung}, \citenamefont {Lo}, \citenamefont {Chen}, \citenamefont
  {Shen}, \citenamefont {Kafenda}, \citenamefont {Su}, \citenamefont {Xia},\
  and\ \citenamefont {Yang}}]{chen2020universal}%
  \BibitemOpen
  \bibfield  {author} {\bibinfo {author} {\bibfnamefont {Yifan}\ \bibnamefont
  {Chen}}, \bibinfo {author} {\bibfnamefont {Siu~Fai}\ \bibnamefont {Hung}},
  \bibinfo {author} {\bibfnamefont {Wing~Ki}\ \bibnamefont {Lo}}, \bibinfo
  {author} {\bibfnamefont {Yang}\ \bibnamefont {Chen}}, \bibinfo {author}
  {\bibfnamefont {Yang}\ \bibnamefont {Shen}}, \bibinfo {author} {\bibfnamefont
  {Kim}\ \bibnamefont {Kafenda}}, \bibinfo {author} {\bibfnamefont {Jia}\
  \bibnamefont {Su}}, \bibinfo {author} {\bibfnamefont {Kangwei}\ \bibnamefont
  {Xia}}, \ and\ \bibinfo {author} {\bibfnamefont {Sen}\ \bibnamefont {Yang}},\
  }\bibfield  {title} {\enquote {\bibinfo {title} {A universal method for
  depositing patterned materials in situ},}\ }\href@noop {} {\bibfield
  {journal} {\bibinfo  {journal} {Nat. Commun.}\ }\textbf {\bibinfo {volume}
  {11}},\ \bibinfo {pages} {1--8} (\bibinfo {year} {2020})}\BibitemShut
  {NoStop}%
\bibitem [{\citenamefont {Fukushige}\ \emph {et~al.}(2020)\citenamefont
  {Fukushige}, \citenamefont {Kawaguchi}, \citenamefont {Shimazaki},
  \citenamefont {Tashima}, \citenamefont {Takashima},\ and\ \citenamefont
  {Takeuchi}}]{fukushige2020identification}%
  \BibitemOpen
  \bibfield  {author} {\bibinfo {author} {\bibfnamefont {Kazuki}\ \bibnamefont
  {Fukushige}}, \bibinfo {author} {\bibfnamefont {Hiroki}\ \bibnamefont
  {Kawaguchi}}, \bibinfo {author} {\bibfnamefont {Konosuke}\ \bibnamefont
  {Shimazaki}}, \bibinfo {author} {\bibfnamefont {Toshiyuki}\ \bibnamefont
  {Tashima}}, \bibinfo {author} {\bibfnamefont {Hideaki}\ \bibnamefont
  {Takashima}}, \ and\ \bibinfo {author} {\bibfnamefont {Shigeki}\ \bibnamefont
  {Takeuchi}},\ }\bibfield  {title} {\enquote {\bibinfo {title} {Identification
  of the orientation of a single nv center in a nanodiamond using a
  three-dimensionally controlled magnetic field},}\ }\href@noop {} {\bibfield
  {journal} {\bibinfo  {journal} {Appl. Phys. Lett.}\ }\textbf {\bibinfo
  {volume} {116}},\ \bibinfo {pages} {264002} (\bibinfo {year}
  {2020})}\BibitemShut {NoStop}%
\bibitem [{\citenamefont {Sadzak}\ \emph {et~al.}(2018)\citenamefont {Sadzak},
  \citenamefont {H{\'e}ritier},\ and\ \citenamefont
  {Benson}}]{sadzak2018coupling}%
  \BibitemOpen
  \bibfield  {author} {\bibinfo {author} {\bibfnamefont {Nikola}\ \bibnamefont
  {Sadzak}}, \bibinfo {author} {\bibfnamefont {Martin}\ \bibnamefont
  {H{\'e}ritier}}, \ and\ \bibinfo {author} {\bibfnamefont {Oliver}\
  \bibnamefont {Benson}},\ }\bibfield  {title} {\enquote {\bibinfo {title}
  {Coupling a single nitrogen-vacancy center in nanodiamond to
  superparamagnetic nanoparticles},}\ }\href@noop {} {\bibfield  {journal}
  {\bibinfo  {journal} {Sci. Rep.}\ }\textbf {\bibinfo {volume} {8}},\ \bibinfo
  {pages} {1--8} (\bibinfo {year} {2018})}\BibitemShut {NoStop}%
\bibitem [{\citenamefont {Jia}\ \emph {et~al.}(2018)\citenamefont {Jia},
  \citenamefont {Shi}, \citenamefont {Qin}, \citenamefont {Rong},\ and\
  \citenamefont {Du}}]{jia2018ultra}%
  \BibitemOpen
  \bibfield  {author} {\bibinfo {author} {\bibfnamefont {Wenfei}\ \bibnamefont
  {Jia}}, \bibinfo {author} {\bibfnamefont {Zhifu}\ \bibnamefont {Shi}},
  \bibinfo {author} {\bibfnamefont {Xi}~\bibnamefont {Qin}}, \bibinfo {author}
  {\bibfnamefont {Xing}\ \bibnamefont {Rong}}, \ and\ \bibinfo {author}
  {\bibfnamefont {Jiangfeng}\ \bibnamefont {Du}},\ }\bibfield  {title}
  {\enquote {\bibinfo {title} {Ultra-broadband coplanar waveguide for optically
  detected magnetic resonance of nitrogen-vacancy centers in diamond},}\
  }\href@noop {} {\bibfield  {journal} {\bibinfo  {journal} {Rev. Sci.
  Instrum.}\ }\textbf {\bibinfo {volume} {89}},\ \bibinfo {pages} {064705}
  (\bibinfo {year} {2018})}\BibitemShut {NoStop}%
\bibitem [{\citenamefont {Igarashi}\ \emph {et~al.}(2012)\citenamefont
  {Igarashi}, \citenamefont {Yoshinari}, \citenamefont {Yokota}, \citenamefont
  {Sugi}, \citenamefont {Sugihara}, \citenamefont {Ikeda}, \citenamefont
  {Sumiya}, \citenamefont {Tsuji}, \citenamefont {Mori}, \citenamefont {Tochio}
  \emph {et~al.}}]{igarashi2012real}%
  \BibitemOpen
  \bibfield  {author} {\bibinfo {author} {\bibfnamefont {Ryuji}\ \bibnamefont
  {Igarashi}}, \bibinfo {author} {\bibfnamefont {Yohsuke}\ \bibnamefont
  {Yoshinari}}, \bibinfo {author} {\bibfnamefont {Hiroaki}\ \bibnamefont
  {Yokota}}, \bibinfo {author} {\bibfnamefont {Takuma}\ \bibnamefont {Sugi}},
  \bibinfo {author} {\bibfnamefont {Fuminori}\ \bibnamefont {Sugihara}},
  \bibinfo {author} {\bibfnamefont {Kazuhiro}\ \bibnamefont {Ikeda}}, \bibinfo
  {author} {\bibfnamefont {Hitoshi}\ \bibnamefont {Sumiya}}, \bibinfo {author}
  {\bibfnamefont {Shigenori}\ \bibnamefont {Tsuji}}, \bibinfo {author}
  {\bibfnamefont {Ikue}\ \bibnamefont {Mori}}, \bibinfo {author} {\bibfnamefont
  {Hidehito}\ \bibnamefont {Tochio}},  \emph {et~al.},\ }\bibfield  {title}
  {\enquote {\bibinfo {title} {Real-time background-free selective imaging of
  fluorescent nanodiamonds in vivo},}\ }\href@noop {} {\bibfield  {journal}
  {\bibinfo  {journal} {Nano Lett.}\ }\textbf {\bibinfo {volume} {12}},\
  \bibinfo {pages} {5726--5732} (\bibinfo {year} {2012})}\BibitemShut {NoStop}%
\bibitem [{\citenamefont {Ajoy}\ \emph {et~al.}(2020)\citenamefont {Ajoy},
  \citenamefont {Nazaryan}, \citenamefont {Druga}, \citenamefont {Liu},
  \citenamefont {Aguilar}, \citenamefont {Han}, \citenamefont {Gierth},
  \citenamefont {Oon}, \citenamefont {Safvati}, \citenamefont {Tsang} \emph
  {et~al.}}]{ajoy2020room}%
  \BibitemOpen
  \bibfield  {author} {\bibinfo {author} {\bibfnamefont {Ashok}\ \bibnamefont
  {Ajoy}}, \bibinfo {author} {\bibfnamefont {Raffi}\ \bibnamefont {Nazaryan}},
  \bibinfo {author} {\bibfnamefont {Emanuel}\ \bibnamefont {Druga}}, \bibinfo
  {author} {\bibfnamefont {Kristina}\ \bibnamefont {Liu}}, \bibinfo {author}
  {\bibfnamefont {Alessandra}\ \bibnamefont {Aguilar}}, \bibinfo {author}
  {\bibfnamefont {Ben}\ \bibnamefont {Han}}, \bibinfo {author} {\bibfnamefont
  {Max}\ \bibnamefont {Gierth}}, \bibinfo {author} {\bibfnamefont {Jner~T}\
  \bibnamefont {Oon}}, \bibinfo {author} {\bibfnamefont {Ben}\ \bibnamefont
  {Safvati}}, \bibinfo {author} {\bibfnamefont {Ryan}\ \bibnamefont {Tsang}},
  \emph {et~al.},\ }\bibfield  {title} {\enquote {\bibinfo {title} {Room
  temperature “optical nanodiamond hyperpolarizer”: Physics, design, and
  operation},}\ }\href@noop {} {\bibfield  {journal} {\bibinfo  {journal} {Rev.
  Sci. Instrum.}\ }\textbf {\bibinfo {volume} {91}},\ \bibinfo {pages} {023106}
  (\bibinfo {year} {2020})}\BibitemShut {NoStop}%
\bibitem [{\citenamefont {Choi}\ \emph {et~al.}(2020)\citenamefont {Choi},
  \citenamefont {Zhou}, \citenamefont {Landig}, \citenamefont {Wu},
  \citenamefont {Yu}, \citenamefont {Von~Stetina}, \citenamefont {Kucsko},
  \citenamefont {Mango}, \citenamefont {Needleman}, \citenamefont {Samuel}
  \emph {et~al.}}]{choi2020probing}%
  \BibitemOpen
  \bibfield  {author} {\bibinfo {author} {\bibfnamefont {Joonhee}\ \bibnamefont
  {Choi}}, \bibinfo {author} {\bibfnamefont {Hengyun}\ \bibnamefont {Zhou}},
  \bibinfo {author} {\bibfnamefont {Renate}\ \bibnamefont {Landig}}, \bibinfo
  {author} {\bibfnamefont {Hai-Yin}\ \bibnamefont {Wu}}, \bibinfo {author}
  {\bibfnamefont {Xiaofei}\ \bibnamefont {Yu}}, \bibinfo {author}
  {\bibfnamefont {Stephen~E}\ \bibnamefont {Von~Stetina}}, \bibinfo {author}
  {\bibfnamefont {Georg}\ \bibnamefont {Kucsko}}, \bibinfo {author}
  {\bibfnamefont {Susan~E}\ \bibnamefont {Mango}}, \bibinfo {author}
  {\bibfnamefont {Daniel~J}\ \bibnamefont {Needleman}}, \bibinfo {author}
  {\bibfnamefont {Aravinthan~DT}\ \bibnamefont {Samuel}},  \emph {et~al.},\
  }\bibfield  {title} {\enquote {\bibinfo {title} {Probing and manipulating
  embryogenesis via nanoscale thermometry and temperature control},}\
  }\href@noop {} {\bibfield  {journal} {\bibinfo  {journal} {Proc. Natl. Acad.
  Sci. U.S.A.}\ }\textbf {\bibinfo {volume} {117}},\ \bibinfo {pages}
  {14636--14641} (\bibinfo {year} {2020})}\BibitemShut {NoStop}%
\bibitem [{\citenamefont {Bolshedvorskii}\ \emph {et~al.}(2017)\citenamefont
  {Bolshedvorskii}, \citenamefont {Vorobyov}, \citenamefont {Soshenko},
  \citenamefont {Shershulin}, \citenamefont {Javadzade}, \citenamefont
  {Zeleneev}, \citenamefont {Komrakova}, \citenamefont {Sorokin}, \citenamefont
  {Belobrov}, \citenamefont {Smolyaninov} \emph
  {et~al.}}]{bolshedvorskii2017single}%
  \BibitemOpen
  \bibfield  {author} {\bibinfo {author} {\bibfnamefont {Stepan~V}\
  \bibnamefont {Bolshedvorskii}}, \bibinfo {author} {\bibfnamefont {Vadim~V}\
  \bibnamefont {Vorobyov}}, \bibinfo {author} {\bibfnamefont {Vladimir~V}\
  \bibnamefont {Soshenko}}, \bibinfo {author} {\bibfnamefont {Vladimir~A}\
  \bibnamefont {Shershulin}}, \bibinfo {author} {\bibfnamefont {Javid}\
  \bibnamefont {Javadzade}}, \bibinfo {author} {\bibfnamefont {Anton~I}\
  \bibnamefont {Zeleneev}}, \bibinfo {author} {\bibfnamefont {Sofya~A}\
  \bibnamefont {Komrakova}}, \bibinfo {author} {\bibfnamefont {Vadim~N}\
  \bibnamefont {Sorokin}}, \bibinfo {author} {\bibfnamefont {Peter~I}\
  \bibnamefont {Belobrov}}, \bibinfo {author} {\bibfnamefont {Andrey~N}\
  \bibnamefont {Smolyaninov}},  \emph {et~al.},\ }\bibfield  {title} {\enquote
  {\bibinfo {title} {Single bright nv centers in aggregates of detonation
  nanodiamonds},}\ }\href@noop {} {\bibfield  {journal} {\bibinfo  {journal}
  {Opt. Mater. Express}\ }\textbf {\bibinfo {volume} {7}},\ \bibinfo {pages}
  {4038--4049} (\bibinfo {year} {2017})}\BibitemShut {NoStop}%
\bibitem [{\citenamefont {Schloss}\ \emph {et~al.}(2018)\citenamefont
  {Schloss}, \citenamefont {Barry}, \citenamefont {Turner},\ and\ \citenamefont
  {Walsworth}}]{schloss2018simultaneous}%
  \BibitemOpen
  \bibfield  {author} {\bibinfo {author} {\bibfnamefont {Jennifer~M}\
  \bibnamefont {Schloss}}, \bibinfo {author} {\bibfnamefont {John~F}\
  \bibnamefont {Barry}}, \bibinfo {author} {\bibfnamefont {Matthew~J}\
  \bibnamefont {Turner}}, \ and\ \bibinfo {author} {\bibfnamefont {Ronald~L}\
  \bibnamefont {Walsworth}},\ }\bibfield  {title} {\enquote {\bibinfo {title}
  {Simultaneous broadband vector magnetometry using solid-state spins},}\
  }\href@noop {} {\bibfield  {journal} {\bibinfo  {journal} {Phys. Rev. Appl.}\
  }\textbf {\bibinfo {volume} {10}},\ \bibinfo {pages} {034044} (\bibinfo
  {year} {2018})}\BibitemShut {NoStop}%
\bibitem [{\citenamefont {Clevenson}\ \emph {et~al.}(2018)\citenamefont
  {Clevenson}, \citenamefont {Pham}, \citenamefont {Teale}, \citenamefont
  {Johnson}, \citenamefont {Englund},\ and\ \citenamefont
  {Braje}}]{clevenson2018robust}%
  \BibitemOpen
  \bibfield  {author} {\bibinfo {author} {\bibfnamefont {Hannah}\ \bibnamefont
  {Clevenson}}, \bibinfo {author} {\bibfnamefont {Linh~M}\ \bibnamefont
  {Pham}}, \bibinfo {author} {\bibfnamefont {Carson}\ \bibnamefont {Teale}},
  \bibinfo {author} {\bibfnamefont {Kerry}\ \bibnamefont {Johnson}}, \bibinfo
  {author} {\bibfnamefont {Dirk}\ \bibnamefont {Englund}}, \ and\ \bibinfo
  {author} {\bibfnamefont {Danielle}\ \bibnamefont {Braje}},\ }\bibfield
  {title} {\enquote {\bibinfo {title} {Robust high-dynamic-range vector
  magnetometry with nitrogen-vacancy centers in diamond},}\ }\href@noop {}
  {\bibfield  {journal} {\bibinfo  {journal} {Appl. Phys. Lett.}\ }\textbf
  {\bibinfo {volume} {112}},\ \bibinfo {pages} {252406} (\bibinfo {year}
  {2018})}\BibitemShut {NoStop}%
\bibitem [{\citenamefont {Igarashi}\ \emph {et~al.}(2020)\citenamefont
  {Igarashi}, \citenamefont {Sugi}, \citenamefont {Sotoma}, \citenamefont
  {Genjo}, \citenamefont {Kumiya}, \citenamefont {Walinda}, \citenamefont
  {Ueno}, \citenamefont {Ikeda}, \citenamefont {Sumiya}, \citenamefont
  {Tochio}, \citenamefont {Yoshinari}, \citenamefont {Harada},\ and\
  \citenamefont {Shirakawa}}]{igarashi2020threeD}%
  \BibitemOpen
  \bibfield  {author} {\bibinfo {author} {\bibfnamefont {Ryuji}\ \bibnamefont
  {Igarashi}}, \bibinfo {author} {\bibfnamefont {Takuma}\ \bibnamefont {Sugi}},
  \bibinfo {author} {\bibfnamefont {Shingo}\ \bibnamefont {Sotoma}}, \bibinfo
  {author} {\bibfnamefont {Takuya}\ \bibnamefont {Genjo}}, \bibinfo {author}
  {\bibfnamefont {Yuta}\ \bibnamefont {Kumiya}}, \bibinfo {author}
  {\bibfnamefont {Erik}\ \bibnamefont {Walinda}}, \bibinfo {author}
  {\bibfnamefont {Hiroshi}\ \bibnamefont {Ueno}}, \bibinfo {author}
  {\bibfnamefont {Kazuhiro}\ \bibnamefont {Ikeda}}, \bibinfo {author}
  {\bibfnamefont {Hitoshi}\ \bibnamefont {Sumiya}}, \bibinfo {author}
  {\bibfnamefont {Hidehito}\ \bibnamefont {Tochio}}, \bibinfo {author}
  {\bibfnamefont {Yohsuke}\ \bibnamefont {Yoshinari}}, \bibinfo {author}
  {\bibfnamefont {Yoshie}\ \bibnamefont {Harada}}, \ and\ \bibinfo {author}
  {\bibfnamefont {Masahiro}\ \bibnamefont {Shirakawa}},\ }\bibfield  {title}
  {\enquote {\bibinfo {title} {Tracking the 3d rotational dynamics in
  nanoscopic biological systems},}\ }\href {\doibase 10.1021/jacs.0c01191}
  {\bibfield  {journal} {\bibinfo  {journal} {J. Am. Chem. Soc.}\ }\textbf
  {\bibinfo {volume} {142}},\ \bibinfo {pages} {7542--7554} (\bibinfo {year}
  {2020})},\ \bibinfo {note} {pMID: 32285668},\ \Eprint
  {http://arxiv.org/abs/https://doi.org/10.1021/jacs.0c01191}
  {https://doi.org/10.1021/jacs.0c01191} \BibitemShut {NoStop}%
\bibitem [{\citenamefont {Tsukamoto}\ \emph {et~al.}(2021)\citenamefont
  {Tsukamoto}, \citenamefont {Ogawa}, \citenamefont {Ozawa}, \citenamefont
  {Iwasaki}, \citenamefont {Hatano}, \citenamefont {Sasaki},\ and\
  \citenamefont {Kobayashi}}]{tsukamoto2021vector}%
  \BibitemOpen
  \bibfield  {author} {\bibinfo {author} {\bibfnamefont {Moeta}\ \bibnamefont
  {Tsukamoto}}, \bibinfo {author} {\bibfnamefont {Kensuke}\ \bibnamefont
  {Ogawa}}, \bibinfo {author} {\bibfnamefont {Hayato}\ \bibnamefont {Ozawa}},
  \bibinfo {author} {\bibfnamefont {Takayuki}\ \bibnamefont {Iwasaki}},
  \bibinfo {author} {\bibfnamefont {Mutsuko}\ \bibnamefont {Hatano}}, \bibinfo
  {author} {\bibfnamefont {Kento}\ \bibnamefont {Sasaki}}, \ and\ \bibinfo
  {author} {\bibfnamefont {Kensuke}\ \bibnamefont {Kobayashi}},\ }\bibfield
  {title} {\enquote {\bibinfo {title} {Vector magnetometry using perfectly
  aligned nitrogen-vacancy center ensemble in diamond},}\ }\href@noop {}
  {\bibfield  {journal} {\bibinfo  {journal} {Appl. Phys. Lett.}\ }\textbf
  {\bibinfo {volume} {118}},\ \bibinfo {pages} {264002} (\bibinfo {year}
  {2021})}\BibitemShut {NoStop}%
\bibitem [{\citenamefont {DeVience}\ \emph {et~al.}(2015)\citenamefont
  {DeVience}, \citenamefont {Pham}, \citenamefont {Lovchinsky}, \citenamefont
  {Sushkov}, \citenamefont {Bar-Gill}, \citenamefont {Belthangady},
  \citenamefont {Casola}, \citenamefont {Corbett}, \citenamefont {Zhang},
  \citenamefont {Lukin} \emph {et~al.}}]{devience2015nanoscale}%
  \BibitemOpen
  \bibfield  {author} {\bibinfo {author} {\bibfnamefont {Stephen~J}\
  \bibnamefont {DeVience}}, \bibinfo {author} {\bibfnamefont {Linh~M}\
  \bibnamefont {Pham}}, \bibinfo {author} {\bibfnamefont {Igor}\ \bibnamefont
  {Lovchinsky}}, \bibinfo {author} {\bibfnamefont {Alexander~O}\ \bibnamefont
  {Sushkov}}, \bibinfo {author} {\bibfnamefont {Nir}\ \bibnamefont {Bar-Gill}},
  \bibinfo {author} {\bibfnamefont {Chinmay}\ \bibnamefont {Belthangady}},
  \bibinfo {author} {\bibfnamefont {Francesco}\ \bibnamefont {Casola}},
  \bibinfo {author} {\bibfnamefont {Madeleine}\ \bibnamefont {Corbett}},
  \bibinfo {author} {\bibfnamefont {Huiliang}\ \bibnamefont {Zhang}}, \bibinfo
  {author} {\bibfnamefont {Mikhail}\ \bibnamefont {Lukin}},  \emph {et~al.},\
  }\bibfield  {title} {\enquote {\bibinfo {title} {Nanoscale nmr spectroscopy
  and imaging of multiple nuclear species},}\ }\href@noop {} {\bibfield
  {journal} {\bibinfo  {journal} {Nat. Nanotechnol.}\ }\textbf {\bibinfo
  {volume} {10}},\ \bibinfo {pages} {129--134} (\bibinfo {year}
  {2015})}\BibitemShut {NoStop}%
\bibitem [{\citenamefont {Smits}\ \emph {et~al.}(2019)\citenamefont {Smits},
  \citenamefont {Damron}, \citenamefont {Kehayias}, \citenamefont {McDowell},
  \citenamefont {Mosavian}, \citenamefont {Fescenko}, \citenamefont {Ristoff},
  \citenamefont {Laraoui}, \citenamefont {Jarmola},\ and\ \citenamefont
  {Acosta}}]{smits2019two}%
  \BibitemOpen
  \bibfield  {author} {\bibinfo {author} {\bibfnamefont {Janis}\ \bibnamefont
  {Smits}}, \bibinfo {author} {\bibfnamefont {Joshua~T}\ \bibnamefont
  {Damron}}, \bibinfo {author} {\bibfnamefont {Pauli}\ \bibnamefont
  {Kehayias}}, \bibinfo {author} {\bibfnamefont {Andrew~F}\ \bibnamefont
  {McDowell}}, \bibinfo {author} {\bibfnamefont {Nazanin}\ \bibnamefont
  {Mosavian}}, \bibinfo {author} {\bibfnamefont {Ilja}\ \bibnamefont
  {Fescenko}}, \bibinfo {author} {\bibfnamefont {Nathaniel}\ \bibnamefont
  {Ristoff}}, \bibinfo {author} {\bibfnamefont {Abdelghani}\ \bibnamefont
  {Laraoui}}, \bibinfo {author} {\bibfnamefont {Andrey}\ \bibnamefont
  {Jarmola}}, \ and\ \bibinfo {author} {\bibfnamefont {Victor~M}\ \bibnamefont
  {Acosta}},\ }\bibfield  {title} {\enquote {\bibinfo {title} {Two-dimensional
  nuclear magnetic resonance spectroscopy with a microfluidic diamond quantum
  sensor},}\ }\href@noop {} {\bibfield  {journal} {\bibinfo  {journal} {Sci.
  Adv.}\ }\textbf {\bibinfo {volume} {5}},\ \bibinfo {pages} {eaaw7895}
  (\bibinfo {year} {2019})}\BibitemShut {NoStop}%
\bibitem [{\citenamefont {Holzgrafe}\ \emph {et~al.}(2020)\citenamefont
  {Holzgrafe}, \citenamefont {Gu}, \citenamefont {Beitner}, \citenamefont
  {Kara}, \citenamefont {Knowles},\ and\ \citenamefont
  {Atat{\"u}re}}]{holzgrafe2020nanoscale}%
  \BibitemOpen
  \bibfield  {author} {\bibinfo {author} {\bibfnamefont {Jeffrey}\ \bibnamefont
  {Holzgrafe}}, \bibinfo {author} {\bibfnamefont {Qiushi}\ \bibnamefont {Gu}},
  \bibinfo {author} {\bibfnamefont {Jan}\ \bibnamefont {Beitner}}, \bibinfo
  {author} {\bibfnamefont {Dhiren~M}\ \bibnamefont {Kara}}, \bibinfo {author}
  {\bibfnamefont {Helena~S}\ \bibnamefont {Knowles}}, \ and\ \bibinfo {author}
  {\bibfnamefont {Mete}\ \bibnamefont {Atat{\"u}re}},\ }\bibfield  {title}
  {\enquote {\bibinfo {title} {Nanoscale nmr spectroscopy using nanodiamond
  quantum sensors},}\ }\href@noop {} {\bibfield  {journal} {\bibinfo  {journal}
  {Phys. Rev. Appl.}\ }\textbf {\bibinfo {volume} {13}},\ \bibinfo {pages}
  {044004} (\bibinfo {year} {2020})}\BibitemShut {NoStop}%
\bibitem [{\citenamefont {Pozar}(2005)}]{pozar2005}%
  \BibitemOpen
  \bibfield  {author} {\bibinfo {author} {\bibfnamefont {David~M.}\
  \bibnamefont {Pozar}},\ }\href@noop {} {\emph {\bibinfo {title} {Microwave
  Engineering}}},\ \bibinfo {edition} {3rd}\ ed.\ (\bibinfo  {publisher} {John
  wiley \& sons},\ \bibinfo {year} {2005})\BibitemShut {NoStop}%
\bibitem [{\citenamefont {Nishimura}\ \emph {et~al.}(2021)\citenamefont
  {Nishimura}, \citenamefont {Oshimi}, \citenamefont {Umehara}, \citenamefont
  {Kumon}, \citenamefont {Miyaji}, \citenamefont {Yukawa}, \citenamefont
  {Shikano}, \citenamefont {Matsubara}, \citenamefont {Fujiwara}, \citenamefont
  {Baba} \emph {et~al.}}]{nishimura2021wide}%
  \BibitemOpen
  \bibfield  {author} {\bibinfo {author} {\bibfnamefont {Yushi}\ \bibnamefont
  {Nishimura}}, \bibinfo {author} {\bibfnamefont {Keisuke}\ \bibnamefont
  {Oshimi}}, \bibinfo {author} {\bibfnamefont {Yumi}\ \bibnamefont {Umehara}},
  \bibinfo {author} {\bibfnamefont {Yuka}\ \bibnamefont {Kumon}}, \bibinfo
  {author} {\bibfnamefont {Kazu}\ \bibnamefont {Miyaji}}, \bibinfo {author}
  {\bibfnamefont {Hiroshi}\ \bibnamefont {Yukawa}}, \bibinfo {author}
  {\bibfnamefont {Yutaka}\ \bibnamefont {Shikano}}, \bibinfo {author}
  {\bibfnamefont {Tsutomu}\ \bibnamefont {Matsubara}}, \bibinfo {author}
  {\bibfnamefont {Masazumi}\ \bibnamefont {Fujiwara}}, \bibinfo {author}
  {\bibfnamefont {Yoshinobu}\ \bibnamefont {Baba}},  \emph {et~al.},\
  }\bibfield  {title} {\enquote {\bibinfo {title} {Wide-field fluorescent
  nanodiamond spin measurements toward real-time large-area intracellular
  thermometry},}\ }\href@noop {} {\bibfield  {journal} {\bibinfo  {journal}
  {Sci. Rep.}\ }\textbf {\bibinfo {volume} {11}},\ \bibinfo {pages} {1--12}
  (\bibinfo {year} {2021})}\BibitemShut {NoStop}%
\bibitem [{\citenamefont {Yukawa}\ \emph {et~al.}(2020)\citenamefont {Yukawa},
  \citenamefont {Fujiwara}, \citenamefont {Kobayashi}, \citenamefont {Kumon},
  \citenamefont {Miyaji}, \citenamefont {Nishimura}, \citenamefont {Oshimi},
  \citenamefont {Umehara}, \citenamefont {Teki}, \citenamefont {Iwasaki} \emph
  {et~al.}}]{yukawa2020quantum}%
  \BibitemOpen
  \bibfield  {author} {\bibinfo {author} {\bibfnamefont {Hiroshi}\ \bibnamefont
  {Yukawa}}, \bibinfo {author} {\bibfnamefont {Masazumi}\ \bibnamefont
  {Fujiwara}}, \bibinfo {author} {\bibfnamefont {Kaori}\ \bibnamefont
  {Kobayashi}}, \bibinfo {author} {\bibfnamefont {Yuka}\ \bibnamefont {Kumon}},
  \bibinfo {author} {\bibfnamefont {Kazu}\ \bibnamefont {Miyaji}}, \bibinfo
  {author} {\bibfnamefont {Yushi}\ \bibnamefont {Nishimura}}, \bibinfo {author}
  {\bibfnamefont {Keisuke}\ \bibnamefont {Oshimi}}, \bibinfo {author}
  {\bibfnamefont {Yumi}\ \bibnamefont {Umehara}}, \bibinfo {author}
  {\bibfnamefont {Yoshio}\ \bibnamefont {Teki}}, \bibinfo {author}
  {\bibfnamefont {Takayuki}\ \bibnamefont {Iwasaki}},  \emph {et~al.},\
  }\bibfield  {title} {\enquote {\bibinfo {title} {A quantum thermometric
  sensing and analysis system using fluorescent nanodiamonds for the evaluation
  of living stem cell functions according to intracellular temperature},}\
  }\href@noop {} {\bibfield  {journal} {\bibinfo  {journal} {Nanoscale Adv.}\
  }\textbf {\bibinfo {volume} {2}},\ \bibinfo {pages} {1859--1868} (\bibinfo
  {year} {2020})}\BibitemShut {NoStop}%
\bibitem [{\citenamefont {Izutsu}(2019)}]{izutsu2019skin}%
  \BibitemOpen
  \bibfield  {author} {\bibinfo {author} {\bibfnamefont {Yumi}\ \bibnamefont
  {Izutsu}},\ }\bibfield  {title} {\enquote {\bibinfo {title} {Skin grafting in
  xenopus laevis: a technique for assessing development and immunological
  disparity},}\ }\href@noop {} {\bibfield  {journal} {\bibinfo  {journal} {Cold
  Spring Harb. Protoc.}\ }\textbf {\bibinfo {volume} {2019}},\ \bibinfo {pages}
  {pdb--prot099788} (\bibinfo {year} {2019})}\BibitemShut {NoStop}%
\bibitem [{\citenamefont {Izutsu}\ and\ \citenamefont
  {Ma{\'e}no}(2005)}]{izutsu2005analyses}%
  \BibitemOpen
  \bibfield  {author} {\bibinfo {author} {\bibfnamefont {Yumi}\ \bibnamefont
  {Izutsu}}\ and\ \bibinfo {author} {\bibfnamefont {Mitsugu}\ \bibnamefont
  {Ma{\'e}no}},\ }\bibfield  {title} {\enquote {\bibinfo {title} {Analyses of
  immune responses to ontogeny-specific antigens using an inbred strain of
  xenopus laevis (j strain)},}\ }\href@noop {} {\bibfield  {journal} {\bibinfo
  {journal} {Methods Mol. Med.}\ ,\ \bibinfo {pages} {149--158}} (\bibinfo
  {year} {2005})}\BibitemShut {NoStop}%
\bibitem [{\citenamefont {Zhao}\ \emph {et~al.}(2011)\citenamefont {Zhao},
  \citenamefont {Takimoto}, \citenamefont {Ito}, \citenamefont {Kitagawa},
  \citenamefont {Kimura},\ and\ \citenamefont
  {Komatsu}}]{zhao2011chromatographic}%
  \BibitemOpen
  \bibfield  {author} {\bibinfo {author} {\bibfnamefont {Li}~\bibnamefont
  {Zhao}}, \bibinfo {author} {\bibfnamefont {Tatsuya}\ \bibnamefont
  {Takimoto}}, \bibinfo {author} {\bibfnamefont {Masaaki}\ \bibnamefont {Ito}},
  \bibinfo {author} {\bibfnamefont {Naoko}\ \bibnamefont {Kitagawa}}, \bibinfo
  {author} {\bibfnamefont {Takahide}\ \bibnamefont {Kimura}}, \ and\ \bibinfo
  {author} {\bibfnamefont {Naoki}\ \bibnamefont {Komatsu}},\ }\bibfield
  {title} {\enquote {\bibinfo {title} {Chromatographic separation of highly
  soluble diamond nanoparticles prepared by polyglycerol grafting},}\
  }\href@noop {} {\bibfield  {journal} {\bibinfo  {journal} {Angew. Chem. Int.
  Ed.}\ }\textbf {\bibinfo {volume} {50}},\ \bibinfo {pages} {1388--1392}
  (\bibinfo {year} {2011})}\BibitemShut {NoStop}%
\bibitem [{\citenamefont {Brenner}(1974)}]{brenner1974genetics}%
  \BibitemOpen
  \bibfield  {author} {\bibinfo {author} {\bibfnamefont {Sydney}\ \bibnamefont
  {Brenner}},\ }\bibfield  {title} {\enquote {\bibinfo {title} {The genetics of
  caenorhabditis elegans},}\ }\href@noop {} {\bibfield  {journal} {\bibinfo
  {journal} {Genetics}\ }\textbf {\bibinfo {volume} {77}},\ \bibinfo {pages}
  {71--94} (\bibinfo {year} {1974})}\BibitemShut {NoStop}%
\bibitem [{\citenamefont {Tsukahara}\ \emph {et~al.}(2019)\citenamefont
  {Tsukahara}, \citenamefont {Fujiwara}, \citenamefont {Sera}, \citenamefont
  {Nishimura}, \citenamefont {Sugai}, \citenamefont {Jentgens}, \citenamefont
  {Teki}, \citenamefont {Hashimoto},\ and\ \citenamefont
  {Shikata}}]{tsukahara2019removing}%
  \BibitemOpen
  \bibfield  {author} {\bibinfo {author} {\bibfnamefont {Ryuta}\ \bibnamefont
  {Tsukahara}}, \bibinfo {author} {\bibfnamefont {Masazumi}\ \bibnamefont
  {Fujiwara}}, \bibinfo {author} {\bibfnamefont {Yoshihiko}\ \bibnamefont
  {Sera}}, \bibinfo {author} {\bibfnamefont {Yushi}\ \bibnamefont {Nishimura}},
  \bibinfo {author} {\bibfnamefont {Yuko}\ \bibnamefont {Sugai}}, \bibinfo
  {author} {\bibfnamefont {Christian}\ \bibnamefont {Jentgens}}, \bibinfo
  {author} {\bibfnamefont {Yoshio}\ \bibnamefont {Teki}}, \bibinfo {author}
  {\bibfnamefont {Hideki}\ \bibnamefont {Hashimoto}}, \ and\ \bibinfo {author}
  {\bibfnamefont {Shinichi}\ \bibnamefont {Shikata}},\ }\bibfield  {title}
  {\enquote {\bibinfo {title} {Removing non-size-dependent electron spin
  decoherence of nanodiamond quantum sensors by aerobic oxidation},}\
  }\href@noop {} {\bibfield  {journal} {\bibinfo  {journal} {ACS Appl. Nano
  Mater.}\ }\textbf {\bibinfo {volume} {2}},\ \bibinfo {pages} {3701--3710}
  (\bibinfo {year} {2019})}\BibitemShut {NoStop}%
\bibitem [{\citenamefont {Fujiwara}\ \emph
  {et~al.}(2020{\natexlab{b}})\citenamefont {Fujiwara}, \citenamefont {Dohms},
  \citenamefont {Suto}, \citenamefont {Nishimura}, \citenamefont {Oshimi},
  \citenamefont {Teki}, \citenamefont {Cai}, \citenamefont {Benson},\ and\
  \citenamefont {Shikano}}]{fujiwara2020real-1}%
  \BibitemOpen
  \bibfield  {author} {\bibinfo {author} {\bibfnamefont {Masazumi}\
  \bibnamefont {Fujiwara}}, \bibinfo {author} {\bibfnamefont {Alexander}\
  \bibnamefont {Dohms}}, \bibinfo {author} {\bibfnamefont {Ken}\ \bibnamefont
  {Suto}}, \bibinfo {author} {\bibfnamefont {Yushi}\ \bibnamefont {Nishimura}},
  \bibinfo {author} {\bibfnamefont {Keisuke}\ \bibnamefont {Oshimi}}, \bibinfo
  {author} {\bibfnamefont {Yoshio}\ \bibnamefont {Teki}}, \bibinfo {author}
  {\bibfnamefont {Kai}\ \bibnamefont {Cai}}, \bibinfo {author} {\bibfnamefont
  {Oliver}\ \bibnamefont {Benson}}, \ and\ \bibinfo {author} {\bibfnamefont
  {Yutaka}\ \bibnamefont {Shikano}},\ }\bibfield  {title} {\enquote {\bibinfo
  {title} {Real-time estimation of the optically detected magnetic resonance
  shift in diamond quantum thermometry toward biological applications},}\
  }\href@noop {} {\bibfield  {journal} {\bibinfo  {journal} {Phys. Rev. Res.}\
  }\textbf {\bibinfo {volume} {2}},\ \bibinfo {pages} {043415} (\bibinfo {year}
  {2020}{\natexlab{b}})}\BibitemShut {NoStop}%
\bibitem [{\citenamefont {Fujiwara}\ \emph {et~al.}(2019)\citenamefont
  {Fujiwara}, \citenamefont {Tsukahara}, \citenamefont {Sera}, \citenamefont
  {Yukawa}, \citenamefont {Baba}, \citenamefont {Shikata},\ and\ \citenamefont
  {Hashimoto}}]{fujiwara2019monitoring}%
  \BibitemOpen
  \bibfield  {author} {\bibinfo {author} {\bibfnamefont {Masazumi}\
  \bibnamefont {Fujiwara}}, \bibinfo {author} {\bibfnamefont {Ryuta}\
  \bibnamefont {Tsukahara}}, \bibinfo {author} {\bibfnamefont {Yoshihiko}\
  \bibnamefont {Sera}}, \bibinfo {author} {\bibfnamefont {Hiroshi}\
  \bibnamefont {Yukawa}}, \bibinfo {author} {\bibfnamefont {Yoshinobu}\
  \bibnamefont {Baba}}, \bibinfo {author} {\bibfnamefont {Shinichi}\
  \bibnamefont {Shikata}}, \ and\ \bibinfo {author} {\bibfnamefont {Hideki}\
  \bibnamefont {Hashimoto}},\ }\bibfield  {title} {\enquote {\bibinfo {title}
  {Monitoring spin coherence of single nitrogen-vacancy centers in nanodiamonds
  during ph changes in aqueous buffer solutions},}\ }\href@noop {} {\bibfield
  {journal} {\bibinfo  {journal} {RSC Adv.}\ }\textbf {\bibinfo {volume} {9}},\
  \bibinfo {pages} {12606--12614} (\bibinfo {year} {2019})}\BibitemShut
  {NoStop}%
\bibitem [{\citenamefont {Sefa}\ and\ \citenamefont
  {Maraj}(2011)}]{sefa2011analysis}%
  \BibitemOpen
  \bibfield  {author} {\bibinfo {author} {\bibfnamefont {Ruzhdi}\ \bibnamefont
  {Sefa}}\ and\ \bibinfo {author} {\bibfnamefont {Arianit}\ \bibnamefont
  {Maraj}},\ }\bibfield  {title} {\enquote {\bibinfo {title} {Analysis and
  design of microstrip to balanced stripline transitions},}\ }in\ \href@noop {}
  {\emph {\bibinfo {booktitle} {Proceedings of the 10th WSEAS international
  conference on Telecommunications and informatics and microelectronics,
  nanoelectronics, optoelectronics, and WSEAS international conference on
  Signal processing, Canary Islands, Spain}}}\ (\bibinfo {organization}
  {Citeseer},\ \bibinfo {year} {2011})\ pp.\ \bibinfo {pages}
  {27--29}\BibitemShut {NoStop}%
\bibitem [{\citenamefont {P{\'e}rez-Escudero}\ \emph
  {et~al.}(2018)\citenamefont {P{\'e}rez-Escudero}, \citenamefont
  {Torres-Garc{\'\i}a}, \citenamefont {Gonzalo},\ and\ \citenamefont
  {Ederra}}]{perez2018simplified}%
  \BibitemOpen
  \bibfield  {author} {\bibinfo {author} {\bibfnamefont {Jos{\'e}~M}\
  \bibnamefont {P{\'e}rez-Escudero}}, \bibinfo {author} {\bibfnamefont
  {Alicia~E}\ \bibnamefont {Torres-Garc{\'\i}a}}, \bibinfo {author}
  {\bibfnamefont {Ram{\'o}n}\ \bibnamefont {Gonzalo}}, \ and\ \bibinfo {author}
  {\bibfnamefont {I{\~n}igo}\ \bibnamefont {Ederra}},\ }\bibfield  {title}
  {\enquote {\bibinfo {title} {A simplified design inline
  microstrip-to-waveguide transition},}\ }\href@noop {} {\bibfield  {journal}
  {\bibinfo  {journal} {Electronics}\ }\textbf {\bibinfo {volume} {7}},\
  \bibinfo {pages} {215} (\bibinfo {year} {2018})}\BibitemShut {NoStop}%
\bibitem [{\citenamefont {Dr{\'e}au}\ \emph {et~al.}(2011)\citenamefont
  {Dr{\'e}au}, \citenamefont {Lesik}, \citenamefont {Rondin}, \citenamefont
  {Spinicelli}, \citenamefont {Arcizet}, \citenamefont {Roch},\ and\
  \citenamefont {Jacques}}]{dreau2011avoiding}%
  \BibitemOpen
  \bibfield  {author} {\bibinfo {author} {\bibfnamefont {A}~\bibnamefont
  {Dr{\'e}au}}, \bibinfo {author} {\bibfnamefont {M}~\bibnamefont {Lesik}},
  \bibinfo {author} {\bibfnamefont {L}~\bibnamefont {Rondin}}, \bibinfo
  {author} {\bibfnamefont {P}~\bibnamefont {Spinicelli}}, \bibinfo {author}
  {\bibfnamefont {O}~\bibnamefont {Arcizet}}, \bibinfo {author} {\bibfnamefont
  {J-F}\ \bibnamefont {Roch}}, \ and\ \bibinfo {author} {\bibfnamefont
  {V}~\bibnamefont {Jacques}},\ }\bibfield  {title} {\enquote {\bibinfo {title}
  {Avoiding power broadening in optically detected magnetic resonance of single
  nv defects for enhanced dc magnetic field sensitivity},}\ }\href@noop {}
  {\bibfield  {journal} {\bibinfo  {journal} {Phys. Rev. B}\ }\textbf {\bibinfo
  {volume} {84}},\ \bibinfo {pages} {195204} (\bibinfo {year}
  {2011})}\BibitemShut {NoStop}%
\bibitem [{\citenamefont {Bayat}\ \emph {et~al.}(2014)\citenamefont {Bayat},
  \citenamefont {Choy}, \citenamefont {Farrokh~Baroughi}, \citenamefont
  {Meesala},\ and\ \citenamefont {Loncar}}]{bayat2014efficient}%
  \BibitemOpen
  \bibfield  {author} {\bibinfo {author} {\bibfnamefont {Khadijeh}\
  \bibnamefont {Bayat}}, \bibinfo {author} {\bibfnamefont {Jennifer}\
  \bibnamefont {Choy}}, \bibinfo {author} {\bibfnamefont {Mahdi}\ \bibnamefont
  {Farrokh~Baroughi}}, \bibinfo {author} {\bibfnamefont {Srujan}\ \bibnamefont
  {Meesala}}, \ and\ \bibinfo {author} {\bibfnamefont {Marko}\ \bibnamefont
  {Loncar}},\ }\bibfield  {title} {\enquote {\bibinfo {title} {Efficient,
  uniform, and large area microwave magnetic coupling to nv centers in diamond
  using double split-ring resonators},}\ }\href@noop {} {\bibfield  {journal}
  {\bibinfo  {journal} {Nano Lett.}\ }\textbf {\bibinfo {volume} {14}},\
  \bibinfo {pages} {1208--1213} (\bibinfo {year} {2014})}\BibitemShut {NoStop}%
\bibitem [{\citenamefont {Camparo}\ and\ \citenamefont
  {Frueholz}(1988)}]{camparo1988observation}%
  \BibitemOpen
  \bibfield  {author} {\bibinfo {author} {\bibfnamefont {James~C}\ \bibnamefont
  {Camparo}}\ and\ \bibinfo {author} {\bibfnamefont {Robert~P}\ \bibnamefont
  {Frueholz}},\ }\bibfield  {title} {\enquote {\bibinfo {title} {Observation of
  the rabi-resonance spectrum},}\ }\href@noop {} {\bibfield  {journal}
  {\bibinfo  {journal} {Phys. Rev. A}\ }\textbf {\bibinfo {volume} {38}},\
  \bibinfo {pages} {6143} (\bibinfo {year} {1988})}\BibitemShut {NoStop}%
\bibitem [{\citenamefont {Bertaina}\ \emph {et~al.}(2014)\citenamefont
  {Bertaina}, \citenamefont {Dutoit}, \citenamefont {Van~Tol}, \citenamefont
  {Dressel}, \citenamefont {Barbara},\ and\ \citenamefont
  {Stepanov}}]{bertaina2014rabi}%
  \BibitemOpen
  \bibfield  {author} {\bibinfo {author} {\bibfnamefont {Sylvain}\ \bibnamefont
  {Bertaina}}, \bibinfo {author} {\bibfnamefont {C-E}\ \bibnamefont {Dutoit}},
  \bibinfo {author} {\bibfnamefont {Johan}\ \bibnamefont {Van~Tol}}, \bibinfo
  {author} {\bibfnamefont {Martin}\ \bibnamefont {Dressel}}, \bibinfo {author}
  {\bibfnamefont {Bernard}\ \bibnamefont {Barbara}}, \ and\ \bibinfo {author}
  {\bibfnamefont {Anatoli}\ \bibnamefont {Stepanov}},\ }\bibfield  {title}
  {\enquote {\bibinfo {title} {Rabi oscillations of pinned solitons in spin
  chains: A route to quantum computation and communication},}\ }\href@noop {}
  {\bibfield  {journal} {\bibinfo  {journal} {Phys. Rev. B}\ }\textbf {\bibinfo
  {volume} {90}},\ \bibinfo {pages} {060404} (\bibinfo {year}
  {2014})}\BibitemShut {NoStop}%
\bibitem [{\citenamefont {Hanson}\ \emph {et~al.}(2008)\citenamefont {Hanson},
  \citenamefont {Dobrovitski}, \citenamefont {Feiguin}, \citenamefont {Gywat},\
  and\ \citenamefont {Awschalom}}]{hanson2008coherent}%
  \BibitemOpen
  \bibfield  {author} {\bibinfo {author} {\bibfnamefont {R}~\bibnamefont
  {Hanson}}, \bibinfo {author} {\bibfnamefont {VV}~\bibnamefont {Dobrovitski}},
  \bibinfo {author} {\bibfnamefont {AE}~\bibnamefont {Feiguin}}, \bibinfo
  {author} {\bibfnamefont {O}~\bibnamefont {Gywat}}, \ and\ \bibinfo {author}
  {\bibfnamefont {DD}~\bibnamefont {Awschalom}},\ }\bibfield  {title} {\enquote
  {\bibinfo {title} {Coherent dynamics of a single spin interacting with an
  adjustable spin bath},}\ }\href@noop {} {\bibfield  {journal} {\bibinfo
  {journal} {Science}\ }\textbf {\bibinfo {volume} {320}},\ \bibinfo {pages}
  {352--355} (\bibinfo {year} {2008})}\BibitemShut {NoStop}%
\bibitem [{\citenamefont {Sasaki}\ \emph {et~al.}(2016)\citenamefont {Sasaki},
  \citenamefont {Monnai}, \citenamefont {Saijo}, \citenamefont {Fujita},
  \citenamefont {Watanabe}, \citenamefont {Ishi-Hayase}, \citenamefont {Itoh},\
  and\ \citenamefont {Abe}}]{sasaki2016broadband}%
  \BibitemOpen
  \bibfield  {author} {\bibinfo {author} {\bibfnamefont {Kento}\ \bibnamefont
  {Sasaki}}, \bibinfo {author} {\bibfnamefont {Yasuaki}\ \bibnamefont
  {Monnai}}, \bibinfo {author} {\bibfnamefont {Soya}\ \bibnamefont {Saijo}},
  \bibinfo {author} {\bibfnamefont {Ryushiro}\ \bibnamefont {Fujita}}, \bibinfo
  {author} {\bibfnamefont {Hideyuki}\ \bibnamefont {Watanabe}}, \bibinfo
  {author} {\bibfnamefont {Junko}\ \bibnamefont {Ishi-Hayase}}, \bibinfo
  {author} {\bibfnamefont {Kohei~M}\ \bibnamefont {Itoh}}, \ and\ \bibinfo
  {author} {\bibfnamefont {Eisuke}\ \bibnamefont {Abe}},\ }\bibfield  {title}
  {\enquote {\bibinfo {title} {Broadband, large-area microwave antenna for
  optically detected magnetic resonance of nitrogen-vacancy centers in
  diamond},}\ }\href@noop {} {\bibfield  {journal} {\bibinfo  {journal} {Rev.
  Sci. Instrum.}\ }\textbf {\bibinfo {volume} {87}},\ \bibinfo {pages} {053904}
  (\bibinfo {year} {2016})}\BibitemShut {NoStop}%
\bibitem [{\citenamefont {Vallabhapurapu}\ \emph {et~al.}(2021)\citenamefont
  {Vallabhapurapu}, \citenamefont {Slack-Smith}, \citenamefont {Sewani},
  \citenamefont {Adambukulam}, \citenamefont {Morello}, \citenamefont {Pla},\
  and\ \citenamefont {Laucht}}]{vallabhapurapu2021fast}%
  \BibitemOpen
  \bibfield  {author} {\bibinfo {author} {\bibfnamefont {Hyma~H}\ \bibnamefont
  {Vallabhapurapu}}, \bibinfo {author} {\bibfnamefont {James~P}\ \bibnamefont
  {Slack-Smith}}, \bibinfo {author} {\bibfnamefont {Vikas~K}\ \bibnamefont
  {Sewani}}, \bibinfo {author} {\bibfnamefont {Chris}\ \bibnamefont
  {Adambukulam}}, \bibinfo {author} {\bibfnamefont {Andrea}\ \bibnamefont
  {Morello}}, \bibinfo {author} {\bibfnamefont {Jarryd~J}\ \bibnamefont {Pla}},
  \ and\ \bibinfo {author} {\bibfnamefont {Arne}\ \bibnamefont {Laucht}},\
  }\bibfield  {title} {\enquote {\bibinfo {title} {Fast coherent control of a
  nitrogen-vacancy-center spin ensemble using a \ce{KTaO3} dielectric resonator
  at cryogenic temperatures},}\ }\href@noop {} {\bibfield  {journal} {\bibinfo
  {journal} {Phys. Rev. Appl.}\ }\textbf {\bibinfo {volume} {16}},\ \bibinfo
  {pages} {044051} (\bibinfo {year} {2021})}\BibitemShut {NoStop}%
\bibitem [{\citenamefont {Slegerova}\ \emph {et~al.}(2014)\citenamefont
  {Slegerova}, \citenamefont {Rehor}, \citenamefont {Havlik}, \citenamefont
  {Raabova}, \citenamefont {Muchova},\ and\ \citenamefont
  {Cigler}}]{slegerova2014nanodiamonds}%
  \BibitemOpen
  \bibfield  {author} {\bibinfo {author} {\bibfnamefont {Jitka}\ \bibnamefont
  {Slegerova}}, \bibinfo {author} {\bibfnamefont {Ivan}\ \bibnamefont {Rehor}},
  \bibinfo {author} {\bibfnamefont {Jan}\ \bibnamefont {Havlik}}, \bibinfo
  {author} {\bibfnamefont {Helena}\ \bibnamefont {Raabova}}, \bibinfo {author}
  {\bibfnamefont {Eva}\ \bibnamefont {Muchova}}, \ and\ \bibinfo {author}
  {\bibfnamefont {Petr}\ \bibnamefont {Cigler}},\ }\bibfield  {title} {\enquote
  {\bibinfo {title} {Nanodiamonds as intracellular probes for imaging in
  biology and medicine},}\ }in\ \href@noop {} {\emph {\bibinfo {booktitle}
  {Intracellular delivery II}}}\ (\bibinfo  {publisher} {Springer},\ \bibinfo
  {year} {2014})\ pp.\ \bibinfo {pages} {363--401}\BibitemShut {NoStop}%
\bibitem [{\citenamefont {Sater}\ and\ \citenamefont
  {Moody}(2017)}]{sater2017using}%
  \BibitemOpen
  \bibfield  {author} {\bibinfo {author} {\bibfnamefont {Amy~K}\ \bibnamefont
  {Sater}}\ and\ \bibinfo {author} {\bibfnamefont {Sally~A}\ \bibnamefont
  {Moody}},\ }\bibfield  {title} {\enquote {\bibinfo {title} {Using xenopus to
  understand human disease and developmental disorders},}\ }\href@noop {}
  {\bibfield  {journal} {\bibinfo  {journal} {Genesis}\ }\textbf {\bibinfo
  {volume} {55}},\ \bibinfo {pages} {e22997} (\bibinfo {year}
  {2017})}\BibitemShut {NoStop}%
\bibitem [{\citenamefont {Straka}\ and\ \citenamefont
  {Simmers}(2012)}]{straka2012xenopus}%
  \BibitemOpen
  \bibfield  {author} {\bibinfo {author} {\bibfnamefont {Hans}\ \bibnamefont
  {Straka}}\ and\ \bibinfo {author} {\bibfnamefont {John}\ \bibnamefont
  {Simmers}},\ }\bibfield  {title} {\enquote {\bibinfo {title} {Xenopus laevis:
  An ideal experimental model for studying the developmental dynamics of neural
  network assembly and sensory-motor computations},}\ }\href@noop {} {\bibfield
   {journal} {\bibinfo  {journal} {Dev. Neurobiol.}\ }\textbf {\bibinfo
  {volume} {72}},\ \bibinfo {pages} {649--663} (\bibinfo {year}
  {2012})}\BibitemShut {NoStop}%
\bibitem [{\citenamefont {Nakai}\ \emph {et~al.}(2017)\citenamefont {Nakai},
  \citenamefont {Nakajima},\ and\ \citenamefont
  {Yaoita}}]{nakai2017mechanisms}%
  \BibitemOpen
  \bibfield  {author} {\bibinfo {author} {\bibfnamefont {Yuya}\ \bibnamefont
  {Nakai}}, \bibinfo {author} {\bibfnamefont {Keisuke}\ \bibnamefont
  {Nakajima}}, \ and\ \bibinfo {author} {\bibfnamefont {Yoshio}\ \bibnamefont
  {Yaoita}},\ }\bibfield  {title} {\enquote {\bibinfo {title} {Mechanisms of
  tail resorption during anuran metamorphosis},}\ }\href@noop {} {\bibfield
  {journal} {\bibinfo  {journal} {Biomol Concepts}\ }\textbf {\bibinfo {volume}
  {8}},\ \bibinfo {pages} {179--183} (\bibinfo {year} {2017})}\BibitemShut
  {NoStop}%
\bibitem [{\citenamefont {Slack}\ \emph {et~al.}(2008)\citenamefont {Slack},
  \citenamefont {Lin},\ and\ \citenamefont {Chen}}]{slack2008xenopus}%
  \BibitemOpen
  \bibfield  {author} {\bibinfo {author} {\bibfnamefont {JM}~\bibnamefont
  {Slack}}, \bibinfo {author} {\bibfnamefont {G}~\bibnamefont {Lin}}, \ and\
  \bibinfo {author} {\bibfnamefont {Y}~\bibnamefont {Chen}},\ }\bibfield
  {title} {\enquote {\bibinfo {title} {The xenopus tadpole: a new model for
  regeneration research.}}\ }\href@noop {} {\bibfield  {journal} {\bibinfo
  {journal} {Cell. Mol. Life Sci.}\ }\textbf {\bibinfo {volume} {65}},\
  \bibinfo {pages} {54--63} (\bibinfo {year} {2008})}\BibitemShut {NoStop}%
\bibitem [{\citenamefont {Phipps}\ \emph {et~al.}(2020)\citenamefont {Phipps},
  \citenamefont {Marshall}, \citenamefont {Dorey},\ and\ \citenamefont
  {Amaya}}]{phipps2020model}%
  \BibitemOpen
  \bibfield  {author} {\bibinfo {author} {\bibfnamefont {Lauren~S}\
  \bibnamefont {Phipps}}, \bibinfo {author} {\bibfnamefont {Lindsey}\
  \bibnamefont {Marshall}}, \bibinfo {author} {\bibfnamefont {Karel}\
  \bibnamefont {Dorey}}, \ and\ \bibinfo {author} {\bibfnamefont {Enrique}\
  \bibnamefont {Amaya}},\ }\bibfield  {title} {\enquote {\bibinfo {title}
  {Model systems for regeneration: Xenopus},}\ }\href@noop {} {\bibfield
  {journal} {\bibinfo  {journal} {Development}\ }\textbf {\bibinfo {volume}
  {147}},\ \bibinfo {pages} {dev180844} (\bibinfo {year} {2020})}\BibitemShut
  {NoStop}%
\bibitem [{\citenamefont {King}\ \emph {et~al.}(2012)\citenamefont {King},
  \citenamefont {Neff},\ and\ \citenamefont {Mescher}}]{king2012developing}%
  \BibitemOpen
  \bibfield  {author} {\bibinfo {author} {\bibfnamefont {Michael~W}\
  \bibnamefont {King}}, \bibinfo {author} {\bibfnamefont {Anton~W}\
  \bibnamefont {Neff}}, \ and\ \bibinfo {author} {\bibfnamefont {Anthony~L}\
  \bibnamefont {Mescher}},\ }\bibfield  {title} {\enquote {\bibinfo {title}
  {The developing xenopus limb as a model for studies on the balance between
  inflammation and regeneration},}\ }\href@noop {} {\bibfield  {journal}
  {\bibinfo  {journal} {Anat. Rec. (Hoboken)}\ }\textbf {\bibinfo {volume}
  {295}},\ \bibinfo {pages} {1552--1561} (\bibinfo {year} {2012})}\BibitemShut
  {NoStop}%
\bibitem [{\citenamefont {Davaapil}\ \emph {et~al.}(2017)\citenamefont
  {Davaapil}, \citenamefont {Brockes},\ and\ \citenamefont
  {Yun}}]{davaapil2017conserved}%
  \BibitemOpen
  \bibfield  {author} {\bibinfo {author} {\bibfnamefont {Hongorzul}\
  \bibnamefont {Davaapil}}, \bibinfo {author} {\bibfnamefont {Jeremy~P}\
  \bibnamefont {Brockes}}, \ and\ \bibinfo {author} {\bibfnamefont
  {Maximina~H}\ \bibnamefont {Yun}},\ }\bibfield  {title} {\enquote {\bibinfo
  {title} {Conserved and novel functions of programmed cellular senescence
  during vertebrate development},}\ }\href@noop {} {\bibfield  {journal}
  {\bibinfo  {journal} {Development}\ }\textbf {\bibinfo {volume} {144}},\
  \bibinfo {pages} {106--114} (\bibinfo {year} {2017})}\BibitemShut {NoStop}%
\bibitem [{\citenamefont {Weber}(1969)}]{weber1969isolated}%
  \BibitemOpen
  \bibfield  {author} {\bibinfo {author} {\bibfnamefont {Rudolf}\ \bibnamefont
  {Weber}},\ }\bibfield  {title} {\enquote {\bibinfo {title} {The isolated
  tadpole tail as a model system for studies on the mechanism of
  hormone-dependent tissue involution},}\ }\href@noop {} {\bibfield  {journal}
  {\bibinfo  {journal} {Gen. Comp. Endocrinol.}\ }\textbf {\bibinfo {volume}
  {2}},\ \bibinfo {pages} {408--416} (\bibinfo {year} {1969})}\BibitemShut
  {NoStop}%
\bibitem [{\citenamefont {Niki}\ \emph {et~al.}(1982)\citenamefont {Niki},
  \citenamefont {Namiki}, \citenamefont {Kikuyama},\ and\ \citenamefont
  {Yoshizato}}]{niki1982epidermal}%
  \BibitemOpen
  \bibfield  {author} {\bibinfo {author} {\bibfnamefont {Kaoru}\ \bibnamefont
  {Niki}}, \bibinfo {author} {\bibfnamefont {Hideo}\ \bibnamefont {Namiki}},
  \bibinfo {author} {\bibfnamefont {Sakae}\ \bibnamefont {Kikuyama}}, \ and\
  \bibinfo {author} {\bibfnamefont {Katsutoshi}\ \bibnamefont {Yoshizato}},\
  }\bibfield  {title} {\enquote {\bibinfo {title} {Epidermal tissue requirement
  for tadpole tail regression induced by thyroid hormone},}\ }\href@noop {}
  {\bibfield  {journal} {\bibinfo  {journal} {Dev. Biol.}\ }\textbf {\bibinfo
  {volume} {94}},\ \bibinfo {pages} {116--120} (\bibinfo {year}
  {1982})}\BibitemShut {NoStop}%
\bibitem [{\citenamefont {Izutsu}\ \emph {et~al.}(1996)\citenamefont {Izutsu},
  \citenamefont {Yoshizato},\ and\ \citenamefont {Tochinai}}]{izutsu1996adult}%
  \BibitemOpen
  \bibfield  {author} {\bibinfo {author} {\bibfnamefont {Yumi}\ \bibnamefont
  {Izutsu}}, \bibinfo {author} {\bibfnamefont {Katsutoshi}\ \bibnamefont
  {Yoshizato}}, \ and\ \bibinfo {author} {\bibfnamefont {Shin}\ \bibnamefont
  {Tochinai}},\ }\bibfield  {title} {\enquote {\bibinfo {title} {Adult-type
  splenocytes of xenopus induce apoptosis of histocompatible larval tail cells
  in vitro},}\ }\href@noop {} {\bibfield  {journal} {\bibinfo  {journal}
  {Differentiation}\ }\textbf {\bibinfo {volume} {60}},\ \bibinfo {pages}
  {277--286} (\bibinfo {year} {1996})}\BibitemShut {NoStop}%
\bibitem [{\citenamefont {Nakajima}\ \emph {et~al.}(2019)\citenamefont
  {Nakajima}, \citenamefont {Tazawa},\ and\ \citenamefont
  {Shi}}]{nakajima2019unique}%
  \BibitemOpen
  \bibfield  {author} {\bibinfo {author} {\bibfnamefont {Keisuke}\ \bibnamefont
  {Nakajima}}, \bibinfo {author} {\bibfnamefont {Ichiro}\ \bibnamefont
  {Tazawa}}, \ and\ \bibinfo {author} {\bibfnamefont {Yun-Bo}\ \bibnamefont
  {Shi}},\ }\bibfield  {title} {\enquote {\bibinfo {title} {A unique role of
  thyroid hormone receptor $\beta$ in regulating notochord resorption during
  xenopus metamorphosis},}\ }\href@noop {} {\bibfield  {journal} {\bibinfo
  {journal} {Gen. Comp. Endocrinol.}\ }\textbf {\bibinfo {volume} {277}},\
  \bibinfo {pages} {66--72} (\bibinfo {year} {2019})}\BibitemShut {NoStop}%
\bibitem [{\citenamefont {Tsujioka}\ \emph {et~al.}(2017)\citenamefont
  {Tsujioka}, \citenamefont {Kunieda}, \citenamefont {Katou}, \citenamefont
  {Shirahige}, \citenamefont {Fukazawa},\ and\ \citenamefont
  {Kubo}}]{tsujioka2017interleukin}%
  \BibitemOpen
  \bibfield  {author} {\bibinfo {author} {\bibfnamefont {Hiroshi}\ \bibnamefont
  {Tsujioka}}, \bibinfo {author} {\bibfnamefont {Takekazu}\ \bibnamefont
  {Kunieda}}, \bibinfo {author} {\bibfnamefont {Yuki}\ \bibnamefont {Katou}},
  \bibinfo {author} {\bibfnamefont {Katsuhiko}\ \bibnamefont {Shirahige}},
  \bibinfo {author} {\bibfnamefont {Taro}\ \bibnamefont {Fukazawa}}, \ and\
  \bibinfo {author} {\bibfnamefont {Takeo}\ \bibnamefont {Kubo}},\ }\bibfield
  {title} {\enquote {\bibinfo {title} {interleukin-11 induces and maintains
  progenitors of different cell lineages during xenopus tadpole tail
  regeneration},}\ }\href@noop {} {\bibfield  {journal} {\bibinfo  {journal}
  {Nat. Commun.}\ }\textbf {\bibinfo {volume} {8}},\ \bibinfo {pages} {1--12}
  (\bibinfo {year} {2017})}\BibitemShut {NoStop}%
\bibitem [{\citenamefont {Mukaigasa}\ \emph {et~al.}(2009)\citenamefont
  {Mukaigasa}, \citenamefont {Hanasaki}, \citenamefont {Ma{\'e}no},
  \citenamefont {Fujii}, \citenamefont {Hayashida}, \citenamefont {Itoh},
  \citenamefont {Kobayashi}, \citenamefont {Tochinai}, \citenamefont {Hatta},
  \citenamefont {Iwabuchi} \emph {et~al.}}]{mukaigasa2009keratin}%
  \BibitemOpen
  \bibfield  {author} {\bibinfo {author} {\bibfnamefont {Katsuki}\ \bibnamefont
  {Mukaigasa}}, \bibinfo {author} {\bibfnamefont {Akira}\ \bibnamefont
  {Hanasaki}}, \bibinfo {author} {\bibfnamefont {Mitsugu}\ \bibnamefont
  {Ma{\'e}no}}, \bibinfo {author} {\bibfnamefont {Hiroshi}\ \bibnamefont
  {Fujii}}, \bibinfo {author} {\bibfnamefont {Shin-ichiro}\ \bibnamefont
  {Hayashida}}, \bibinfo {author} {\bibfnamefont {Mari}\ \bibnamefont {Itoh}},
  \bibinfo {author} {\bibfnamefont {Makoto}\ \bibnamefont {Kobayashi}},
  \bibinfo {author} {\bibfnamefont {Shin}\ \bibnamefont {Tochinai}}, \bibinfo
  {author} {\bibfnamefont {Masayuki}\ \bibnamefont {Hatta}}, \bibinfo {author}
  {\bibfnamefont {Kazuya}\ \bibnamefont {Iwabuchi}},  \emph {et~al.},\
  }\bibfield  {title} {\enquote {\bibinfo {title} {The keratin-related
  ouroboros proteins function as immune antigens mediating tail regression in
  xenopus metamorphosis},}\ }\href@noop {} {\bibfield  {journal} {\bibinfo
  {journal} {Proc. Natl. Acad. Sci. U.S.A.}\ }\textbf {\bibinfo {volume}
  {106}},\ \bibinfo {pages} {18309--18314} (\bibinfo {year}
  {2009})}\BibitemShut {NoStop}%
\bibitem [{\citenamefont {Le~Sage}\ \emph {et~al.}(2013)\citenamefont
  {Le~Sage}, \citenamefont {Arai}, \citenamefont {Glenn}, \citenamefont
  {DeVience}, \citenamefont {Pham}, \citenamefont {Rahn-Lee}, \citenamefont
  {Lukin}, \citenamefont {Yacoby}, \citenamefont {Komeili},\ and\ \citenamefont
  {Walsworth}}]{le2013optical}%
  \BibitemOpen
  \bibfield  {author} {\bibinfo {author} {\bibfnamefont {David}\ \bibnamefont
  {Le~Sage}}, \bibinfo {author} {\bibfnamefont {Koji}\ \bibnamefont {Arai}},
  \bibinfo {author} {\bibfnamefont {David~R}\ \bibnamefont {Glenn}}, \bibinfo
  {author} {\bibfnamefont {Stephen~J}\ \bibnamefont {DeVience}}, \bibinfo
  {author} {\bibfnamefont {Linh~M}\ \bibnamefont {Pham}}, \bibinfo {author}
  {\bibfnamefont {Lilah}\ \bibnamefont {Rahn-Lee}}, \bibinfo {author}
  {\bibfnamefont {Mikhail~D}\ \bibnamefont {Lukin}}, \bibinfo {author}
  {\bibfnamefont {Amir}\ \bibnamefont {Yacoby}}, \bibinfo {author}
  {\bibfnamefont {Arash}\ \bibnamefont {Komeili}}, \ and\ \bibinfo {author}
  {\bibfnamefont {Ronald~L}\ \bibnamefont {Walsworth}},\ }\bibfield  {title}
  {\enquote {\bibinfo {title} {Optical magnetic imaging of living cells},}\
  }\href@noop {} {\bibfield  {journal} {\bibinfo  {journal} {Nature}\ }\textbf
  {\bibinfo {volume} {496}},\ \bibinfo {pages} {486--489} (\bibinfo {year}
  {2013})}\BibitemShut {NoStop}%
\end{thebibliography}%

\appendix
\renewcommand{\thefigure}{S\arabic{figure}}
\setcounter{figure}{0}  

\onecolumngrid
\newpage
\section*{Appendix}

\section{S-parameter comparison with and without plastic supports (dishes)}
\begin{figure*}[th!]
    \centering
    \includegraphics[width=15.8cm]{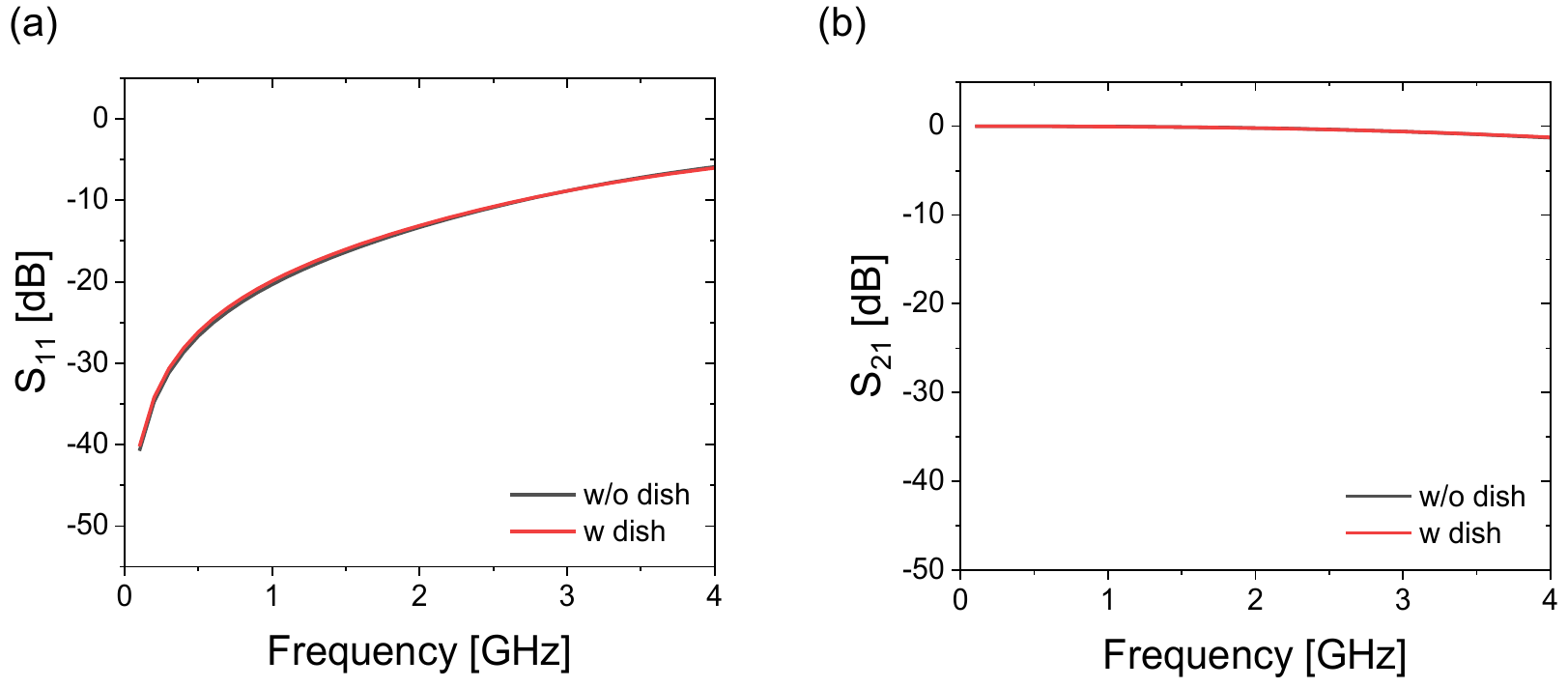}
    \caption{(a), (b) $S$-parameter comparison with and without a plastic dish on the chip device. 
    }
    \label{figS1}
\end{figure*}

\section{Convergence of the numerical S-parameter for the different mesh sizes}
\begin{figure*}[th!]
    \centering
    \includegraphics[width=15.8cm]{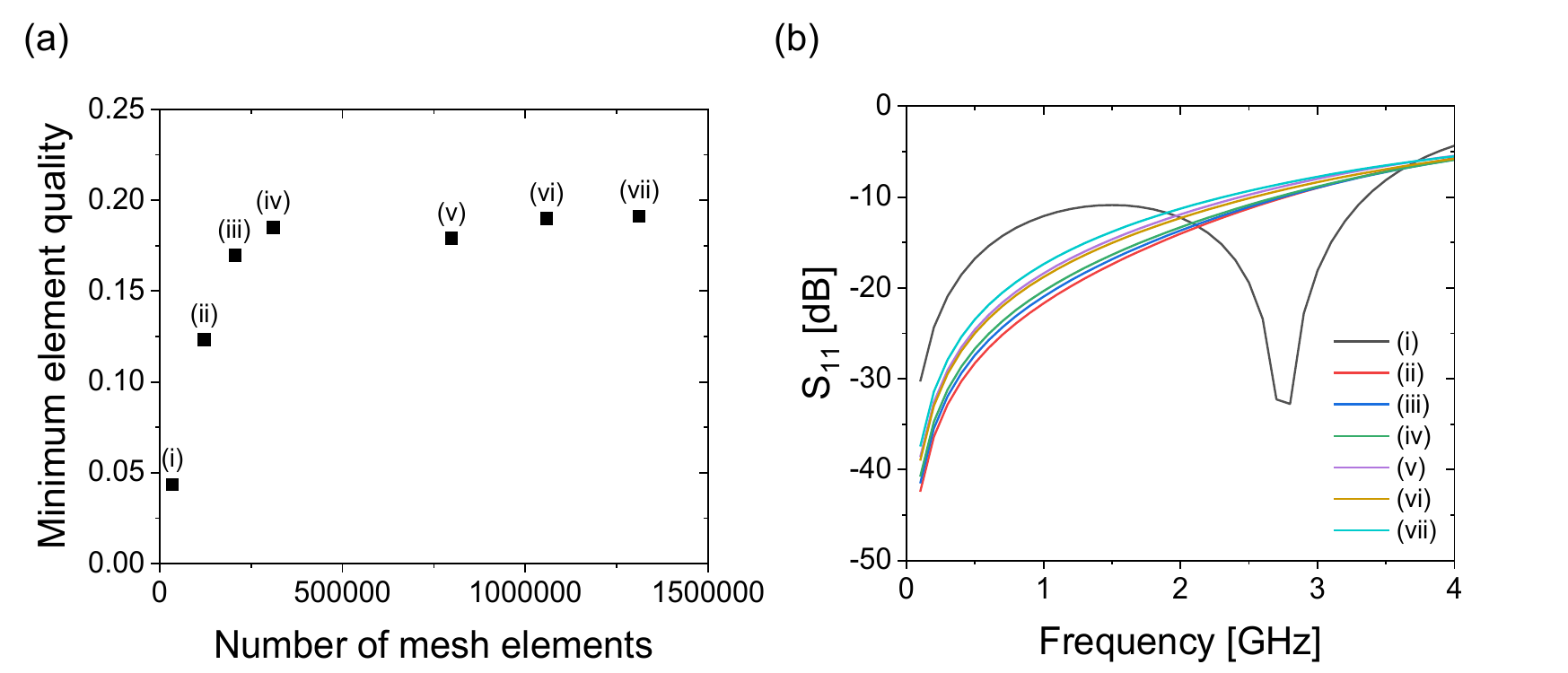}
    \caption{(a) Minimum element quality versus number of mesh elements in the simulations. In COMSOL, condition (i) is called `extremely coarse', (ii) `coarse', (iii) `normal', (iv) `fine', and (vii) `extremely fine'. The conditions (v) and (vi) are set to provide intermediates between (iv) and (vii) by varying maximum element growth rate. (b) Convergence of the $S_{\rm 11}$ spectrum for different mesh sizes for (i)--(vii). The extremely coarse simulation reveals a peak that disappears with decreasing mesh size, yielding approximately the same $S_{\rm 11}$ spectrum. 
    }
    \label{figS1-1}
\end{figure*}

\section{Parameter sweeps of the notch shaped antenna}

\begin{figure*}[bh!]
    \centering
    \includegraphics[width=15.2cm]{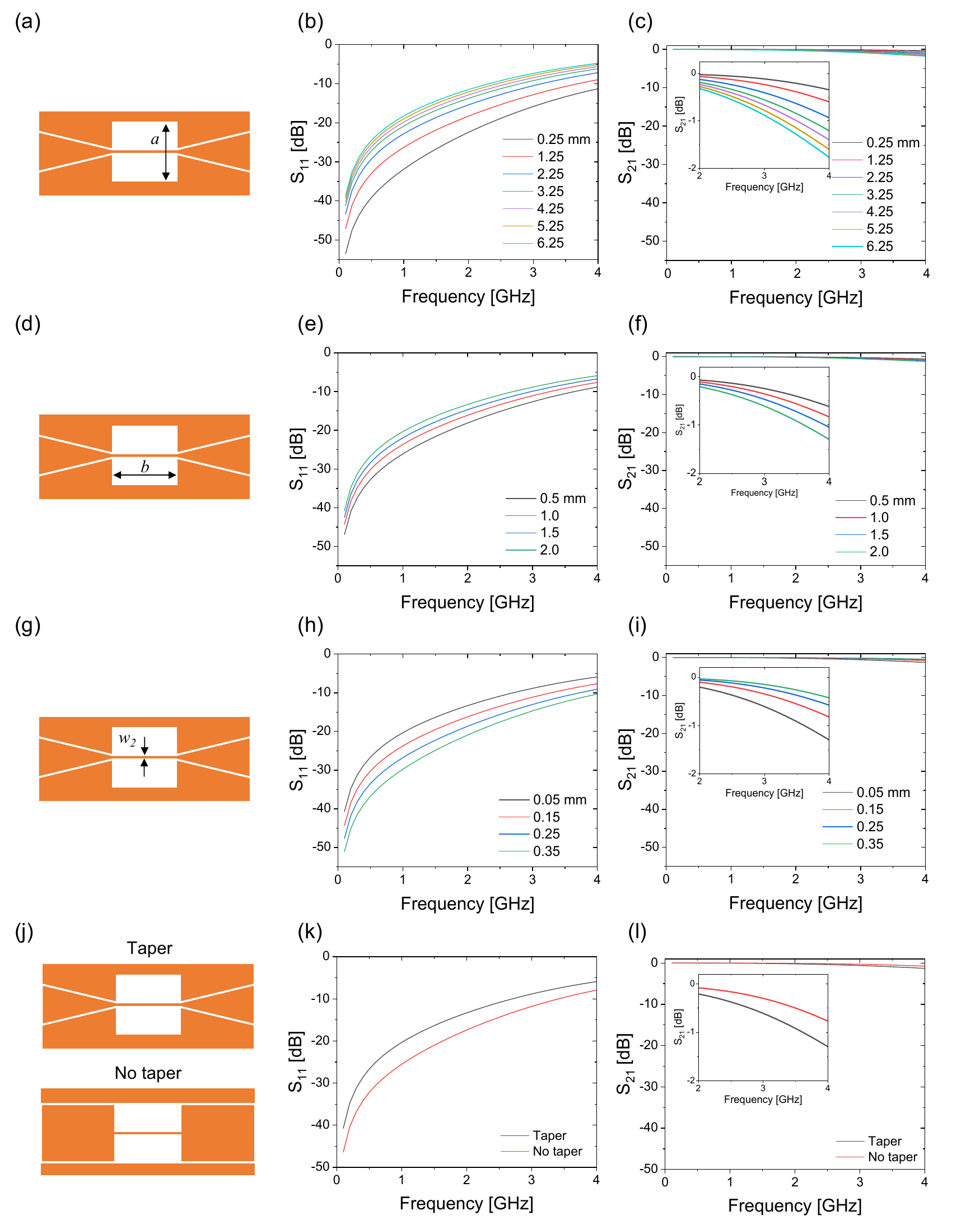}
    \caption{(a), (b), (c) Schematic diagrams of the notch-shaped geometry, with notch height $a$ and the corresponding $S_{\rm 11}$ and $S_{\rm 21}$ spectra, respectively. Inset: enlarged $S_{\rm 21}$ spectra at 2--4 GHz.
    (d), (e), (f) Schematic of the notch-shape geometry with a notch width of $b$ and the corresponding $S_{\rm 11}$ and $S_{\rm 21}$ spectra for different values of $b$. 
    (g), (h), (i) Schematic of the central thin wire with a width of $w_{2}$ in the notch shape and the corresponding $S_{\rm 11}$ and $S_{\rm 21}$ spectra. 
    (j) Schematics for notch-shape geometry with and without tapers. (k), (l) Simulated $S_{\rm 11}$ and $S_{\rm 21}$ spectra for taper and no-taper structures. 
    }
    \label{figS2}
\end{figure*}

\section{Comparison of notch shaped antenna with other antenna types on the S-parameters}

\begin{figure*}[th!]
    \centering
    \includegraphics[width=15.8cm]{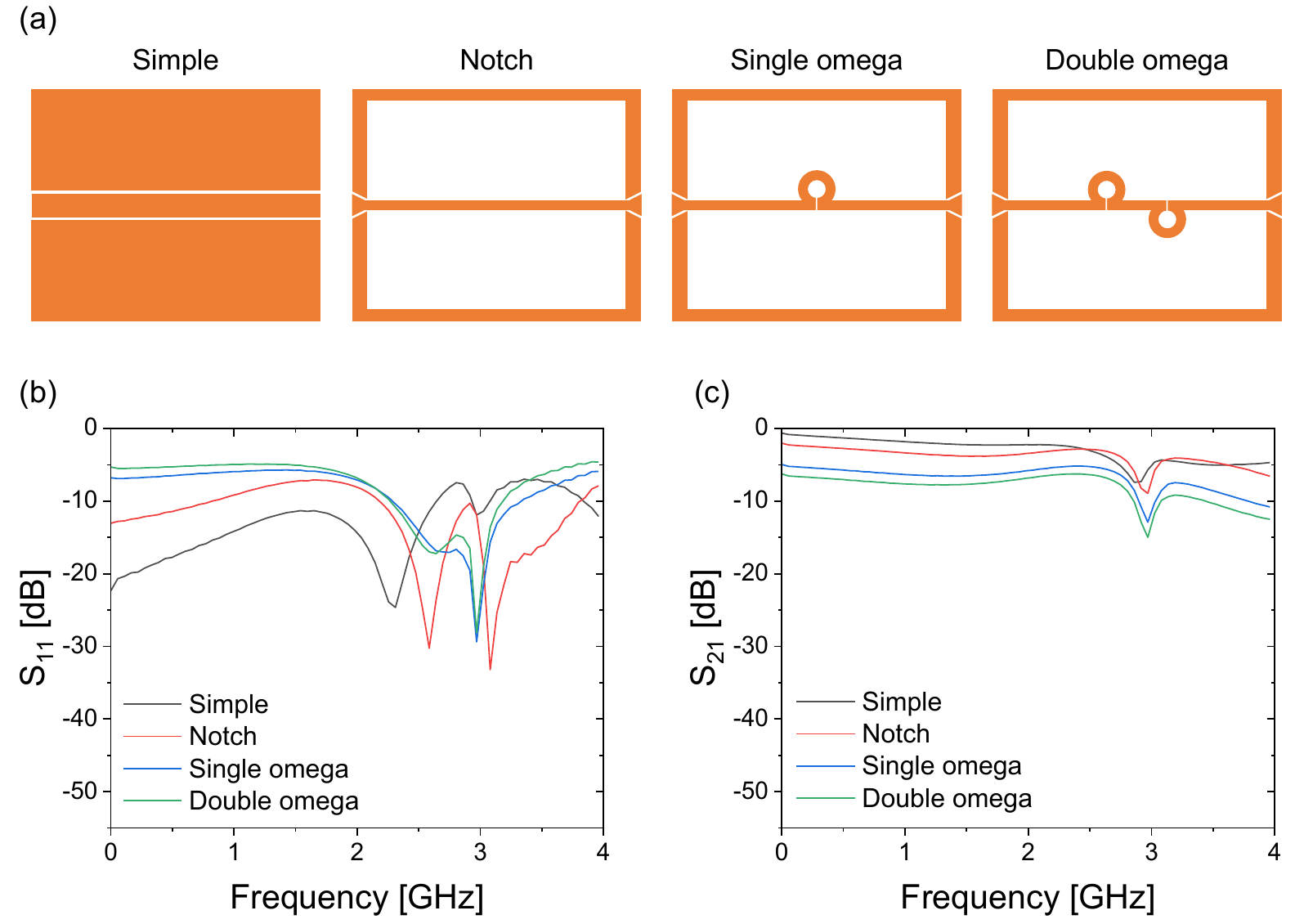}
    \caption{(a) Schematic drawings for different antenna structures: conventional coplanar (simple), notch (present device), non-resonant single omega (single omega), and non-resonant double omega (double omega). (b), (c) Experimental data of corresponding $S$-parameter spectra. $S_{\rm 11}$ and $S_{\rm 21}$ display one or two dips. These dips are absent in the $S$-parameter simulations (see Fig. 3 in the main text), indicating a cavity resonance by impedance-mismatched reflections at the copper junctions between the coverslip antenna and the PCB.}
    \label{figS3}
\end{figure*}

\newpage
\section{Application to the multi-well dishes and S-parameter characteristics of the conductive rubber connectors}

\begin{figure*}[th!]
    \centering
    \includegraphics[width=15.3cm]{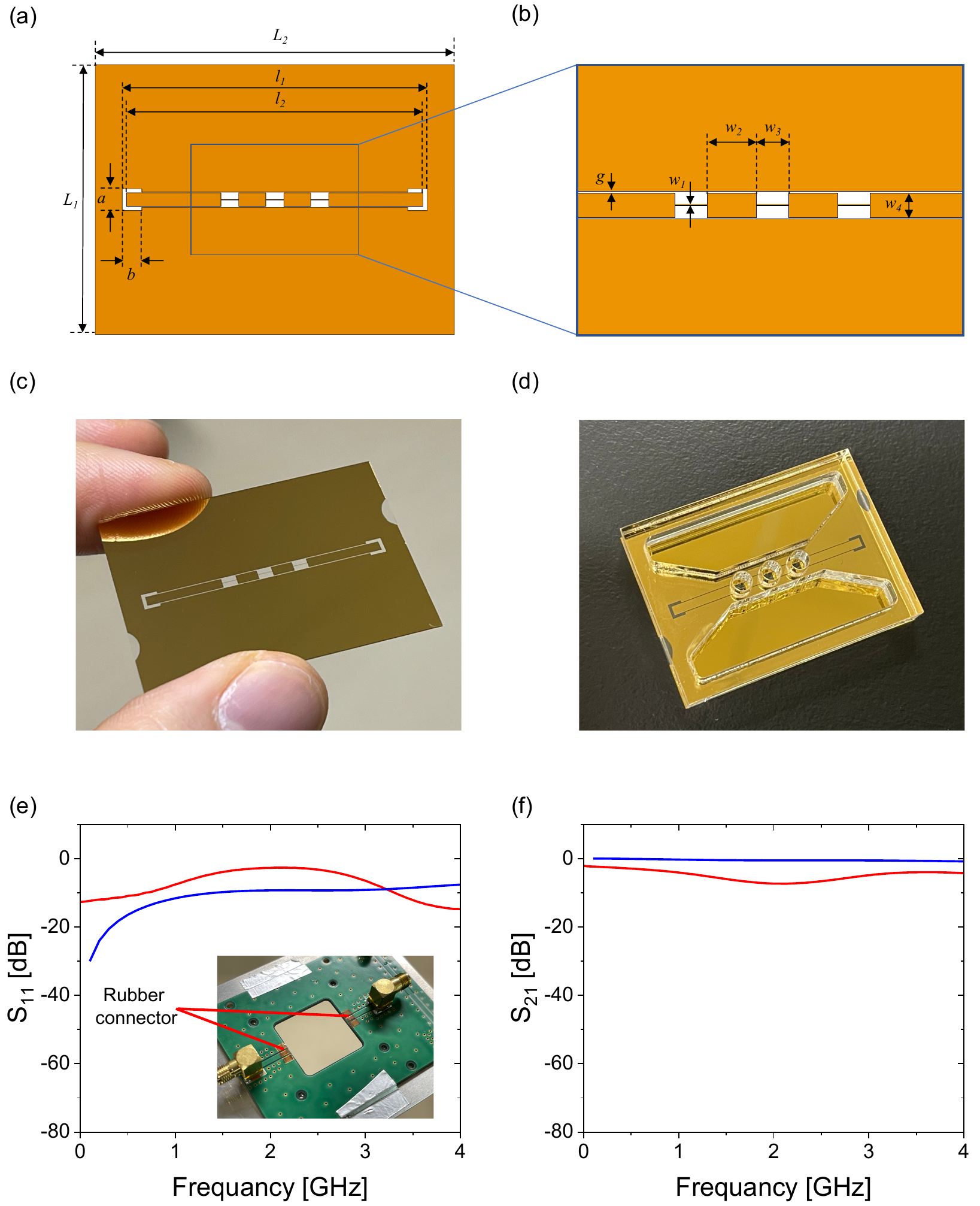}
    \caption{\Rev{(a), (b) Geometrical structure of the antenna coverslip for multiwell-dishes with dimensions as follows: $L_1 = 30 \ \si{\mm}$, $L_2 = 40 \ \si{\mm}$, $l_1 = 33.94 \ \si{\mm}$, $l_2 = 33 \ \si{\mm}$, $a = 2.44 \ \si{\mm}$, $b = 2.17 \ \si{\mm}$, $g = 0.1 \ \si{\mm}$,$w_1 = 50 \ \si{\um}$, $w_2 = 3.0 \ \si{\mm}$, $w_3 = 2.0 \ \si{\mm}$,  and $w_4 = 1.5 \ \si{\mm}$.} (c) Photograph of an antenna chip device, comprising three observation areas along the central transmission line. (b) Photograph of a multi-well dish. (d), (e) Experimentally measured (red) and simulated (blue) $S_{11}$ and $S_{21}$ spectra for the multi-well-plate chip device. An anisotropic conductive rubber connector (inset) is used instead of the spring fingers (Fig. 1(b) in the main text) to connect the device to the PCB, in which the finger-derived resonance peaks disappeared.}
    \label{figS4}
\end{figure*}

\newpage
\section{Microwave system and total gain}
\begin{figure*}[th!]
    \centering
    \includegraphics[width=13cm]{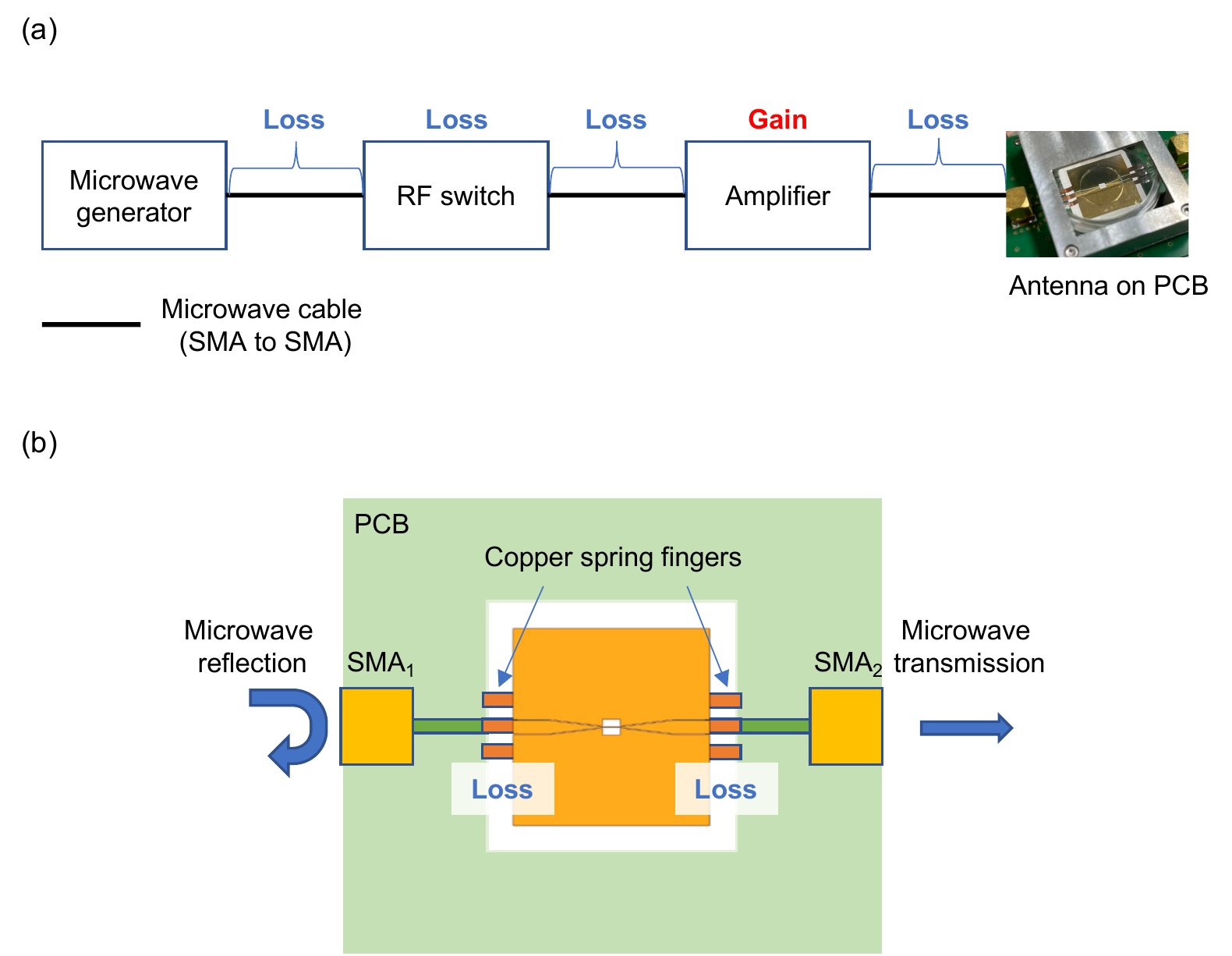}
    \caption{(a) Schematic of the microwave system from a microwave generator to an antenna on the PCB in our ODMR experiments. Microwaves are amplified by an amplifier while the microwave loss are caused by the insertion losses of SMA cables and microwave components.
    (b) Schematic diagram of the microwave reflection and loss on the PCB.
    }
    \label{figS5}
\end{figure*}

Figure~\ref{figS5} shows the schematic diagrams of the present microwave system. 
Table~\ref{tabel1} summarizes the losses and gains of the individual components including SMA cables (taken from their specification sheets) and resultant microwave power.
For the amplifier gain, we assumed a minimum gain of 40 dB considering the slight impedance mismatch between the cables and the antenna on PCB.
The loss of the copper spring fingers were calculated from $S_{11}$ and $S_{21}$ for the antenna on the PCB as provided in Fig. 3 in the main text.
The loss of the incident microwave power at the $\rm SMA_{1}$ on the PCB is provided by the following equation~\cite{vallabhapurapu2021fast}:
\begin{equation}
    \begin{aligned}
    Loss_{in} &= 10\log_{10}(1-10^{\frac{S_{11}}{10}}).\\ 
    \end{aligned}
    \label{eq:P_in}
\end{equation}
\noindent
$S_{21}$ is an indicator related to the microwave transmission on the PCB ($\rm SMA_{1}$ $\rightarrow$ antenna $\rightarrow$ $\rm SMA_{2}$).
Therefore, the insertion loss at the first copper fingers (between the $\rm SMA_{1}$ and the antenna) is provided by the following equation:
\begin{equation}
    \begin{aligned}
    Loss_{\rm SMA_{1} \rightarrow antenna} &= \frac{S_{21}}{2}.\\ 
    \end{aligned}
    \label{eq:P_trans}
\end{equation}
A factor of two is determined from the device symmetry. 
Note that the formulation of Eq. 2 is not affected by losses that we did not explicitly assume as it includes the entire loss from $\rm SMA_{1}$ to $\rm SMA_{2}$.
With these parameters, we find the total irradiation power in the notch shaped area to be 18.9 dBm in the present ODMR experiments (Sec. 3.2, 3.3 in the main text).

\begin{table}[th!]
  \caption{List of microwave power, loss and gain in the corresponding components. The final power of 18.9 dBm is used for the calculation.}
  \label{tbls1}
  \centering
  \begin{tabular}{wc{40mm}wc{25mm}wc{25mm}wc{30mm}}
	Components  & Loss or Gain [dB]  & Power [dBm]  & Notes \\ 
    \hline \hline 
    Microwave generator     &      &   $-7.0 $ & Take from Sec. 3.2, 3.3 \\
    RF switches 			&	$-3.5$     &   $-10.5$  & Total insertion loss	\\
    Microwave amplifier		&   $+40.0$	   &  29.5   & Amplifier gain  \\
    Microwave cables    	&	$-7.0$    	 &  22.5  &Total insertion loss \\
    Antenna reflection loss &	$-0.4$    & 22.1   &  $S_{11} = -11.0$, using Eq.~\ref{eq:P_in}\\
    Antenna insertion loss  &   $-3.2$    & 18.9 &  $S_{21} = -6.32$, using Eq.~\ref{eq:P_trans}\\
     \hline 
    Total                   &       &   $18.9$ [dBm]   & 
  \end{tabular}
  \label{tabel1}
\end{table}

\newpage
\section{Theoretical model of ODMR contrast}
In Sec. 3.2, we overlaid the theoretical curves of ODMR contrast on the plots of the experimental ODMR.
These curves are calculated using the following theoretical equation~\cite{dreau2011avoiding}:

\begin{equation}
    \begin{aligned}
    C &= \Theta\frac{f_{R}^2}{f_{R}^2 + \Gamma_{p}^{\infty}\Gamma_{c}^{\infty}(\dfrac{s}{1+s})^2} \\ 
    f_R &= \frac{\gamma |\boldsymbol{B}|}{\sqrt{2}}
    \end{aligned}
    \label{eq:contrasr model}
\end{equation}
where $\Theta$, $\Gamma_{p}^{\infty}$, $\Gamma_{c}^{\infty}$, and $s$ are the overall normalization factor, polarization rate at saturation, rate of optical cycles at saturation, and saturation parameter of the radiative transition, respectively. Here, $\Theta = 0.18$, $\Gamma_{p}^{\infty} = 7\times10^6~\si{s^{-1}}$, and $\Gamma_{c}^{\infty} = 9 \times 10^7~\si{s^{-1}}$ were obtained from Ref.~\citenum{dreau2011avoiding}. 
$s = 0.4$ was determined based on the saturation parameter of the present experimental setup using NA = 0.65 in comparison with that in Ref.~\citenum{fujiwara2020real-1}.
$\gamma$ = 2.8 MHz/G is the NV gyromagnetic ratio.
For $|\boldsymbol{B}|$, we used simulated $|\boldsymbol{B}|$ generated for the microwave input power defined with a peak voltage of 3.06 V (which effectively corresponds to 18.9 dBm, considering the microwave reflection in the simulation).

Note that $|\boldsymbol{B}|$ in Eq.~\ref{eq:contrasr model} should take account of the NV axis in a more rigorous manner.
$f_R$ should be proportional to the magnetic field perpendicular to the NV axis ($|\boldsymbol{B}|_\perp$)~\cite{hanson2008coherent, sasaki2016broadband, vallabhapurapu2021fast}, and $|\boldsymbol{B}|$ can be $\frac{|\boldsymbol{B}|_\perp}{\sin \theta}$, where $\theta$ is the angle between $\boldsymbol{B}$ (microwave magnetic field) and the NV axis, which we ignored in the present simulation.
This discrepancy appears as a relatively large variation of the ODMR depth and Rabi nutation frequency.

\section{Mcrowave magnetic field near the transmission line}

The four Rabi nutation frequency data and the corresponding $|\boldsymbol{B}|$ are summarized in Table~\ref{Rabi}.
These NDs were located at the center close to the transmission line.
The distance of NDs from the transmission line are also listed.
\begin{table}[h!]
  \caption{Rabi nutation frequency and $|\boldsymbol{B}|$ near the transmission line}
  \label{tbls1}
  \centering
  \begin{tabular}{ccccc}
	Distance [\si{\um}]  & $f_R$ [MHz] & $|\boldsymbol{B}|$ [G] \\
    \hline
    25  			&	3.81	& 1.93	\\
    36   			&	4.53	& 2.29	\\
    44  			&	5.67	& 2.86  \\
    57  			&	3.60	& 1.82  \\
     \hline
  \end{tabular}
  \label{Rabi}
\end{table}

\newpage
\Rev{\section{other data for the biological applications}}
\begin{figure*}[th!]
    \centering
    \includegraphics[width=15.8cm]{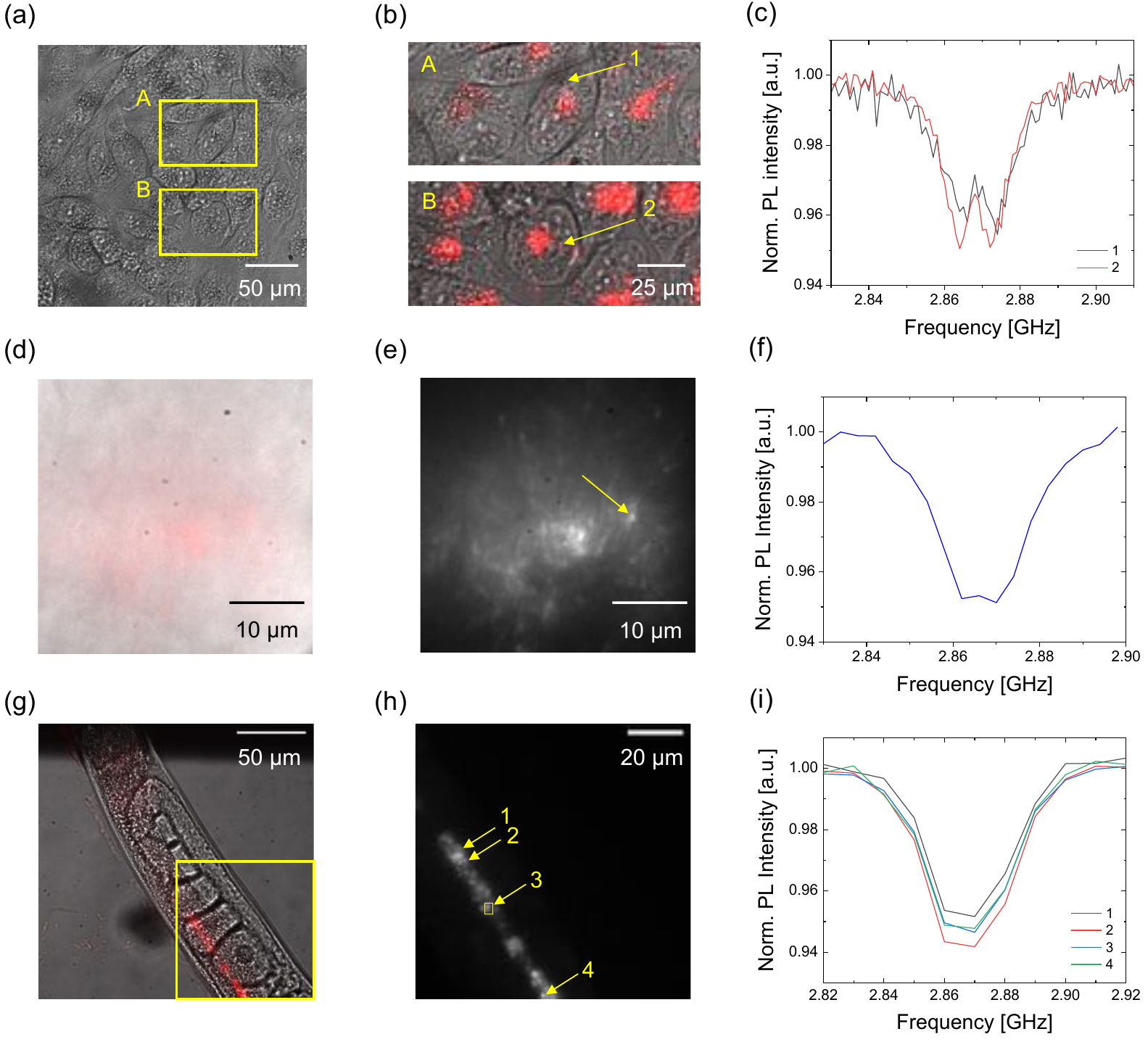}
    \caption{\Rev{(a) Bright-field image of living HeLa cells and (b) merged bright-field images with red fluorescence of ND-labeled HeLa cells inside yellow frames (A and B) in Fig.~\ref{figS9}(a). 
    (c) ODMR spectrum of the NDs, indicated by a yellow arrow in Fig.~\ref{figS9}(b). The NDs in other cells distributed over $\sim 100~\si{\um}$ of the cell in Fig. 6(b) showed comparable ODMR depths. (d) Merged bright-field image of a tissue fragment of the \textit{Xenopus} tadpole tail and (e) red fluorescence image of NDs inside the tissue. (f) ODMR spectrum of the NDs, indicated by a yellow arrow in  Fig.~\ref{figS9}(e). The NDs in the other tissue from other tadpoles exhibited ODMR depths similar to the data in the main text (Fig. 6(g)). 
    (g) Merged bright-field image of \textit{C. elegans} and (h) red fluorescence image of NDs corresponding to the yellow square in Fig.~\ref{figS9}(g). 
    (i) ODMR spectrum of the NDs, indicated by a yellow arrow in  Fig.~\ref{figS9}(h). NDs distributed over $100~\si{\um}$ in this worm presented ODMR depths within $\pm 0.01$, confirming the $|B|$ uniformity shown in the main text (see Figs.~6 (j)--(l)).} Note that the ODMR data for the worm were measured by a camera-based widefield ODMR method, as reported in Ref.~\cite{nishimura2021wide} (cells and tadpole tissue were measured by the confocal method).}
    \label{figS9}
\end{figure*}

\newpage
\section{Magnetic field distribution of coil and omega shaped antenna}
\begin{figure*}[th!]
    \centering
    \includegraphics[width=15.8cm]{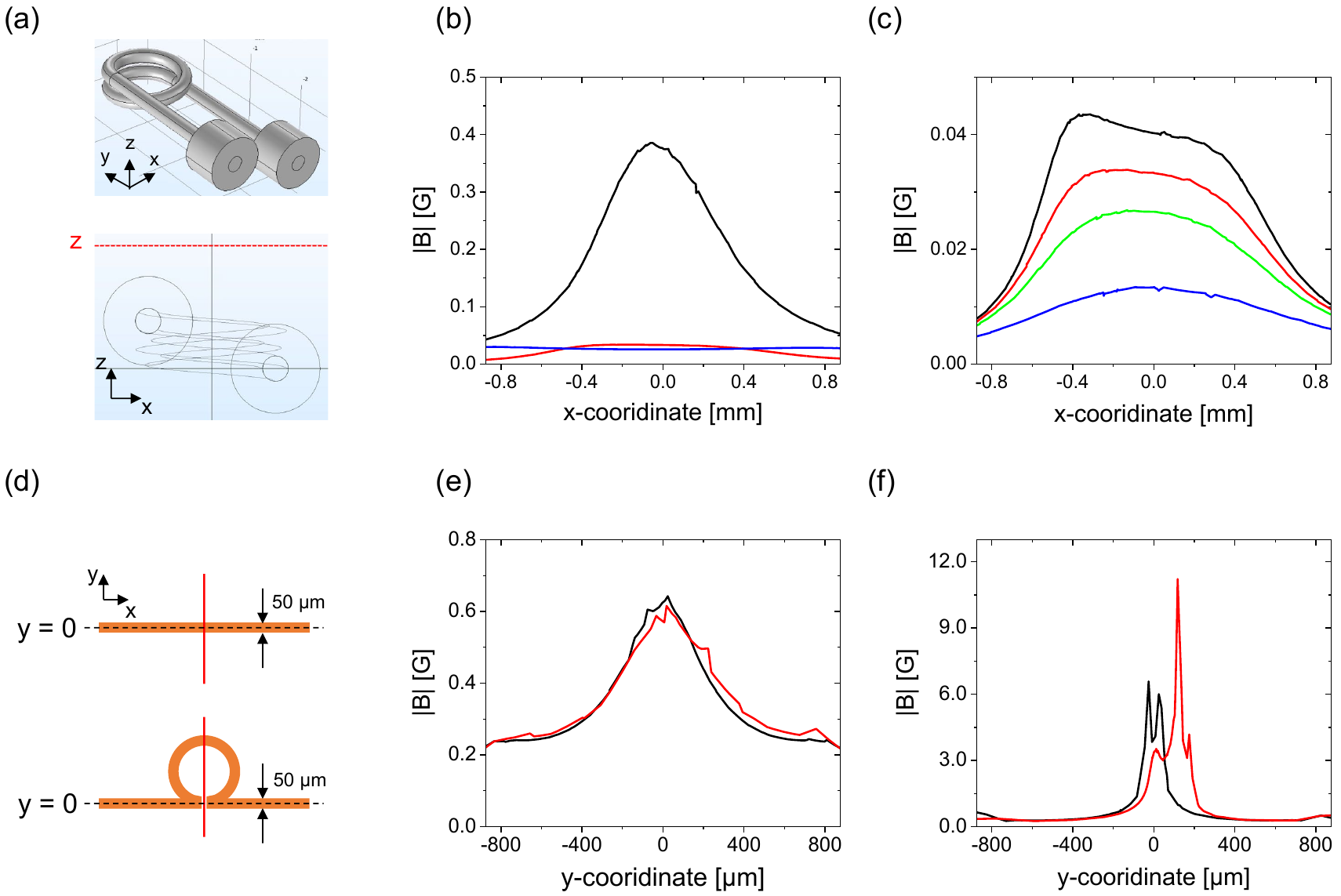}
    \caption{(a) 3D image of the 1.5-turn coil with the 0.1-mm-diameter wire (upper diagram) and 2D image from y direction (bottom). The dotted line in the z direction identifies the position of the cut for the $|\boldsymbol{B}|$ distribution at $f = 2.87$ GHz and $V = 3.06$ V (peak voltage), as determined by COMSOL. (b) Comparison of the simulated $|\boldsymbol{B}|$ distributions of the coil with diameters of 0.5 mm (black), 1 mm (red), and 2 mm (blue) at $z = 0.17$ mm above the coil. (c) Comparison of the simulated $|\boldsymbol{B}|$ distribution at $z = 0.1$ mm (black), $0.17$ mm (red), $0.25$ mm (green), and $0.5$ mm (blue) above the 1-mm-diameter coil.
    (d) Images of a straight line (original) and an 100-$\si{\um}$ diameter omega-shaped transmission line with our notch antenna. Red lines indicate the cross-section of the $|\boldsymbol{B}|$ distribution; (e) and (f) compare the simulated $|\boldsymbol{B}|$ distribution along the red cross section in Fig. S7(d).
    The black and red solid lines refer, respectively, to the straight and omega-shaped lines. (e) Results on the surface 0.17 mm (the coverslip thickness) above the antenna plane. (f) Results on the same plane as the antenna. }
    \label{figS6}
\end{figure*}

\newpage
\section{Live/Dead assay of tissue-derived stem cells (ASCs) on the antenna chip device}
\begin{figure*}[th!]
    \centering
    \includegraphics[width=14cm]{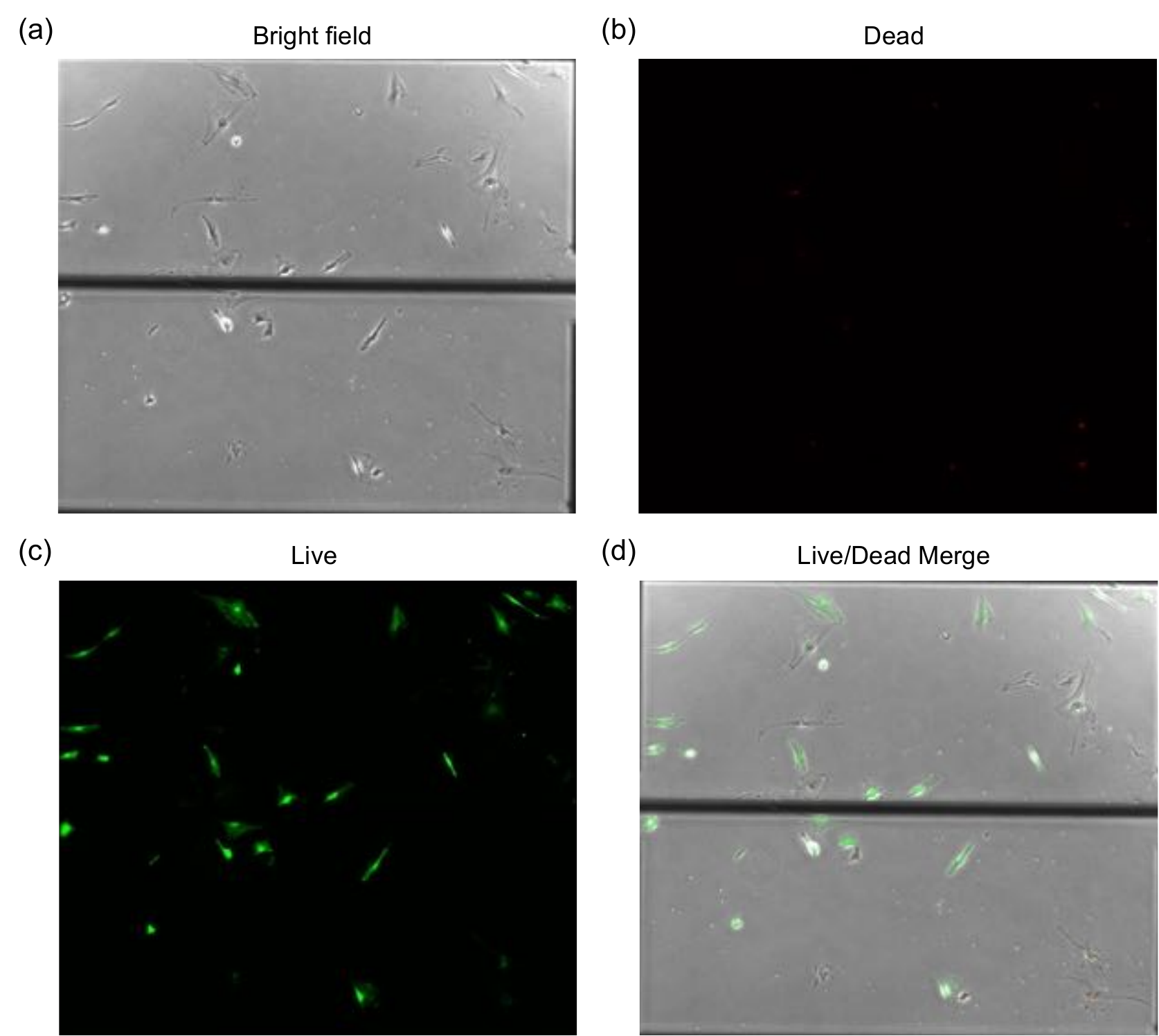}
    \caption{(a) Bright-field image of adipose tissue-derived stem cells (ASCs) on the antenna chip device. The central black line shows the thin transmission line of the microwaves. (b) Dead, (c) live stained, and (d) merged images of ASCs on the device, captured using the Live/Dead Viability/Cytotoxicity Kit for mammalian cells (Molecular Probes, L3224). Delicate stem cells such as ASCs can be cultivated in our chip devices. }
    \label{figS7}
\end{figure*}




\end{document}